\documentclass[12pt,preprint]{aastex}

\usepackage{graphicx}
\usepackage{txfonts}
\usepackage{natbib}
\begin{document}

\title{2D MHD and 1D HD models of a solar flare -- a comprehensive comparison of the results}

\author{R. Falewicz\altaffilmark{1}, P. Rudawy\altaffilmark{1}}
\affil{Astronomical Institute, University of Wroc{\l}aw, 51-622
Wroc{\l}aw, ul. Kopernika 11, Poland} \email{falewicz@astro.uni.wroc.pl}\email{rudawy@astro.uni.wroc.pl}

\author{K. Murawski\altaffilmark{2}}
\affil{Group of Astrophysics, UMCS, ul. Radziszewskiego 10, 20-031 Lublin, Poland}
\email{kmur@kft.umcs.lublin.pl}

\author{A. K. Srivastava\altaffilmark{3}}

\affil{Department of Physics, Indian Institute of Technology (Banaras Hindu University), Varanasi-221005, India}
\email{asrivastava.app@iitbhu.ac.in}

\begin{abstract}

Without any doubt solar flaring loops possess a multi-thread internal structure that is poorly resolved and there are no means to observe heating episodes and thermodynamic evolution of the individual threads. These limitations cause fundamental problems in numerical modelling of flaring loops, such as selection of a structure and a number of threads, and an implementation of a proper model of the energy deposition process. A set of 1D hydrodynamic and 2D magnetohydrodynamic models of a flaring loop are developed to compare energy redistribution and plasma dynamics in the course of a prototypical solar flare. Basic parameters of the modeled loop are set according to the progenitor M1.8 flare recorded in the AR10126 on September 20, 2002 between 09:21 UT and 09:50 UT. The non-ideal 1D models include thermal conduction and radiative losses of the optically thin plasma as energy loss mechanisms, while the non-ideal 2D models take into account viscosity and thermal conduction as energy loss mechanisms only. The 2D models have a continuous distribution of the parameters of the plasma across the loop, and are powered by varying in time and space along and across the loop heating flux. We show that such 2D models are a borderline case of a multi-thread internal structure of the flaring loop, with a filling factor equal to one. Despite the assumptions used in applied 2D models, their overall success in replicating the observations suggests that they can be adopted as a correct approximation of the observed flaring structures.

\end{abstract}

\keywords{Sun: flares --- Sun: X-rays, gamma rays --- Sun: corona --- Sun: chromosphere}

\section{Introduction}

Numerical models of solar flares are presently to a large extent consistent with basic observational properties of the observed events. They are widely utilized to understand the main physical processes occurring in solar flares, such as magnetic reconnection, generation of highly variable beams of high-energy non-thermal electrons (NTEs), interactions of the NTEs with flaring plasma, plasma dynamics and evolution, energy emission and many others. It is commonly assumed that the NTEs move along the magnetic field lines towards the chromosphere and photosphere and deposit their energy in cascaded Coulomb collisions with dense plasma. In the course of the bremsstrahlung process part of the energy carried by NTEs is emitted as hard X-rays (HXR) \citep[e.g.][]{Brown71}. Rapid deposition of energy by NTEs inevitably results in the localised accumulation of internal energy that is not radiated away by plasma in sufficiently short time scales. It causes an abrupt increase of the plasma temperature and its expansion dynamically upward into the corona in the course of the process called chromospheric evaporation \citep{Anton84,Anton99,Fis85b,Fis85c}. Highly variable emissions of plasma arise simultaneously over the entire electromagnetic spectrum from gamma-rays up to radio waves.

The temporal evolution of mass density, temperature, gas pressure, velocity, and other macro- and micro-parameters of the flaring plasma depends on various interrelated physical processes of energy deposition, transport, interactions of the plasma with magnetic fields, radiation processes, heat conduction and many others. The variations of the plasma parameters also have an influence on positions and sizes of the energy deposition volumes, the net energy budget of the whole events, plasma dynamics and time scales of variations of properties of solar flares \citep[e.g.][]{Den88, Den93, McTier99, Fal09b}. In particular, soft X-ray (SXR) emissions, mostly thermal in origin, are directly related to momentary distributions of the mass density (emission measure) and temperature of the expanding plasma along the whole flaring loop.

Solar flaring loops possess a multi-thread internal structure, \emph{i.e.} they consist of bundles of fine magnetic ropes, which are poorly resolved by contemporary instruments \citep{Aschwan2001, Aschwan2006}. There are no means to observe and to investigate heating episodes and thermodynamic evolution of the individual threads. These observational limitations cause fundamental problems in accurate numerical modelling of flaring loops, including selection of a relevant structure and number of threads as well as a proper spatial and temporal model of energy deposition in each individual thread.

Most of the former studies based on hydrodynamic simulations of solar flares are not able to account for a temporal evolution of the observed emissions \citep[e.g.,][]{Peres87, Mar91}. Many hydrodynamic simulations indicate that as a result of effective energy loss caused by thermal conduction and broadband thermal radiation, the high-density flaring plasma should cool rapidly, on time-scales of minutes. However, the observed SXR thermal emission of solar flares usually persists for hours \citep{McTier93, Jiang06}, powered by an extended phase of low-energy reconnections \citep[e.g.,][]{Fal11, Fal14} or being a direct result of a multi-thread internal structure of the flaring loops \citep{Warr10}. The relative shares of various mechanisms of a transfer and redistribution of the energy already deposited in the flaring loops are not clear, conduction and various kinds of waves can be present simultaneously in these magnetic structures. While the physical mechanisms and processes involved in abrupt heating of the flaring plasma are still not fully understood, a potentially fruitful method to study the flare physics is a comparison of the evolved observational features of the flaring transient plasma with the data provided by the numerical models of the flares, including relevant processes and physical mechanisms. In the general scenario, the hydrodynamic (HD) and magnetohydrodynamic (MHD) simulations of the impulsively heated solar active region loops, which are the fundamental building blocks of the corona, are important in order to understand plasma evolution, heating, and cooling \citep{Car94,Klim08}. In the ideal MHD approximation, several case studies were also proposed to understand the physics of plasma dynamics in the post-flare loops as well as active region magnetic arches \citep[][and references cited therein]{Sriv12,Sriv14}.

\begin{figure*}[t]
\begin{center}
\includegraphics[angle=0,scale=0.40]{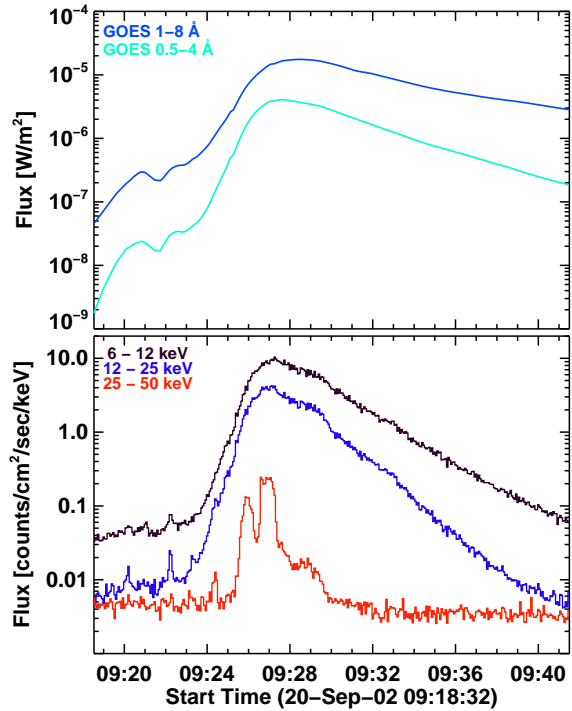}
\end{center}
\vspace{-1.0 cm}
\caption{Upper-panel: \emph{GOES} X-ray 0.5-4 \AA\ and 1-8 \AA\ light curves; lower-panel: \emph{RHESSI} light curves of three energy bands 6-12 keV, 12-25, and 25-50 keV, during the M1.8 \emph{GOES} class solar flare on September $\rm 20^{th}$, 2002.}
\label{Fig1}
\end{figure*}

\begin{figure*}[t]
\includegraphics[angle=0,scale=0.5]{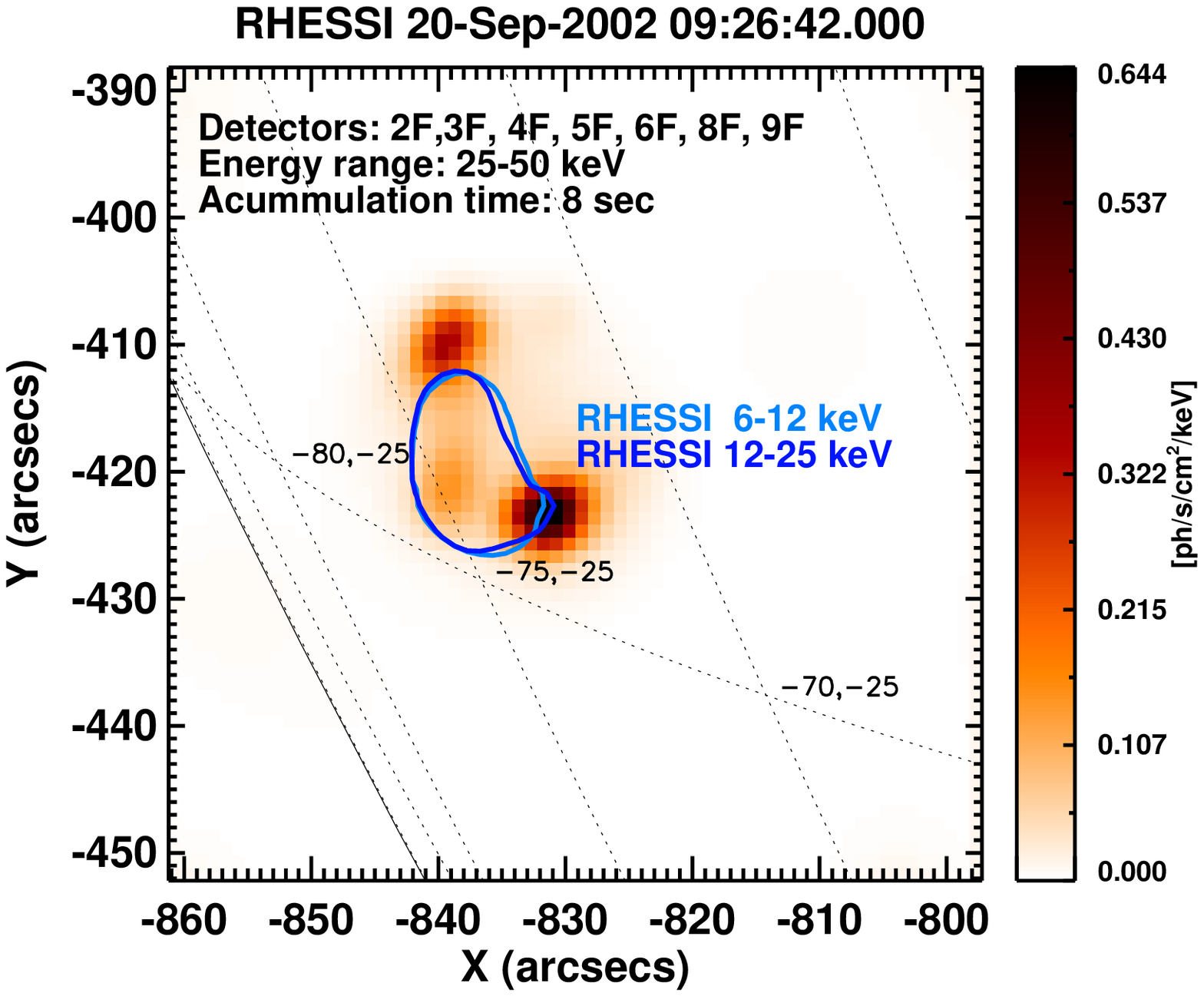}
\includegraphics[angle=0,scale=0.5]{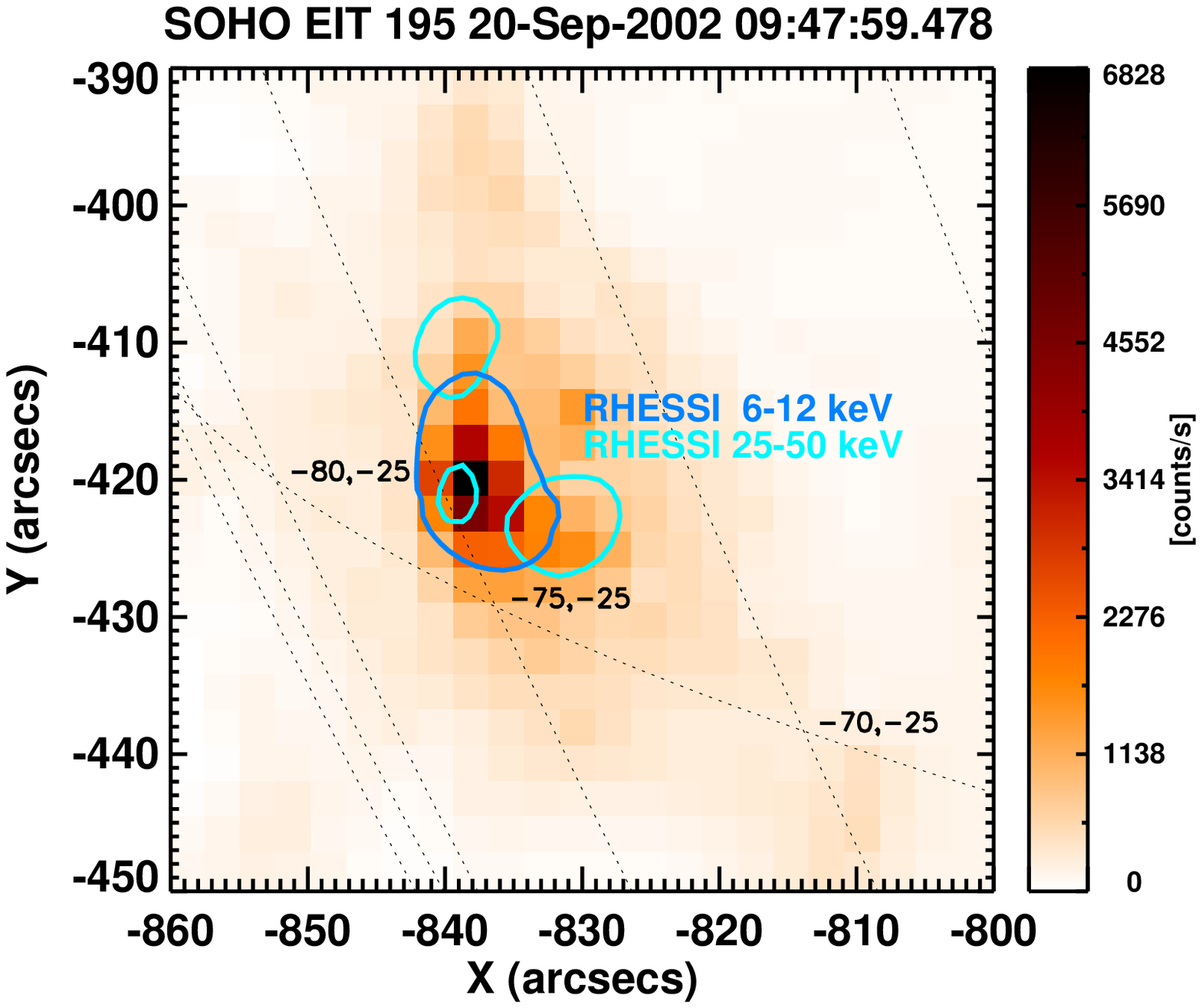}
\vspace{-0.5 cm}
\caption{Images of the M1.8 \emph{GOES} class solar flare on September 20, 2002.
Left panel: the image restored using the PIXON method in 25-50 keV (red temperature scale) and also 6-12 keV and 12-25 keV (contours) energy bands. The signal was accumulated between 09:26:42 UT and 09:26:50 UT. Right panel:
\emph{SOHO/EIT} 195 \AA\ image taken at 09:47:59 UT (red temperature scale) overplotted with the \emph{RHESSI} 6-12 keV and 25-50 keV images registered at 09:26:42 UT (contours).}
\label{Fig2}
\end{figure*}

In this work we present a set of 1D hydrodynamic (HD) and 2D magnetohydrodynamic (MHD) models of a prototypical flaring loop with the aim of comparing plasma parameters variations and flare evolutions. Initial physical and geometrical parameters of the flaring loop are set according to the observations of a progenitor M1.8 flare recorded in the AR10126 active region on September 20, 2002 between 09:21 UT and 09:50 UT (see Figs.~\ref{Fig1} and~\ref{Fig2}). The modeled loop was embedded into the realistic temperature model of the solar atmosphere that is smoothly extended below the photosphere. The non-ideal 1D models include thermal conduction and radiative losses of the optically thin plasma as energy loss mechanisms, while the non-ideal 2D models take into account viscosity and thermal conduction as energy loss mechanisms, which are currently implemented in the applied numerical code. The energy was delivered to the finite regions of the loop, which are selected, for various models, from the upper chromosphere up to the low corona.

In our work we investigate various scenarios of the evolution of a typical flaring loop, of reasonable values of the main physical parameters, such as half-length (\emph{i.e.} the distance between a foot and a top of a loop), apparent diameter, and energy delivered during a heating episode. The applied time evolution of the delivered energy flux mimics only average, typical variations of energy fluxes observed in solar flares, including a gradual growth phase, impulsive phase and gradual decay phase. We developed a broad set of models, for which the energy deposition regions are of a limited size and are located at a selected, fixed height. Such diversity of the position of the energy deposition volumes reflects various parameters of the NTEs beams, particularly various hardness of the NTEs energy spectra. The actual positions of the energy deposition volume, applied in each model are selected so that they overlap the position of the energy deposition region which is estimated with the use of the 1D HD model of a flare (see Falewicz et al., 2011). Thus, synthetic fluxes calculated in any spectral domain for the investigated flare could not be similar to real fluxes recorded for any real solar event, including also the progenitor flare.

This paper is organized as follows. The solar flare prototype is described in Section 2. Numerical models and applied assumptions are presented in Section 3. The obtained results are reported in Section 4. The paper is summarized by discussion of the results and conclusions in Sections 5 and 6.

\section{Observations}

As a prototype of a solar flare, which ensures reasonable geometrical and physical parameters of the modeled flaring loop, we use the M1.8 class solar flare in NOAA AR 10126 (S23E69) on September 20, 2002 at 09:21 UT, according to GOES catalogue (Fig.~\ref{Fig1}). Basic initial physical and geometrical parameters of the flare are estimated using the observational data recorded by the Reuven Ramaty High Energy Solar Spectroscopic Imager (\textit{RHESSI}), EUV Imaging Telescope (EIT) onboard Solar and Heliospheric Observatory (\textit{SOHO}), and Geostationary Operational Environmental Satellite (\textit{GOES}).

The \textit{RHESSI} satellite has nine coaxial germanium detectors, which record an X-ray emission from the full solar-disk in a wide energy range (3~keV--17~MeV) with high temporal and energy resolutions as well as with a high signal sensitivity. The collected data are used for the restoration of images and spectra in the X-ray band, which is crucial  for the investigations of non-thermal emission in solar fares. The images of the progenitor flare are reconstructed using the data collected with sub-collimators 2F, 3F, 4F, 5F, 6F, 8F, and 9F and integrated over 8 s periods with the use of the PIXON imaging algorithm, and with 1 arcsec/pixel size \citep{Met96,Hur02}. These images reveal SXR emission sources in soft (6-12 keV) and intermediate (12-25 keV) energy bands, co-spatial with the flare location. The images reconstructed in the energies above 25 keV show two footpoints and a loop-top source of a single flaring loop (cf., Fig.~\ref{Fig2}, left panel). The superposition of the two hard X-ray emission sources, visible in the RHESSI images restored in the 6-12 keV and 12-25 keV energy bands, suggests a loop-like structure of the flare.

Similar distribution of the emission is also seen in the images obtained with the \textit{SOHO}/EIT telescope \citep{Dela95} in a 195~\AA~band at 09:47:59 UT and 9:59:59 UT, after the impulsive phase of the flare (cf., Fig.~\ref{Fig2}, right panel).

The presence of the sub-resolution structures inside the observed main loop-like flaring structure is very likely, but due to limited spatial resolution of the available images, only the main geometrical properties of the visible main loop were estimated using a method proposed by \citet{Asch99}. The cross-section of the loop, $\rm S = 9.0 \pm 7.6 \times 10^{16} \, cm^2$, is estimated as a mean of areas of both foot points delimited by a flux level equal to 30\% of the maximum flux in the 25-50 keV energy range. The half-length of the loop, $\rm L_0 = 9.3 \pm 1.1 \times 10^8\, cm$, is estimated from a distance between the centers of gravity of the footpoints, seen in the RHESSI images taken in the 25-50 keV energy band, assuming a semi-circular shape of the loop. The volume of the loop then equals to $\rm V = 2~L_0~S$. The geometric parameters of the loop are evaluated under the assumption that an observational error of the position of the observed structure is of the order of 1 pixel.

\begin{table*}[t]
\caption{Main parameters of the analyzed solar flare used in the numerical simulations} 
\label{table:1} 
\centering 
\begin{tabular}{c c c c c c c c} 
\hline\hline 
Event     & \multicolumn{2}{c}{Time of}& \emph{GOES}&Active & S              & $L_0$        & $P_0$       \\ %
date      &start     & maximum         & class      &region &                &              &             \\
          & [UT]     &[UT]             &            & AR    &[$10^{17} {\rm cm}^2$]& [$10^{8}{\rm cm}$] & [${\rm dyn/cm}^2$]\\
 \hline
2002 Sep 20 & 09:21    & 9:28           & M1.8       &10126   &1.1            & 9.5          & 34.4        \\
\hline 
\multicolumn{8}{l}{{\footnotesize Note: $\rm S$ is the cross-section of the flaring loop; $\rm L_0$ the half-length of the loop; }} \\
\multicolumn{8}{l}{{\footnotesize $\rm P_0$ a gas pressure at the base of the transition region}} \\
\vspace{-0.6cm}
\label{Tab1}
\end{tabular}
\end{table*}

The X-ray emission of the investigated flare was also recorded with the \emph{GOES-8} X-ray photometers. \emph{GOES-8} was equipped with two photometers, continuously recording the full-disk integrated X-ray emissions in two energy bands 1-8 \AA\ and 0.5-4 \AA\ with 3 s temporal resolution \citep{Donnelly77}. The SXR (1-8 \AA) emission of the flare recorded by \textit{GOES-8} started at 09:18:15 UT, reached its maximum at 09:28:30 UT, and subsided at 10:00 UT. A harder X-ray emission recorded in (0.5-4 \AA) band started to increase at the same time as the soft X-ray emissions, but peaked one minute earlier at 09:27:30 UT.

\section{Numerical models of the solar flare}

A typical flaring loop is plausibly a 3D structure with a multi-thread internal structure, that is surrounded by a complex active region atmosphere and magnetic fields. Currently there are no means of detecting and observing heating episodes and thermodynamic evolution of the individual threads. The observational limitations cause fundamental problems in numerical modelling of flaring loops, including selection of a relevant structure and number of threads as well as a proper model of spatial and temporal energy deposition. Some crucial aspects of plasma dynamics in the flaring loops were already successfully modeled in a framework of various 1D HD models \citep[see e.g.,][and many others]{Maris85,Fis85a,Fis85b,Fis85c,Maris89,Serio91,Reale97,Siar09,Fal09a,Fal09b, Fal11,Fal14}. However, as a result of inevitable simplifications imposed by the 1D models, an adequacy of these models as well as a precision of the presentation of various involved physical processes is limited. Simplified 2D models, applying a continuous distribution of the physical parameters of the plasma across the loop and powered by a heating flux variable in time as well as along and across the loop, would be recognised as an extreme borderline case of a multi-thread internal structure of the flaring loop with a filling factor equal to one. Therefore such models might mimic to some extent the subtle structure and variations of the plasma parameters better than their 1D counterparts, revealing processes which are inherently absent in 1D models. However, the whole complexity in time and space of the processes occurring in the real multi-thread flaring loops is far beyond their scope.

In order to check general correctness and also to validate an applicability of such simplified multi-dimensional models, we investigate a broad set of models of the benchmark flare. This flare is simulated with the use of the modified NRL 1D HD code \citep{Mar82, Maris89} and of the 2D MHD FLASH code \citep[][see subsection 3.2 for details]{Fry00, Lee09} with the aim of making a comparison between the results of the 1D and 2D models.

\subsection{1D numerical models}

\begin{figure*}[ht]
\begin{center}
\includegraphics[angle=90,scale=0.4]{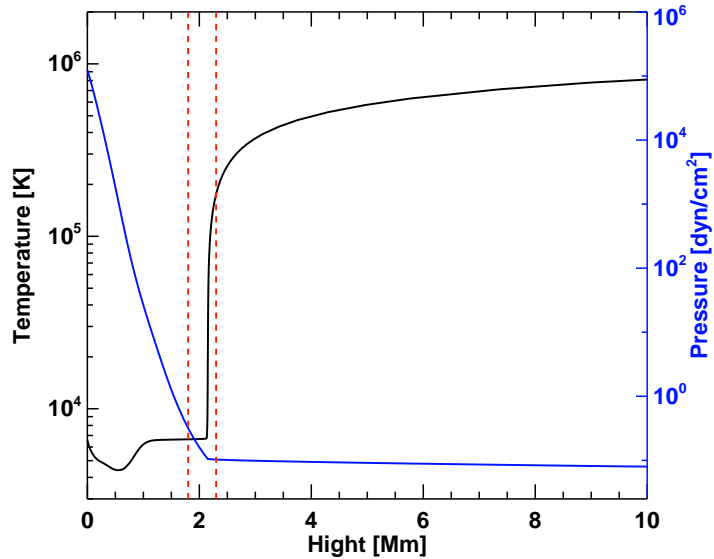}
\end{center}
\vspace{-0.8 cm}
\caption{Temperature (black line) and gas pressure (blue line) profiles of the applied temperature model of the solar atmosphere \citep[C7 of][]{Avrett08}, shown here for $0 \leq y \leq 10$  Mm region only. Vertical red-dashed lines illustrate the regions at which the heating specifically at $h=1.24$ Mm and $h=1.74$ Mm above the temperature minimum is implemented. See the main text for details.}
\label{Figatm}
\end{figure*}

As a 1D numerical code we adopt the modified NRL code \citep{Mar82, Maris89}, in which we implement the new radiative losses and heating functions, with a mesh of the radiative loss rates calculated for the given range of temperatures ($10^4$ - $10^8$ K) and densities ($10^8$ - $10^{14}$ $\rm cm^{-3}$) with the use of the CHIANTI code \citep{Dere97, Landi06}. The original NRL code is written for the isothermal chromosphere, which provides an insufficient amount of plasma available at the feet of the loop in the course of the massive chromospheric evaporation process. Thus, energy fluxes consistent with the medium or large solar flares result in a massive evaporation of this isothermal chromosphere and in a non-physical downward motion of the feet even if the chromosphere was set unrealistically thick. In order to solve the problem, likewise in the 2D model (cf., Section 3.2), we implement the realistic temperature model C7 of the solar atmosphere \citep{Avrett08}, that is smoothly extended downwards by the BP04 model of the solar interior \citep{Bah04} (Fig.~\ref{Figatm}). This realistic model ensures a big enough storage of plasma within the loop. All other aspects of the adopted NRL model are unchanged, among others this code does not include viscosity and diffusion, but it takes into account radiative losses of the optically thin plasma, which can be switched off if necessary, and the whole plasma is assumed to be fully ionized. Energy losses due to emission of the optically thick plasma are not taken into account, but \citet{Millig14} have already shown that the energy losses due to emission of the flaring loop in the optically thick chromospheric lines range only from 3.5 to 15 percent of the total energy deposited to the flaring loop (and even less during the impulsive phase solely). The modified code was thoroughly verified by comparison of the obtained results with those obtained with the original code \citep{Mar82,Maris89}.

We developed a set of models, where in a framework of a particular model the energy deposition region has a limited length and volume and is located at a selected, fixed height. The diversity of the positions of the energy deposition volumes reflects various properties of the NTEs beams, particularly various hardness of the NTEs energy spectra. The actual position of the energy deposition volume along the loop in the 2D models (cf., Section 3.2) is selected in the vicinity of the position of the energy deposition region that was estimated with the use of the 1D HD model of the flare \citep{Fal11}. In case of 2D models, we applied Gaussian distributions of energy deposition in both horizontal and vertical directions. In the horizontal direction a Gaussian distribution of energy deposition is qualitatively similar to the observed spatial distribution of the HXR emission at the feet in real flaring loops. Usually the HXR emission is highest in the centres of these emission sources and decreases toward the edges of the same sources, but an exact distribution of the emission is event-dependent. In the vertical direction a Gaussian distribution of energy deposition rates was adopted both in 1D and 2D models as a simplest model of an effective distribution of the NTEs precipitation depths. The implemented temporal evolution of a delivered energy flux mimics typical variations of energy fluxes observed in solar flares, with gradual growth, and impulsive and gradual decay phases. Such simplification is acceptable as we do not aim to build an exact model of any specific solar flare, but instead our goal is to compare 1D and 2D models of plasma dynamics and evolution of a flaring loop. Guided by this purpose we construct an initial, quasi-stationary pre-flare model of the flaring loop, using basic geometrical and thermodynamic parameters estimated from \emph{RHESSI} and \emph{GOES} data: half-length, cross-section, initial pressure at the base of the transition region, temperature, emission measure, mean electron density, and \emph{GOES}-class (Table~\ref{Tab1}). The initial pressure at the base is set in order to obtain a static plasma having a temperature at the top of the loop that is close to the observed one. The models are calculated for the periods lasting from the very beginning of the impulsive phase to far beyond the maximum of the soft X-ray emissions, lasting usually for about 300 s.

\subsection{2D numerical models}

Two-dimensional numerical models of the solar flare are calculated using the FLASH code \citep{Fry00, Lee09}. In this code a third-order unsplit Godunov and Riemann solvers, various slope limiters, and an adaptive mesh refinement are implemented. It was verified by numerical experiments that numerically induced flows remain at physically acceptable level for the minmod slope limiter and the Roe Riemann solver \citep{Toro06}, and for the refinement strategy based on controlling the gradients in mass density.

The simulation box is set as ($\rm -10$ Mm, $10$ Mm) $\rm \times$ $\rm -10$ Mm, $10$ Mm) which spans $20$ Mm along both the horizontal and vertical directions. In the Cartesian coordinate system, we represent the horizontal axis as $'$x$'$ and the vertical axis as $'$y$'$, and all plasma quantities remain invariant along the third direction ($z$), \emph{i.e.} $\partial/{\partial z}=0$. We impose boundary conditions by fixing in time all plasma quantities at all four boundary layers to their equilibrium values. The only exception is the bottom boundary, where we replace the equilibrium plasma quantities by implementing plasma heating which was modelled by gas pressure variations defined as:
\begin{equation}
\label{eq:presure_heat}
p(x,y,t)=p(x) \left( 1 + A_{p}\sum_{i=1}^{2} \exp\, \left(\frac{-(x-x_{i})^2}{w_{x}^2}\right) \exp\, \left(\frac{-(y-y_{0})^2}{w_{y}^2}\right) f(t) \right) \, .
\end{equation}
Here, $A_{\rm p}$ is the relative amplitude of the pressure signals, and $x_{\rm 1}$ and $x_{\rm 2}$ are the horizontal positions of these two signals, $y_{\rm 0}$ is their vertical position, and $w_{\rm x}$ and $w_{\rm y}$ are the horizontal and vertical widths of the signals. We set $A_{\rm p}=60$, $x_{\rm 1}=-x_{\rm 2}=6$ Mm, $w_{\rm x}=0.5$ Mm, $w_{\rm y}=1$ Mm and hold them fixed but allow $y_{\rm 0}$ to vary.

\begin{figure*}[h!]
\begin{center}
\includegraphics[angle=90,scale=0.4]{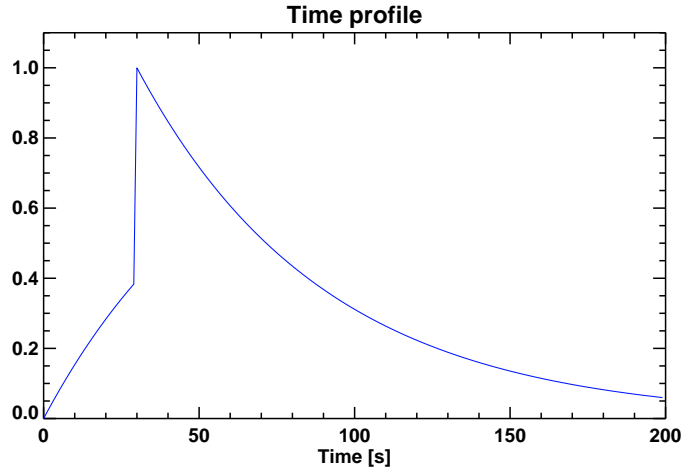}
\end{center}
\vspace{-1.0 cm}
\caption{Temporal profile $f(t)$ of the gas pressure variations applied at the bottom boundary. The $f(t)$ roughly mimics a typical time profile of X-ray emission of solar flares.}
\label{Fig3}
\end{figure*}

Henceforth, time is expressed in seconds and counted from the moment the energy is deposited at the feet of the flaring loop, that is at $t=0$ s. The symbol $f(t)$ denotes the temporal profile of the applied gas pressure signal, that is given by the following formula:
\[f(t) = \left\{
  \begin{array}{lr}
    1-\exp(-t/\tau) & : t < \tau_{max}\\
    \exp(-(t-\tau_{max})/\tau) & : t \geq \tau_{max} \, ,
  \end{array}
\right.
\]
where $\tau=60$ s is the growth and decay time of time profile. $\tau_{max}=30$ s is the time at which $f(t)$ reaches a maximum. The function $f(t)$ is shown in Fig.~\ref{Fig3}.

The FLASH code solves a set of non-ideal and non-adiabatic MHD equations, which are written in the non-conservative form:
\begin{equation}
\label{eq:MHD_rho}
{{\partial \varrho}\over {\partial t}}+\nabla \cdot (\varrho{\bf V})=0\, ,
\end{equation}
\begin{equation}
\label{eq:MHD_V}
\varrho{{\partial {\bf V}}\over {\partial t}} + \varrho\left ({\bf V}\cdot \nabla\right ){\bf V} =
-\nabla p+ \frac{1}{\mu}(\nabla\times{\bf B})\times{\bf B} + \varrho{\bf g}\ + {\bf F_{\rm {{\upsilon}_{\rm v}}}},
\end{equation}
\begin{equation}
\label{eq:MHD_p}
{\partial p\over \partial t} + \nabla\cdot (p{\bf V}) = (1-\gamma)p \nabla \cdot {\bf V} + {\bf F_{\rm {\kappa}}}\, ,
\hspace{3mm}
p = \frac{k_{\rm B}}{m} \varrho T\, ,
\end{equation}
\begin{equation}
\label{eq:MHD_B}
{{\partial {\bf B}}\over {\partial t}}= \eta {{\nabla}^2} {\bf B} + \nabla \times ({\bf V}\times{\bf B})\, ,
\hspace{3mm}
\nabla\cdot{\bf B} = 0\, ,
\end{equation}
where ${\varrho}$ is the mass density, ${\bf V}$ the flow velocity, ${\bf B}$ the magnetic field, $p$ the gas pressure, $T$ temperature, $\gamma=5/3$ the adiabatic index, ${\kappa}$ and ${{\upsilon}_{\rm v}}$ are the coefficients of thermal conduction and plasma viscosity, ${\bf F_{\rm {{\upsilon}_{\rm v}}}}$ the viscous term and ${\bf F_{\rm {\kappa}}}$ the thermal conduction term, ${\bf g}=(0,-g, 0)$ gravitational acceleration with its value $g=274$ $\rm m\:s^{-2}$, $m$ is mean particle mass, and $k_{\rm B}$ is the Boltzmann's constant.

For reasons of simplicity, the medium is assumed to be invariant along the $z$-direction ($\partial/{\partial z}=0$), and $z$-components of the preset background plasma velocity and magnetic field are equal to zero, \emph{i.e.} $V_{\rm z}=B_{\rm z}=0$. As a result of this assumption, Alfv\'{e}n waves are removed from the system which is able to guide the magnetoacoustic-gravity waves only \citep[e.g.][]{Nakar05}.

We assume that the solar atmosphere in its background state is in static equilibrium (${\bf V}={\bf 0}$) with a force-free magnetic field:
\begin{equation}
\label{eq:B}
(\nabla\times{\bf B})\times{\bf B} = \textbf{0}\, ,
\end{equation}
and the pressure gradient is balanced by the gravity force,
\begin{equation}
\label{eq:p}
-\nabla p + \varrho {\bf g} = \textbf{0}\, .
\end{equation}
%

Note, that the temperature profile of the initially static plasma determines directly the initial distribution of mass density and gas pressure (cf., Fig.~\ref{Figatm}). This includes an abrupt decrease of the mass density and pressure in a vicinity of the transition region. Using the ideal gas law and the $y$-component of the hydrostatic balance condition indicated by Eq.~(\ref{eq:p}), the equilibrium gas pressure and mass density are given as follows:
\begin{equation}
\label{eq:pres}
p(y)=p_{\rm 0}~{\rm exp}\left[ -\int_{y_{\rm r}}^{y}\frac{dy^{'}}{\Lambda (y^{'})} \right]\, ,\hspace{3mm}
\label{eq:eq_rho}
\varrho(y)=\frac{p(y)}{g \Lambda(y)}\, ,
\end{equation}
where
\begin{equation}
\Lambda(y) = \frac{k_{\rm B} T(y)}{mg} \, ,
\end{equation}
is the pressure scale-height, $p_{\rm 0}$ denotes the gas pressure at the reference level that we choose in the solar corona at $y_{\rm r}=10$ Mm and $y=0$ corresponds to the base of the photosphere.

\begin{figure*}[t]
\begin{center}
\includegraphics[angle=90,scale=0.6]{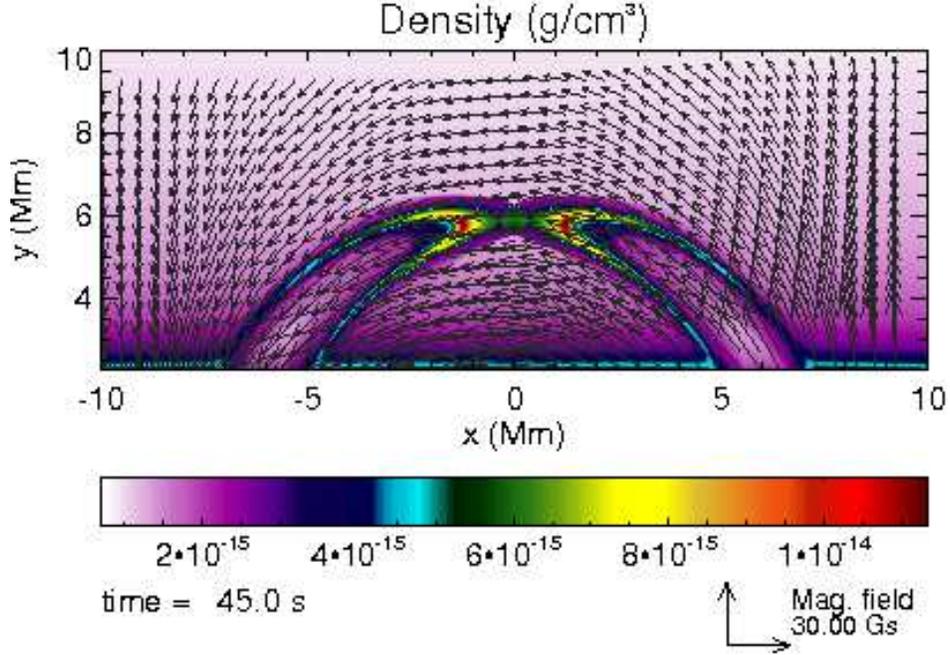}
\end{center}
\vspace{-0.8 cm}
\caption{Magnetic field of an active region (black arrows) overplotted on the mass density map of the evaporating plasma which ascend along the magnetic loop. The position of the loop's feet are defined by position of the regions of external energy deposition.}
\label{Fig5}
\end{figure*}

The force-free condition of Eq.~(\ref{eq:B}) is satisfied by the current-free magnetic field, \emph{i.e.} $\nabla \times {\bf B}={\bf 0}$. The solenoidal condition $\nabla \cdot {\bf B}={\bf 0}$ is automatically fulfilled if we introduce the following magnetic flux function $\textbf{A}=A~\textbf{z}$:
\begin{equation}
\textbf{B} = \nabla \times {\bf A} \left[\frac{\partial A}{\partial y}, -\frac{\partial A}{\partial x},0 \right] \, .
\end{equation}
Here we choose the $z$-component of \textbf{A} as:
\begin{equation}
A(x,y) = B_{\rm 0}{\lambda}_{\rm B} \: {\rm exp}{\left(-\frac{y}{\lambda_{\rm B}}\right)}\cos\left(\frac{x}{\lambda_{\rm B}}\right) \, ,
\end{equation}
where ${\lambda}_{\rm B}=L/\pi$ is the magnetic scale height with $L$ being the half of arcade width. We choose and hold fixed $L=19$ Mm, and specify the magnetic field $B_{\rm 0}$ by requiring that at the reference point ($x=0$ Mm, $y=10$ Mm) the Alfv\'en speed, $c_{\rm A}(x,y)= {\vert {\bf B}(x,y)\vert} / {\sqrt{\mu \varrho(y)}}$, is $10$ times larger than the speed of sound, $c_{\rm s}(y)=\sqrt{{\gamma p(y)} / {\varrho(y)}}$. For these settings $B_{\rm 0} \cong 11.4$ G and the magnetic field is predominantly horizontal around $x=0$ Mm while around ($x=11$, $y=1.75$) Mm it is oblique with $\pi/4$ angle to the solar surface. As a result the topology of the magnetic field mimics the closed and bent field lines of the active region magnetic arcade. The plasma parameter $\beta = P/(B^2/{2\pi})$ is equal to $0.016$ at $y=3$ Mm and $0.028$ at $y=5$ Mm, respectively. As $\beta<1$, the trajectories of the evaporating plasma are fully controlled by the magnetic field lines. The topology of the modeled flaring loop, \textit{i.e.} separation of the feet, half-length and diameter, are not pre-defined but results directly from the actual position and size of the heating regions (Fig.~\ref{Fig5}).

\begin{figure*}[t]
\begin{center}
\includegraphics[angle=90,scale=0.6]{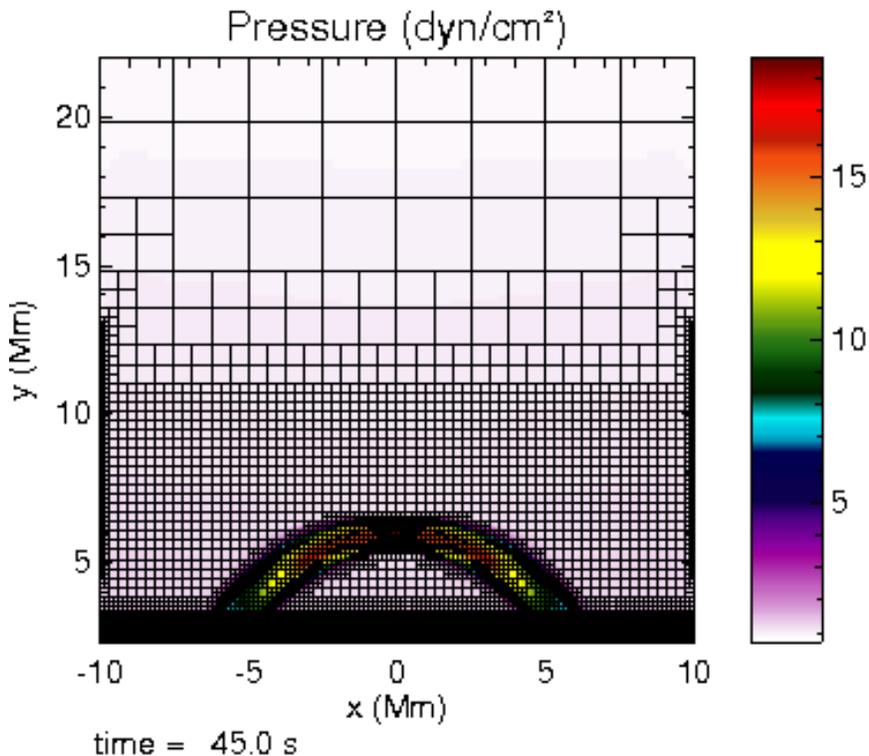}
\end{center}
\vspace{-0.1 cm}
\caption{Multi-scale structure of the calculation blocks, presented as a gas pressure map for $t=0$. The simulation box is set $20$ Mm wide along horizontal and vertical axes. The upper blocks have their size 2.5 Mm in each direction, while the finest one have edges 0.08 Mm only. The mesh was  refined at every few time steps, being particularly dense in the regions of the highest gradients of mass density. A color-coded gas pressure map of the evaporating plasma is overplotted for the comparison of the sizes of the whole mesh and modeled flaring loop only.}
\label{Fig6}
\end{figure*}

A complex, multi-scale structure of the adaptive mesh is applied, which has a finest spatial size in a vicinity of steepest profiles of the mass density and in the region of interest which is chosen below the altitude $y=11$ Mm (Fig.~\ref{Fig6}). As every block is divided into $8 \times 8$ numerical cells, on the first level of grid refinement a spatial resolution is equal to 2.5 Mm, and is refined gradually into $2 \times 2$ twice smaller numerical cells, forming a non-uniform numerical grid. The resulting grid is coarse in the solar corona, but very fine in the low corona and the chromosphere with the finest spatial resolution of $0.08$ Mm (Fig.~\ref{Fig6}).

A transient energy of the solar flare is delivered to the flaring loop by the beams of high-energy NTEs which deposit their energy in the dense plasma, usually in the vicinity of the transition region, in the chromosphere, and sometimes even in the photosphere. The local depositions of energy are defined and modeled by a variable in time, volumetric increase of the plasma pressure, launched in a finite volume of the magnetic structure at the bottom boundary of the numerical box, see Eq.~\ref{eq:presure_heat} and cf., Fig.~\ref{Fig6}. The pressure signal is of Gaussian shape in both horizontal and vertical directions and its amplitude, $A_{\rm p}$, is 60 times higher than the local ambient pressure. Temporal variations of the pressure mimic to a typical time profile of the X-ray emission of a solar flares (Fig.~\ref{Fig3}). However, the amount of the  delivered energy depends on actual physical parameters of the plasma. It is equal to 10~900 erg ${\rm cm}^{-3}$ at the height of $h=1.24$ Mm above the temperature minimum of the model atmosphere and is equal to only 2~685 erg ${\rm cm}^{-3}$ at $h=1.74$ Mm. While the actual deposited energy depends both on the parameters of the NTE beams as well as on the local plasma characteristics, we investigate a set of models, where the altitudes of the energy deposition regions vary between $h=1.2$ Mm and $h=1.8$ Mm above the temperature minimum with the spatial step of $\Delta h=0.1$ Mm. This is in accordance with our former results of our previous investigation of the same flare \citep{Fal11}, in which the height of the heating region was defined as a position of the maximum of the heating caused by NTEs and estimated using Fisher$'$s approximation \citep{Fis89}.

\begin{deluxetable}{|c|c|c|c|ccc|ccc|ccc|}
\rotate
\tabletypesize{\scriptsize}
\tablecolumns{13}
\tablewidth{0pt}
\renewcommand{\arraystretch}{1.2}
\tablecaption{Representative values of plasma temperatures and mass densities in the 1D and 2D models of the flaring loop.
\label{table:results1}}
\startdata
\hline
\hline
\multicolumn{2}{|c|}{\textsc{Model}} & \textsc{HEDA} & \textsc{Time} & \multicolumn{3}{c|}{\textsc{Axial part of the loop 1.5 Mm above TR} } &  \multicolumn{3}{c|}{\textsc{Ambient layer of the loop 1.5 Mm above TR}} & \multicolumn{3}{c|}{\textsc{Top of the loop}}\\
\cline{5-7} \cline{8-10} \cline{11-13}
\multicolumn{2}{|c|}{}&[Mm] &[s] & \textsc{Temperature} & \textsc{Density}   & \textsc{Plasma bulk}  & \textsc{Temperature}  & \textsc{Density}     & \textsc{Plasma bulk}   & \textsc{Temperature}   & \textsc{Density}    & \textsc{Plasma} \\
\multicolumn{2}{|c|}{}&  &   & [MK]        &[$10^{-14}\, g/cm^{3}$] & \textsc{motion}       & [MK]        &[$10^{-14}\, g/cm^{3}$]  & \textsc{motion}        &  [MK]         &[$10^{-14}\, g/cm^{3}$] & \textsc{state} \\
\hline \hline
   &       &     & 100 & 1.3 & 1.6 & up & 0.02 & 16.0 &	up & 1.3 & 0.5	& compression \\
\textsc{2D} & \textsc{IMHD}  & 1.24 &130 & 1.1 & 1.3 &	up & 0.02 & 5.0   & up & 3.5 & 4.0	& compression \\
   &      &      &200 &  1.2 & 2.0 & down&0.02 & 5.0 &	up & 1.0	 & 0.2 & expansion \\
\cline{2-13}
   &      &	    &90 &  0.1 & 6.0 &	up	&0.03 & 2.0   & up & 1.5 & 0.6 & compression \\
   & \textsc{VDC}  &  1.24 &110 & 0.1 & 5.0 &	up	&0.03 &	4.0  & up & 4.0 & 2.0 &	compression \\
   &      &     &200 & 0.2 & 5.0 &down &0.03 &	1.6  & up & 1.8 &1.0 & expansion \\
\hline
            &               &     & 50  &2.5  &6.6 &	up &       &      &    &	2.0 &1.2 &	compression \\
\textsc{1D} & \textsc{NRL}  & 1.24 & 150 &3.7  &12.0 &	up &       &      &    & 	5.2	&14.0 &	compression \\
            &               &     &250 &2.1  &18.0 &	down&      &      &    &	3.2	&10.0 &	expansion \\
\hline
\hline 
            &               &     &30 & 10.0 & 0.06 & up  & 0.1 & 0.32 &	up & 0.5 & 0.14	& not affected \\
\textsc{2D} & \textsc{IMHD} & 1.74 &90 & 20.0 & 0.10  &down & 7.9 & 0.03 &   up & 1.8 & 0.80	& compression \\
            &               &     &200 & 10.0 & 0.10& down& 3.1  & 0.03 &	down & 1.8& 0.40 & expansion \\
\cline{2-13}
            &               &	  &30 & 2.1 & 0.25 &	up  &0.5 & 0.31   & up & 0.5 & 0.14 & not affected \\
            & \textsc{VDC}  & 1.74 &70 & 10.0 & 0.31 &  down  &1.0 &	0.20  & down & 2.5 & 1.00 &	compression \\
            &              &      &200 & 4.5 & 0.16 &  down &1.2 & 0.16   & down & 1.4 & 0.45 & expansion \\
\hline
            &               &     &50  & 5.6  &7.6 & not affected &       &      &    & 10.6 & 1.8 & not affected \\
\textsc{1D} & \textsc{NRL}  & 1.74 & 150 & 7.1  &41.0 & up &       &      &            & 11.5 & 26.0 & compression \\
            &               &     &250 & 4.1  &48.0 & down &       &      &           & 7.3 & 30.0 & expansion \\
\hline
\enddata
\vspace{-0.6 cm}
\tablecomments{\scriptsize IMHD--2D ideal MHD approximation; VDC--2D MHD model with viscosity, diffusivity and thermal conduction; NRL--1D HD model with viscosity and thermal conduction, HEDA--height of the energy deposition area above the temperature minimum.}
\label{Tab2}
\end{deluxetable}

\section{Results of the numerical modelling}

In order to investigate the importance of various energy transfer and redistribution mechanisms acting in a flare, we compare the ideal MHD models with their non-ideal counterparts. The latter include viscosity, magnetic diffusivity and thermal conduction as energy transfer processes. The 1D models which are solved with the use of the NRL HD code, with included thermal conduction and radiative losses of the optically thin plasma are called \emph{NRL} models. All 2D models are solved with the use of a FLASH code. The models based on ideal MHD equations are henceforth called IMHD models. In the IMHD models energy is transported only by magnetoacoustic-gravity waves mixed with upflows and downflows. The 2D models, which take into account non-ideal and non-adiabatic effects such as viscosity, diffusivity, and thermal conduction, are nicknamed \emph{VDC} models. As radiative losses are not implemented in the FLASH code they are not present there. However, analyses of 1D models reveal that the energy losses due to radiation of the modelled loop are a minor factor. \citet{Millig14} have already shown that the energy losses due to emission of the flaring loop in the optically thick chromospheric lines range from a few, up to fifteen percent of the total energy deposited to the flaring loop only.

We perform a number of parametric studies for energy deposition levels which are located around the transition region (TR). With respect to similarities between various resulting models, the whole set of the developed models could be divided into two categories: (i) with the heating region located in the upper chromosphere, just below the TR, that is below the level $h=1.6$ Mm above the temperature minimum, and (ii) with the heating region located just above the TR. The basic properties of all models having heating regions, \textit{i.e.} energy deposition volumes below the TR, are roughly similar. Similarly, the basic properties of all models having a heating region above the TR are also roughly similar. Therefore, in order to shorten our discussion, only two representative models are presented in the forthcoming parts of this paper in more detail. For the first one the energy deposition layer is located at $h=1.24$ Mm above the temperature minimum and thus below the TR, while the second one has its energy deposition layer located at $h=1.74$ Mm above the temperature minimum and thus above the TR. Representative properties of the models are summarized in Table~\ref{Tab2}.

\begin{figure*}[ht!]
\includegraphics[angle=90,width=7.6 cm]{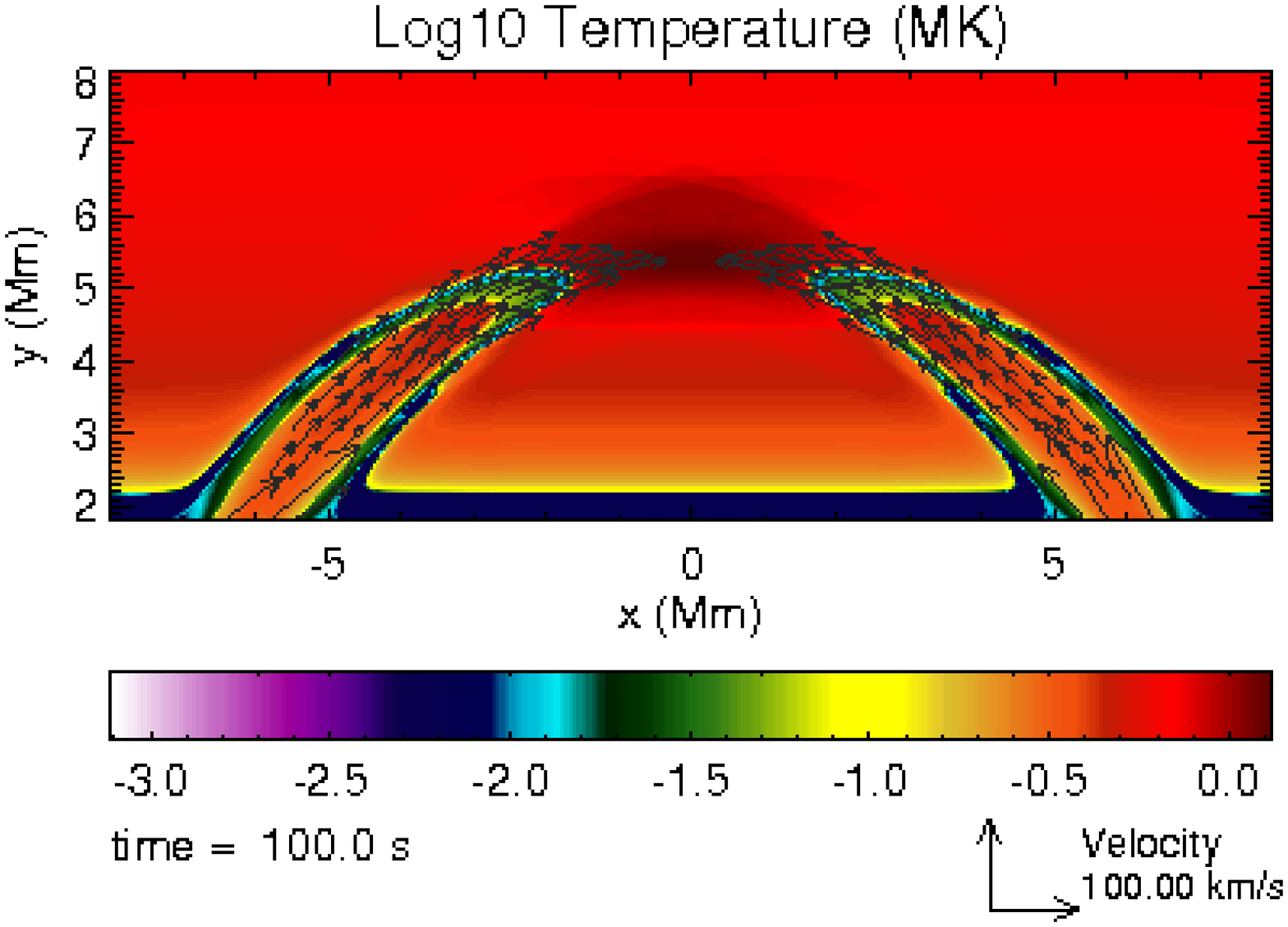}
\includegraphics[angle=90,width=7.6 cm]{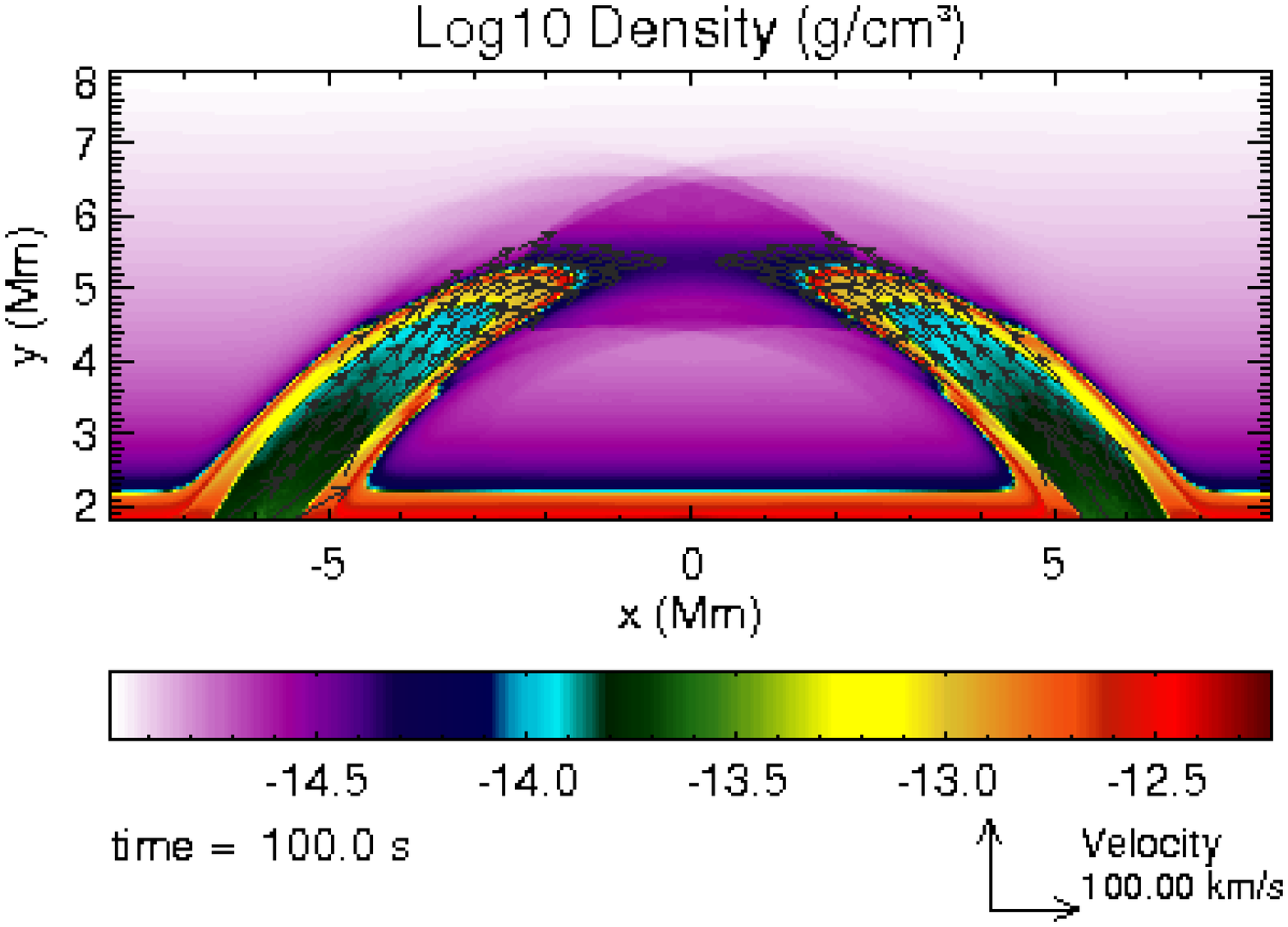}
\includegraphics[angle=90,width=7.6 cm]{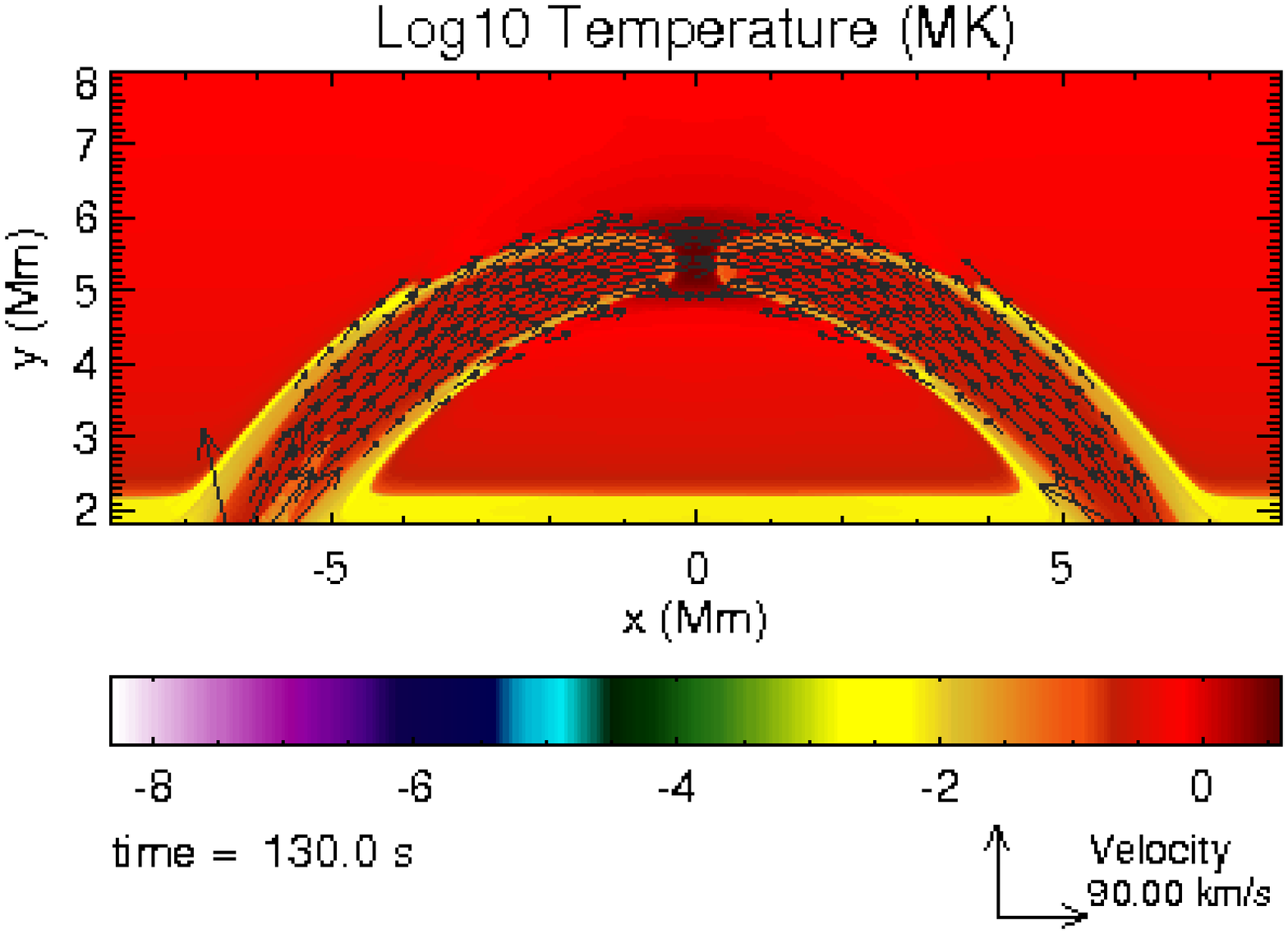}
\includegraphics[angle=90,width=7.6 cm]{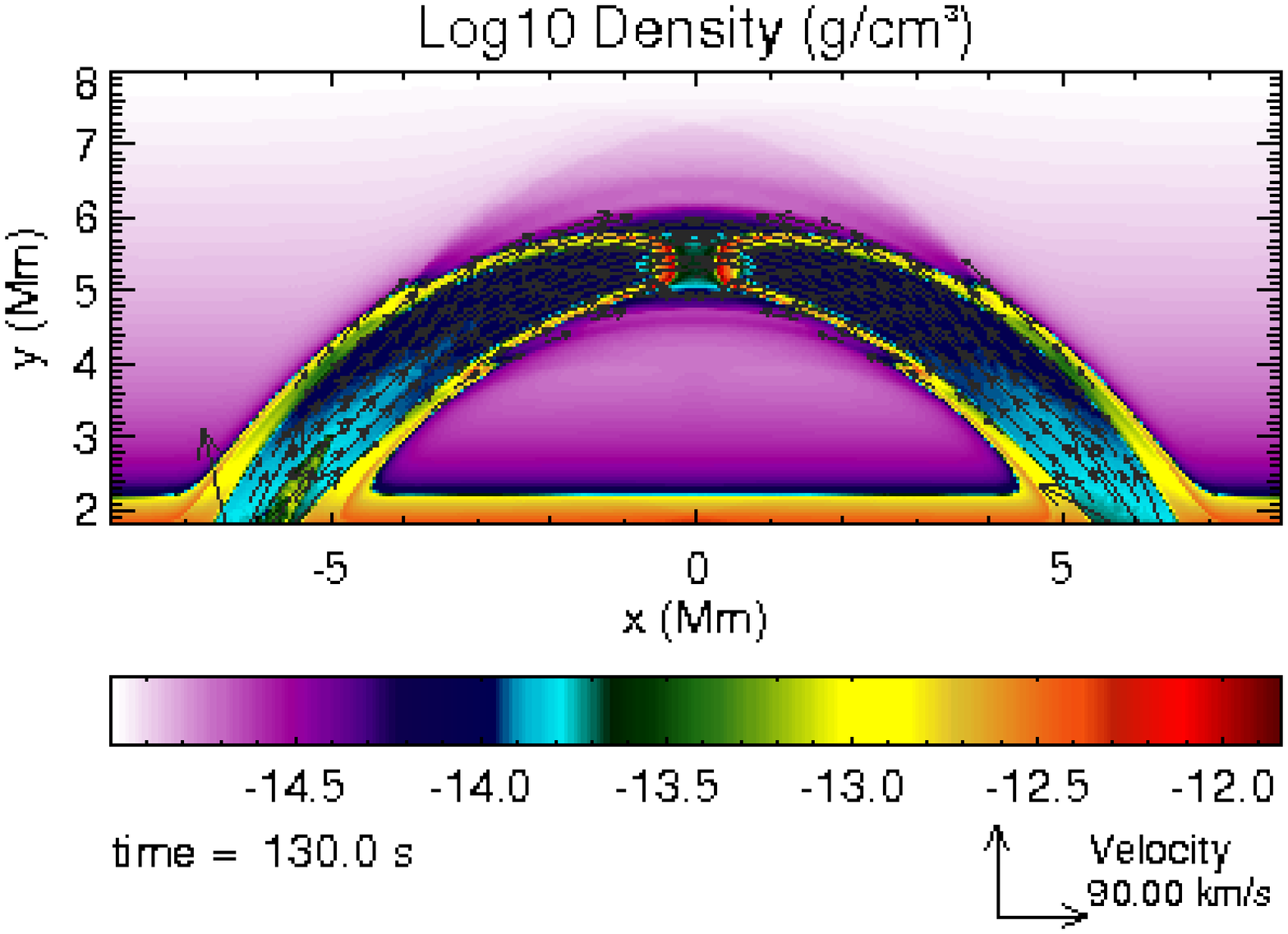}
\includegraphics[angle=90,width=7.6 cm]{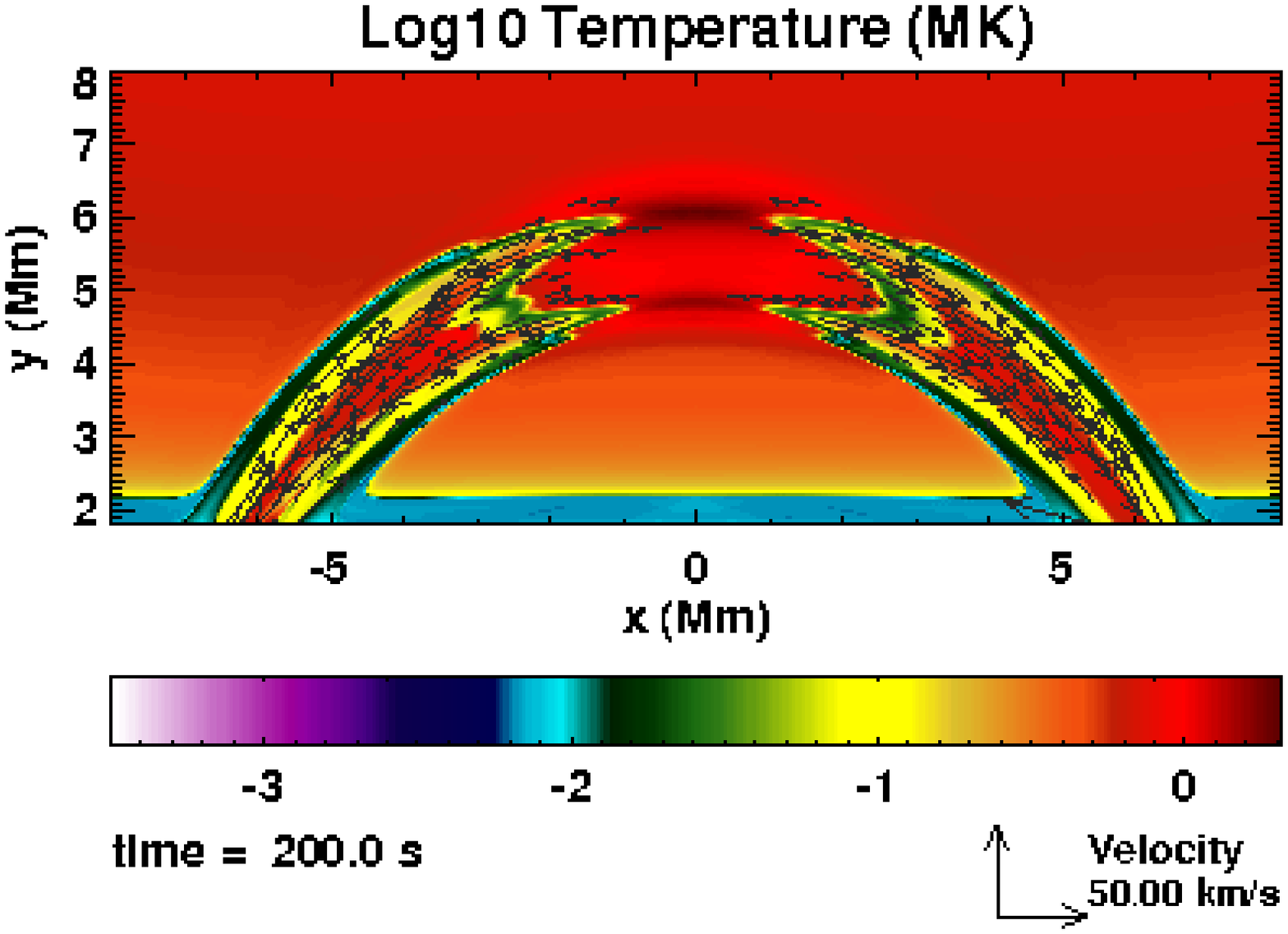}
\hspace{1.00 cm}
\includegraphics[angle=90,width=7.6 cm]{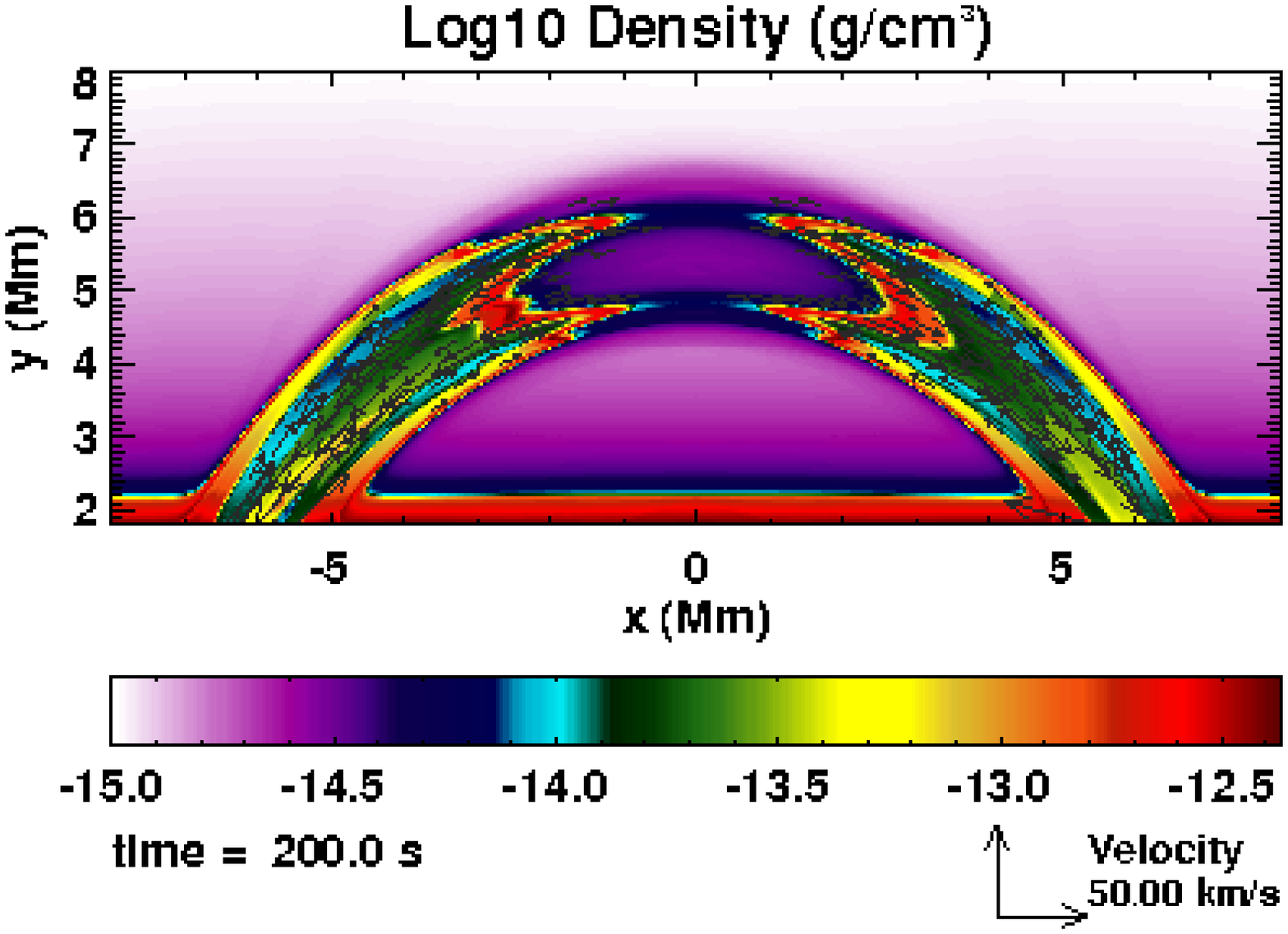}
\vspace{-0.5 cm}
\caption{Temperature (left-column) and mass density (right-column) maps obtained in the framework of the 2D IMHD model in which the energy deposition area lies at the height of $h=1.24$ Mm above the temperature minimum at $t=100$ s (top panels), $t=130$ s (middle panels), and $t=200$ s (bottom panels).}
\label{Fig7}
\end{figure*}

The applied Gaussian spatial distribution of the local gas pressure increases has the horizontal half-width equal to $w_{\rm x}=0.5$ Mm. It was selected and refined by numerous trials in order to obtain the width of the fully filled loop of the VDC model close to the diameter of the 1D model, which was evaluated using \emph{RHESSI} observations of the prototype flare. Spatial distribution of temperatures and densities of the plasma confined in the flaring loop at three representative, specific moments of time are presented for various models in Figures~\ref{Fig7}-~\ref{Fig11} and also summarized in Table~\ref{Tab2} in concise form. Various temperature scales are applied in these figures for best visualization of the temperature distributions at various stages of evolution. For the heating height of $h=1.24$ Mm (chromospheric heating), the results obtained with the IMHD model are  shown in Fig.~\ref{Fig7}, the VDC model is illustrated in Fig.~\ref{Fig8}, and the NRL model in Fig.~\ref{Fig9}. The following moments are presented, illustrating typical states of the plasma (cf., Figs.~\ref{Fig7}-~\ref{Fig9}): (i) the fully developed chromospheric evaporation, defined as a moment when the velocity of the ascending plasma was at its highest for $h=1.5$ Mm above the TR (i.e. in the leg of the loop): $t=100$ s for the IMHD model, $t=90$ s for the VDC model and $t=50$ s for the NRL model, (ii) the maximum of the emissions in the \emph{GOES}~1-8 \AA\ band: $t=130$ s for the IMHD model, $t=110$ s for the VDC model, and $t=150$ s for the NRL model, and (iii) in the middle of the gradual decay phase of the evolution: $t=200$ s for the IMHD and VDC models, and $t=250$ s for the NRL model. For convenient visualisation of the spatial variations of the plasma parameters along the loop in the 1D models, each point of their calculation grid is represented in the Figs.~\ref{Fig9} and~\ref{Fig91} by a segment of a finite width, perpendicular to the loop's axis.

For the heating height of $h=1.74$ Mm (heating above the TR), the results of the IMHD, VDC and NRL models are displayed in Figs.~\ref{Fig10}-~\ref{Fig91}, respectively. The following moments are presented: (i) the fully developed chromospheric evaporation: $t=30$ s for the IMHD and VDC models, and $t=50$ s for the NRL model, (ii) the maximum of the emissions in the \emph{GOES}~1-8 \AA\ band: $t=90$ s for the IMHD model, $t=70$ s for the VDC model, and $t=150$ s for the NRL model, as well as (iii) gradual decay phase of the evolution at $t=200$ s for the IMHD and VDC models and $t=250$ s for the NRL model.

Representative values of temperature, mass density and direction of the bulk motions of the plasma are presented for the same evolution stages of all the models in Table~\ref{Tab2}. The parameters are stated for $h=1.5$ Mm above the TR (\emph{i.e.} the leg of the loop) and for the top of the loop as two representative points. However, in the case of 2D models the parameters are given for the axial part of the loop as well as for the ambient layer of the loop separately, while a as a result of the applied 2D Gaussian spatial distribution of the delivered ambient energy, the plasma parameters vary across the loop. The plasma $\beta$ parameter of all developed models is always much less than 1.0 and the trajectory of moving plasma is determined by the local magnetic field. Therefore, in the developed models the flaring loops indirectly possess the multi-thread internal structure, already proposed by \citet{Asch07} and \citet{Warr06}.

\subsection{Flare with energy deposition region in the upper chromosphere}

In this part of the paper we consider a flare with energy deposition region located in the upper chromosphere, $h=1.24$ Mm above the temperature minimum. Due to initial Gaussian spatial distribution of the local pressure impulse, which mimics the local energy deposition of the NTEs during the early phase of the flare, the evaporating plasma rises toward the loop-top in all the models, forming in the 2D models the heterogeneous columns of hot and dense cores and relatively cool and less dense ambient layers. However, an exclusion of the thermal conduction in the IMHD models exaggerates even more local differences in temperature perpendicular to the loops axis. When the chromospheric evaporation is fully developed, i.e. about 20 s after the onset of the noticeable X-ray emissions of the loop in the 1-8 \AA\ band, the highest temperature of the evaporating plasma at the representative height of 1.5 Mm above the transition region is obtained in the NRL model (about {$\rm T_{NRL}=2.5$ MK), while the temperature in the axial part of the loop is of the order of $\rm T_{IMHD}=1.3$ MK in the case of the IMHD model and $\rm T_{VDC}\cong0.1$ MK in the case of the VDC model. Representative temperatures of the ambient layer of the loops in both 2D models are of the order of chromospheric temperatures only, while these layers of the loop are formed by chromospheric-type plasma pushed from below without any significant heating. Mass density of the hot plasma is $\rm \varrho_{NRL}\cong6\times10^{-14} \,g \,cm^{-3}$ in the NRL and in the axial part of the 2D VDC model, but it is 4 times lower in the axial part of the 2D IMHD model. The latter may be due to neglected viscosity in this model. The raising pillars of plasma caused compression of the plasma contained in the upper part of the loop, gradually forming hot and dense loop-top sources of X-rays.  Twenty seconds after the onset of the X-ray emission, their temperature is still moderate and equal to $\rm T_{IMHD}=1.3$ MK, $\rm T_{VDC}=1.5$ MK and $\rm T_{NRL}=2$ MK in the IMHD, VDC and NRL models, respectively.

\begin{figure*}[h!]
\includegraphics[angle=90,width=7.6 cm]{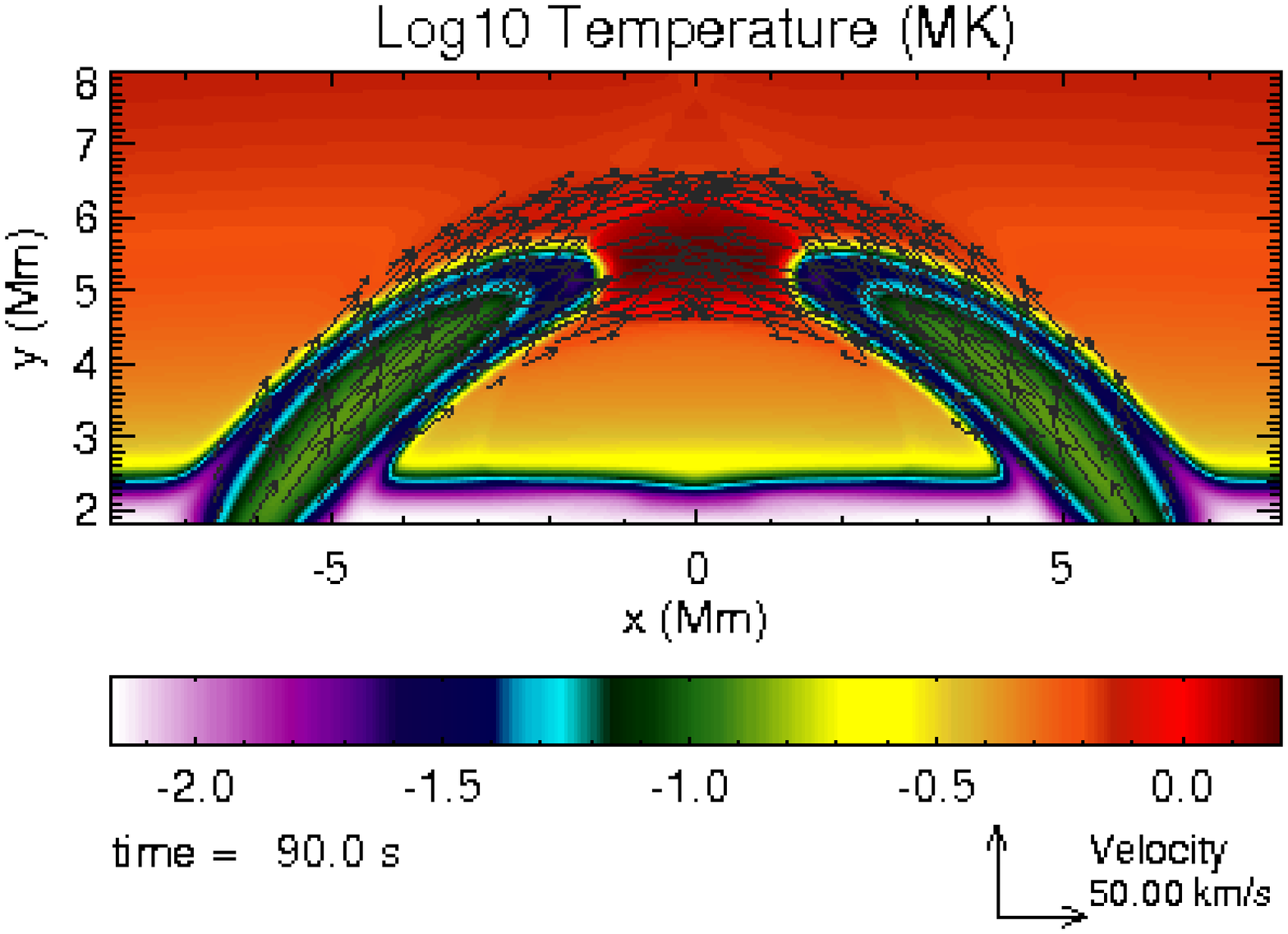}
\includegraphics[angle=90,width=7.6 cm]{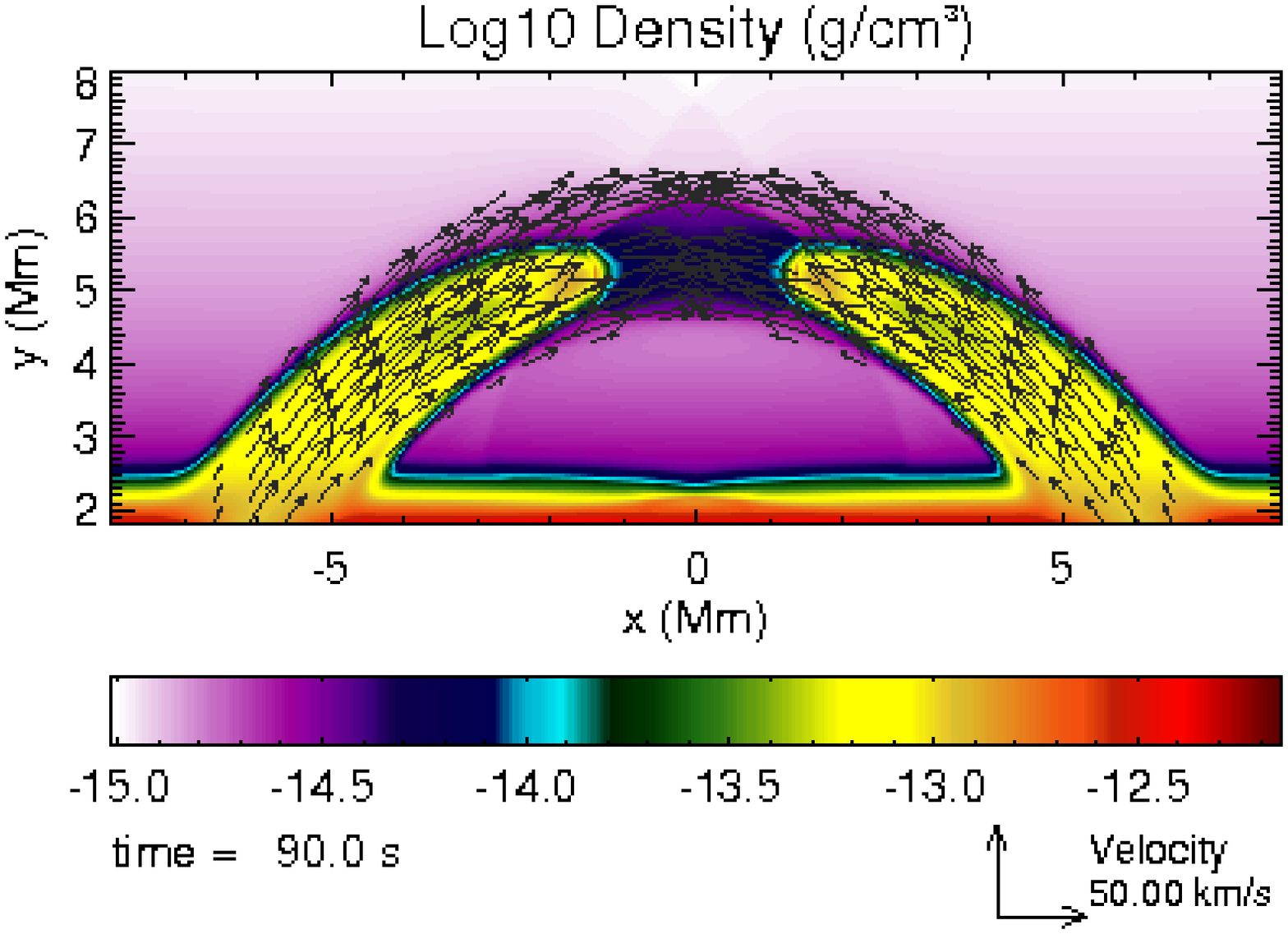}
\includegraphics[angle=90,width=7.6 cm]{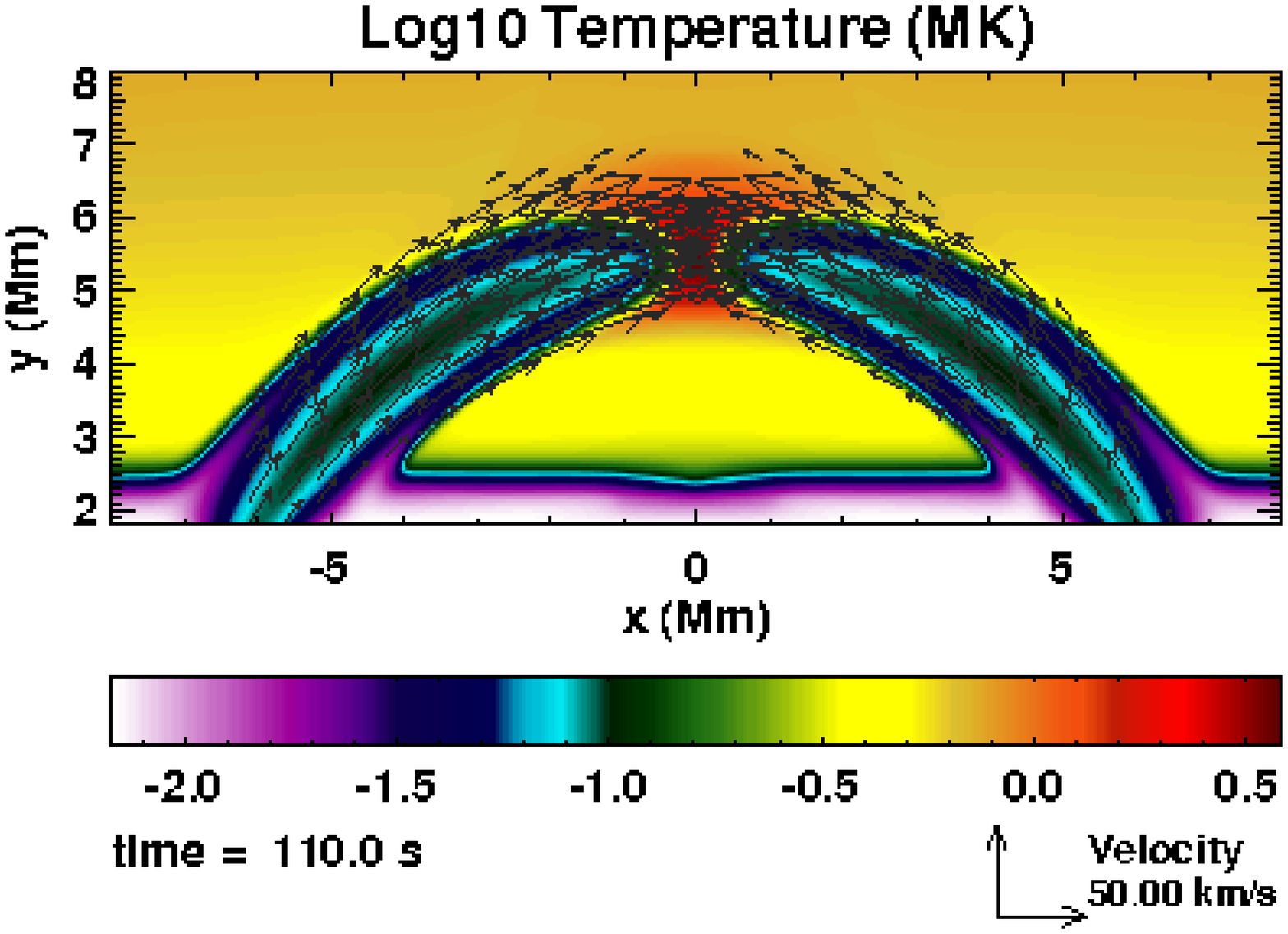}
\includegraphics[angle=90,width=7.6 cm]{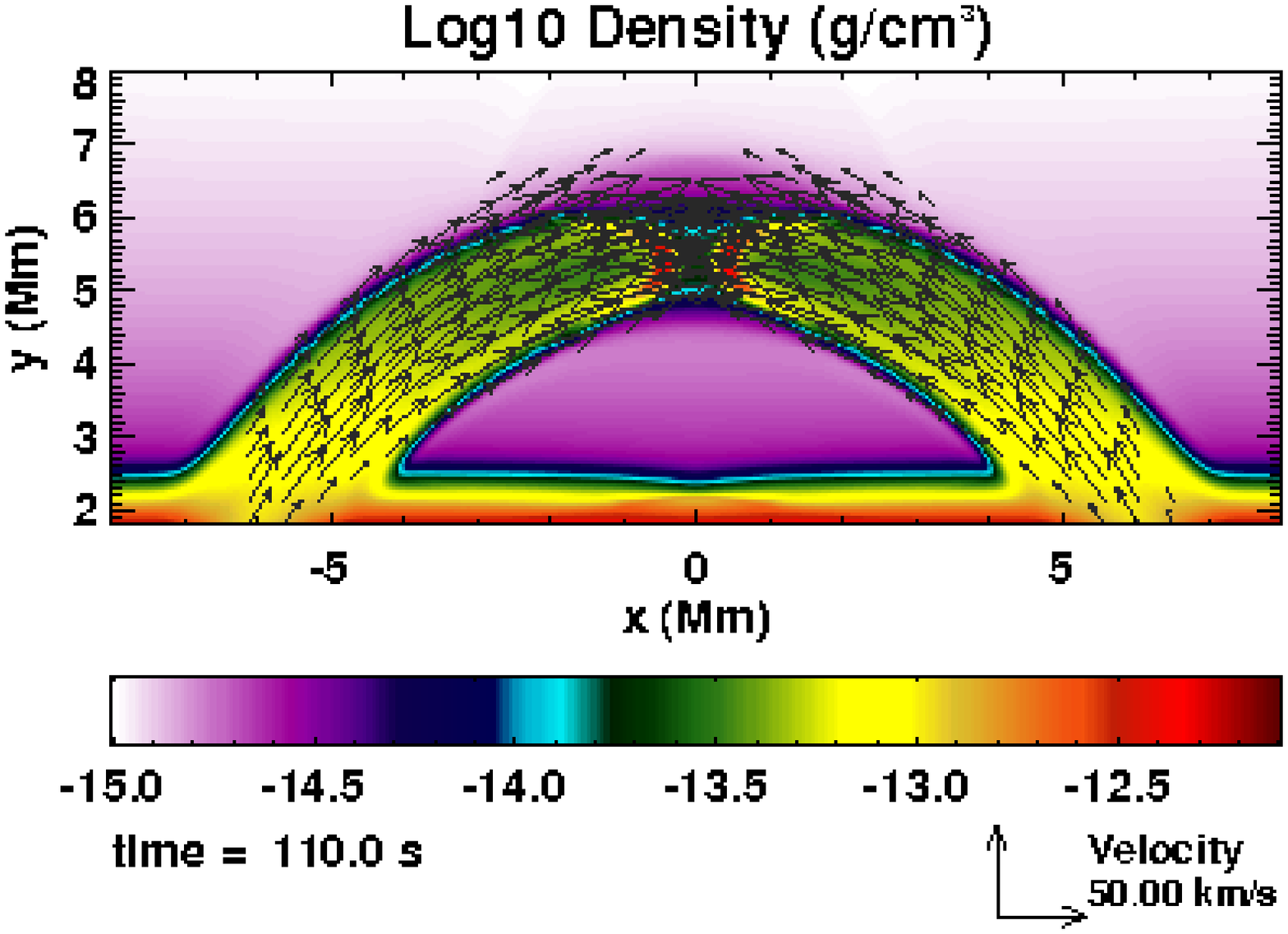}
\includegraphics[angle=90,width=7.6 cm]{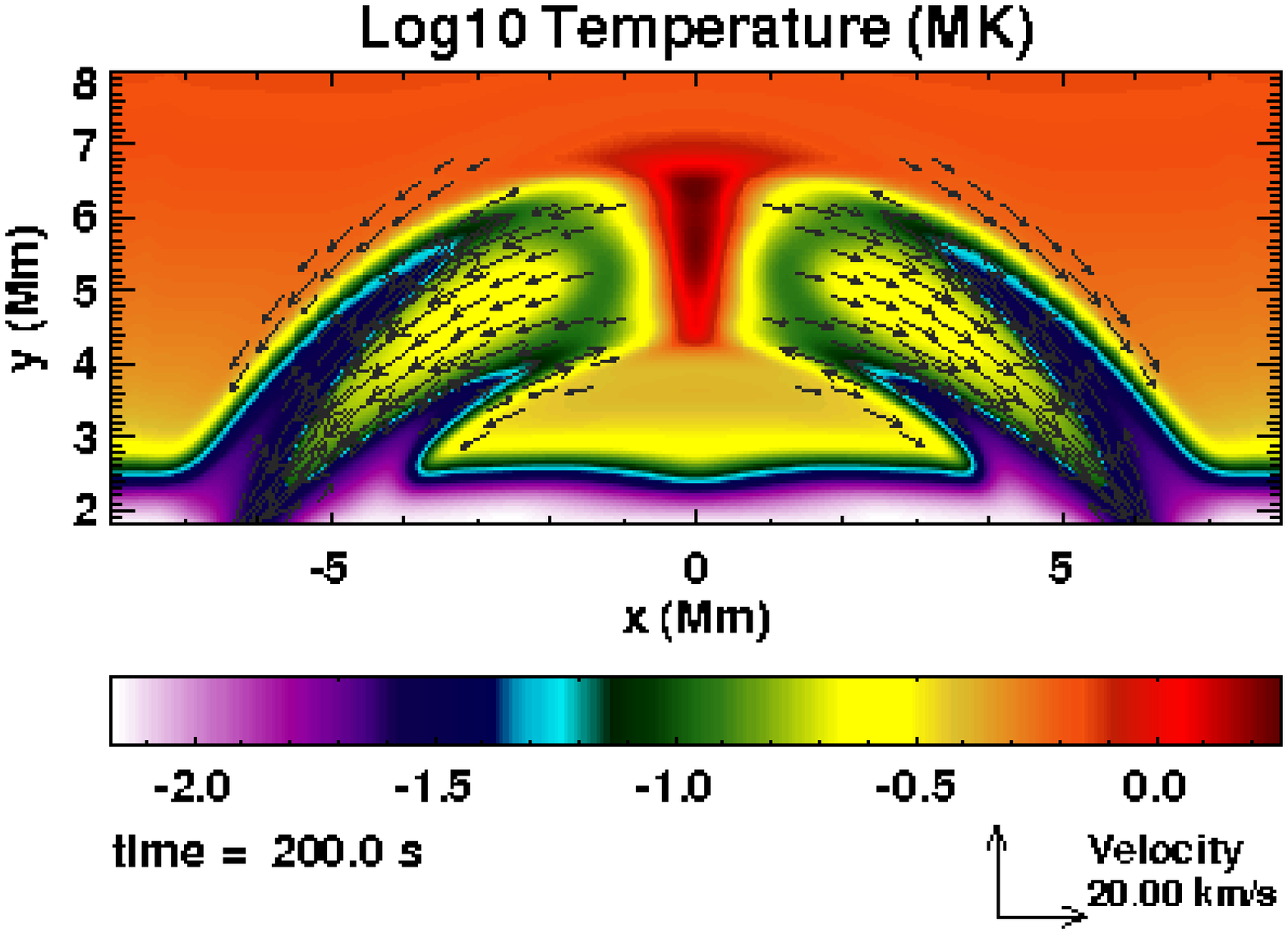}
\hspace{1.00 cm}
\includegraphics[angle=90,width=7.6 cm]{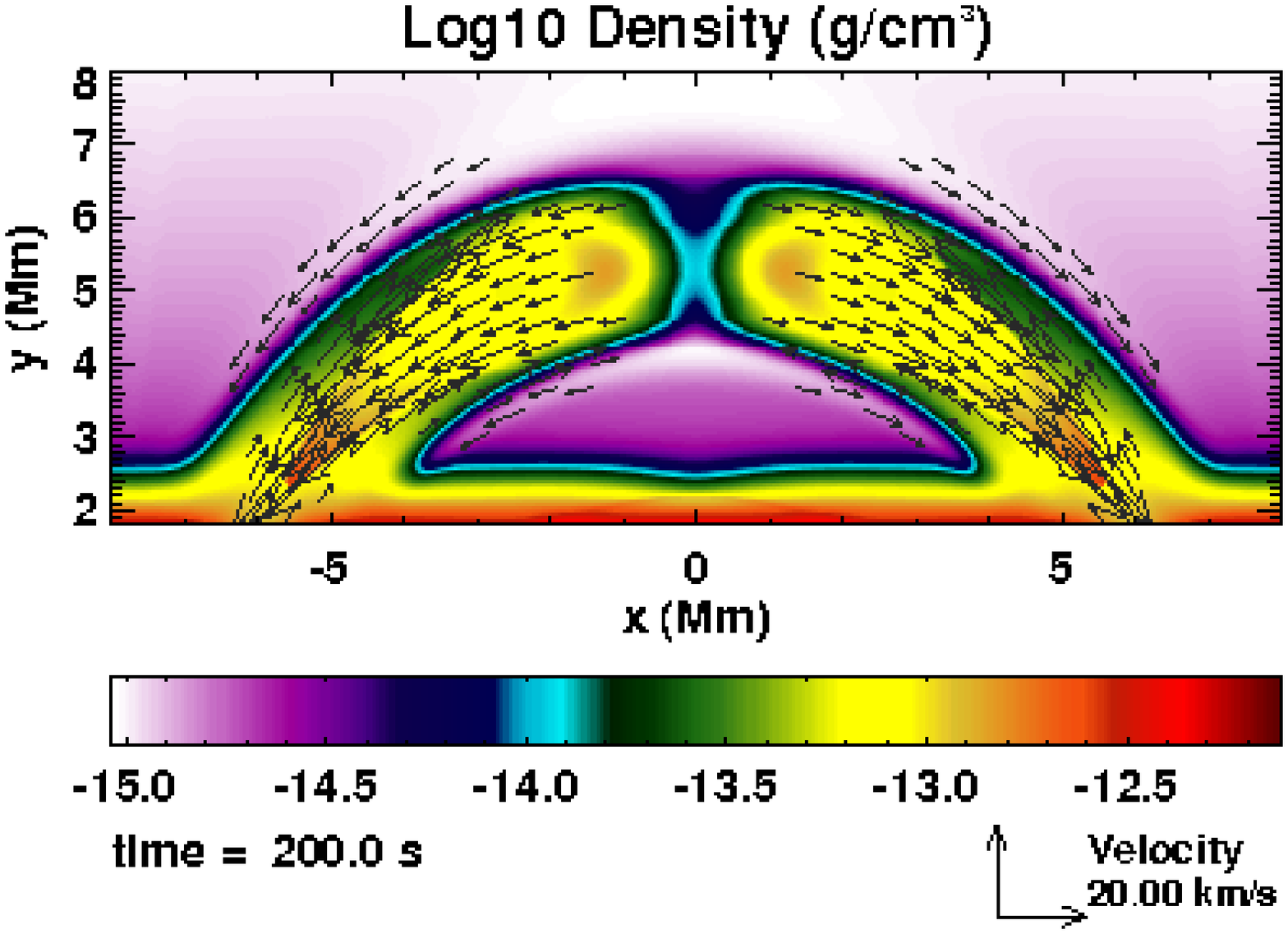}
\vspace{-0.5 cm}
\caption{Temperature (left-column) and mass density (right-column) maps in the framework of the 2D model VDC in which the energy is deposited at the height of $h=1.24$ Mm above the temperature minimum drawn at $t=90$ s (top panels), $t=110$ s (middle panels), and $t=200$ s (bottom panels).}
\label{Fig8}
\end{figure*}

The X-ray 1-8 \AA\ emission attains its maximum at $\rm t_{maxIMHD}=130$ s in the IMHD model, at $\rm t_{maxVDC}=110$ s in the VDC model and at $\rm t_{maxNRL}=150$ s in the NRL model. The overall properties of still ascending pillars of the evaporating plasma at that time remain roughly similar to the plasma properties during the previous stage of evolution. The highest temperature of the evaporating plasma at the representative height of $h=1.5$ Mm above the transition region is again noted in the NRL model (about 3.7 MK), while the temperatures in the axial part of the loop are of the order of $\rm T_{IMHD}=1.1$ MK in the case of the IMHD model and $\rm T_{VDC}=0.1$ MK only in the case of the VDC model. At this evolution stage, the highest temperatures of the plasma occur in the well-formed, but still compressed loop-top regions of the loops, co-spatial with the loop-top sources of the X-ray emissions, where the temperatures reach $\rm T_{IMHD}=3.5$ MK in the IMHD model, $\rm T_{VDC}=1.5$ MK in the case of the VDC model and peaked at $\rm T_{NRL}=5.2$ MK in the NRL model. Plasma densities are $\rm \varrho_{IMHD}=4\times10^{-14}\, g\,cm^{-3}$, $\rm \varrho_{VDC}=2\times10^{-14}\, g\,cm^{-3}$, and $\rm \varrho_{NRL}=1.4\times10^{-13}\, g\,cm^{-3}$. For both 2D models, the calculated temperatures of the loop-top plasma are in a good agreement with the temperature derived from the calculated synthetic X-ray fluxes emitted by the modeled loops, i.e. \textit{GOES}-like temperature, being equal to $\rm T_{IMHD-GOES}=3.5$ MK and $\rm T_{VDC-GOES}=3.4$ MK for the IMHD and the VDC models, respectively. The \textit{GOES}-like temperature describes an average or representative temperature of the flaring plasma. According to the algorithm of White (White, 2005) the \textit{GOES}-like temperature is based on the ratio of the integral X-ray fluxes recorded in the 0.5-4 \AA\ and 1-8 \AA\ bands by the \textit{GOES} satellite. In the NRL model the same temperature peaks 40 s before the maximum of the emitted 1-8 \AA\ flux, reaching $\rm T_{NRL-GOES}=5.2$ MK at $\rm t_{NRL}=110$ s. A nearly equal maximum of the \textit{GOES}-like temperature occurs at $\rm t_{NRL}\cong150$ s, during the maximum of the emission in the 1-8 \AA\ band.

During the late, gradual phase of the evolution, at t=200 s in the case of the IMHD and VDC models and t=250 s in the case of the NRL model, the loop-top regions expanded, and the plasma began to flow-down along the legs of the loops in the NRL model as well as in the central parts of the legs in the 2D models with the velocity of 40-60 $\rm km\,s^{-1}$ for the IMHD model, 10-30 $\rm km\,s^{-1}$ for the VDC model and 20-40 $\rm km\,s^{-1}$ for the NRL model. However, in case of 2D models the plasma in the ambient layer of the loop still moves towards the top of the loop with the velocity of the order of 10 $\rm km\,s^{-1}$. The temperature of the expanding plasma in the loop-top region gradually decreases, reaching at that moment a value of $\rm T_{IMHD}=1.5$ MK in the IMHD model, $\rm T_{VDC}=1.8$ MK in the case of the VDC model and $\rm T_{NRL}=3.2$ MK in the NRL model, which is still the highest amount of the various models we discussed. The X-ray emission of the loop-top source vanished at that moment and the X-ray emission becomes dominant by the emission of the plasma in the legs of the loops. At the height of $h=1.5$ Mm above the TR, the plasma temperature in the central part of the loop reaches $\rm T_{IMHD}\cong1.2$ MK in the case of the IMHD model and $\rm T_{VDC}=0.2$ MK only in the case of the VDC model. Temperature of the plasma at the same height in the NRL model is equal to $\rm T_{NRL}=1.3$ MK. Plasma densities are equal to $\rm \varrho_{IMHD}=2\times10^{-14}\, g\,cm^{-3}$, $\rm \varrho_{VDC}=5\times10^{-14}\, g\,cm^{-3}$, and $\rm \varrho_{NRL}=1.8\times10^{-13}\, g\,cm^{-3}$. The representative temperature of the ambient plasma layer of the loops in both 2D models is of the order of the chromospheric temperatures only. As a result of an isotropic thermal conduction applied in the FLASH code, the total width of the hot loop gradually grows in time when modeled in 2D, reaching a value of about 2500 km after 200 s of its evolution.

\begin{figure*}[h!]
\includegraphics[angle=0,width=7.6 cm]{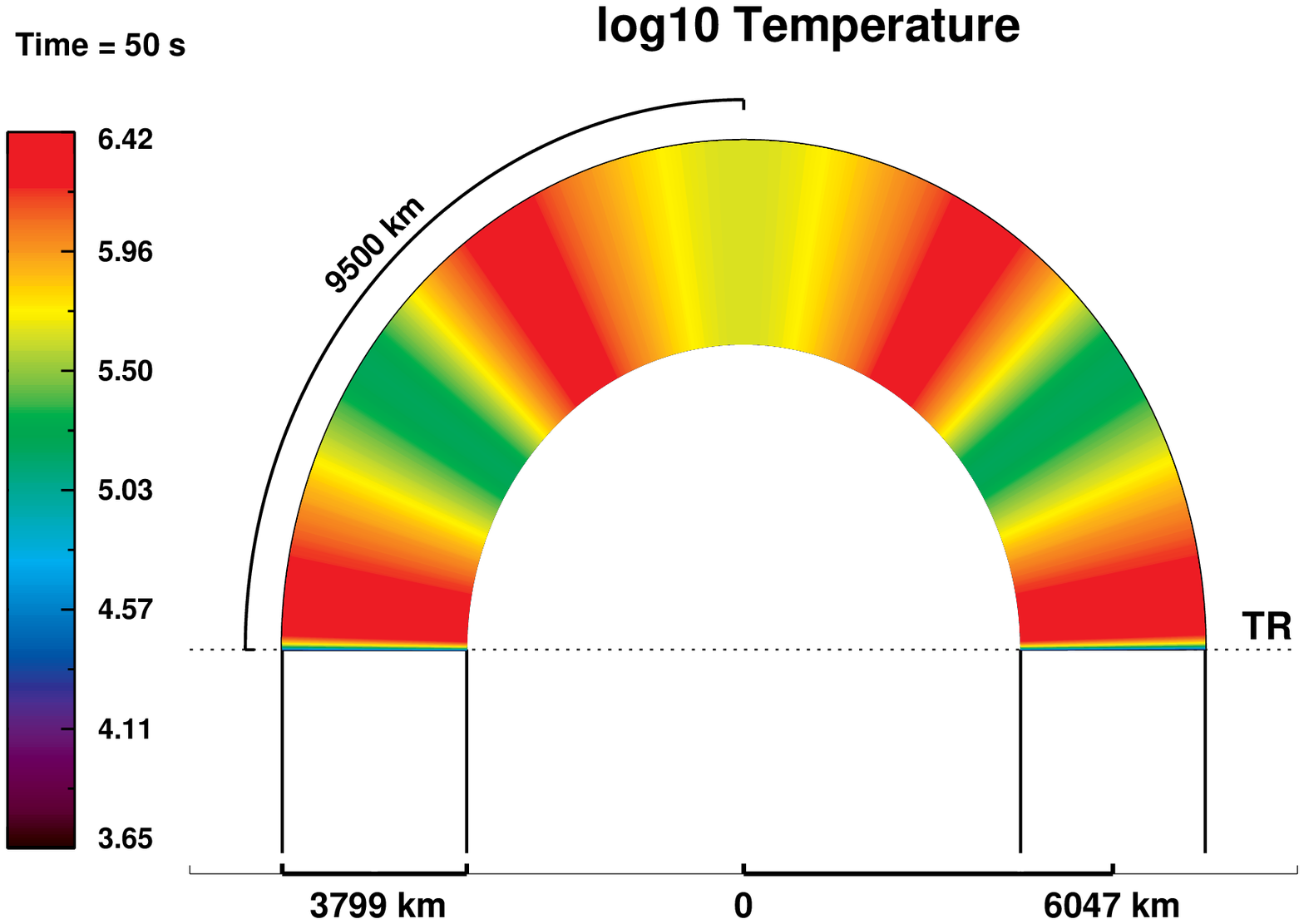}
\includegraphics[angle=0,width=7.6 cm]{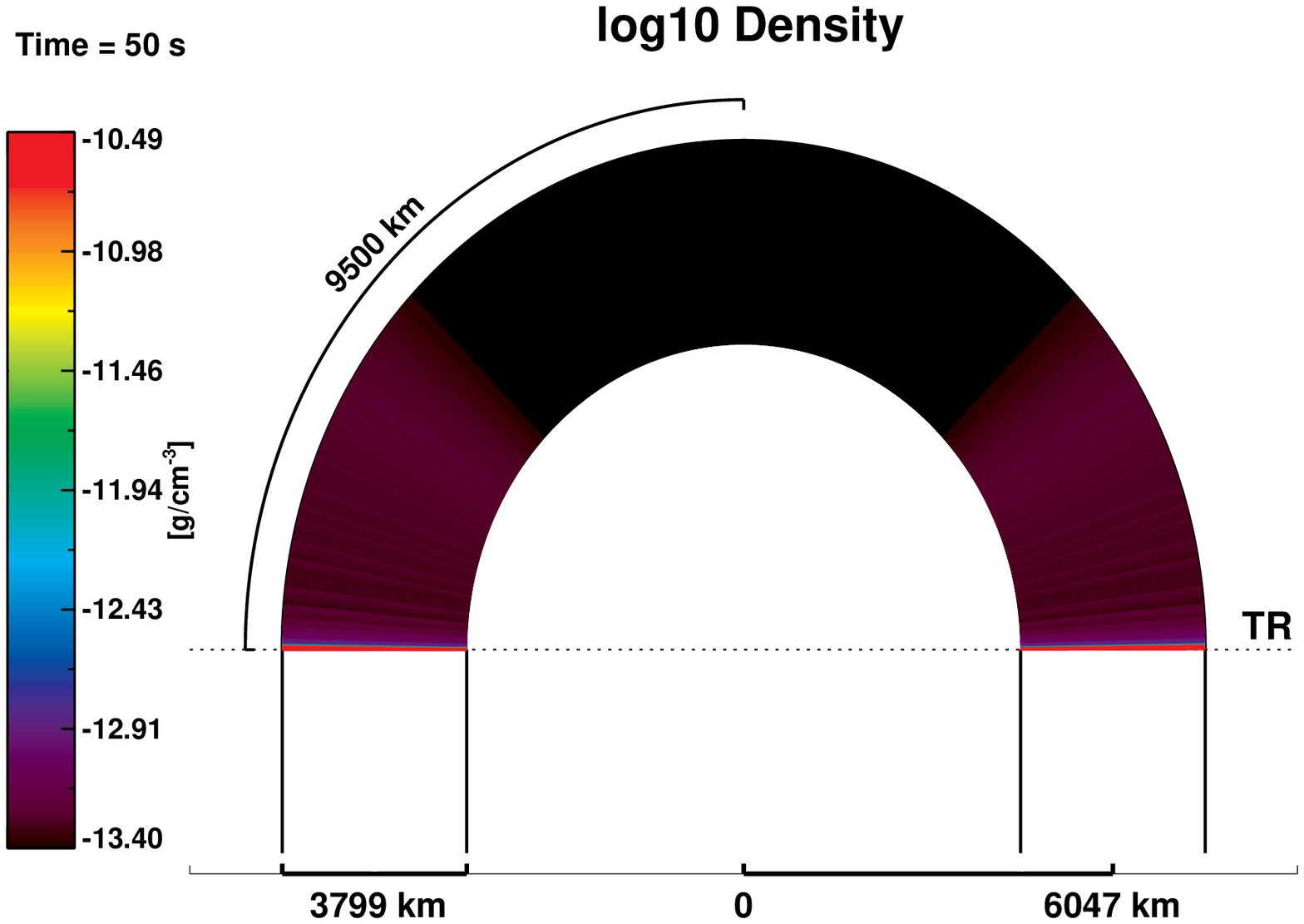}
\includegraphics[angle=0,width=7.6 cm]{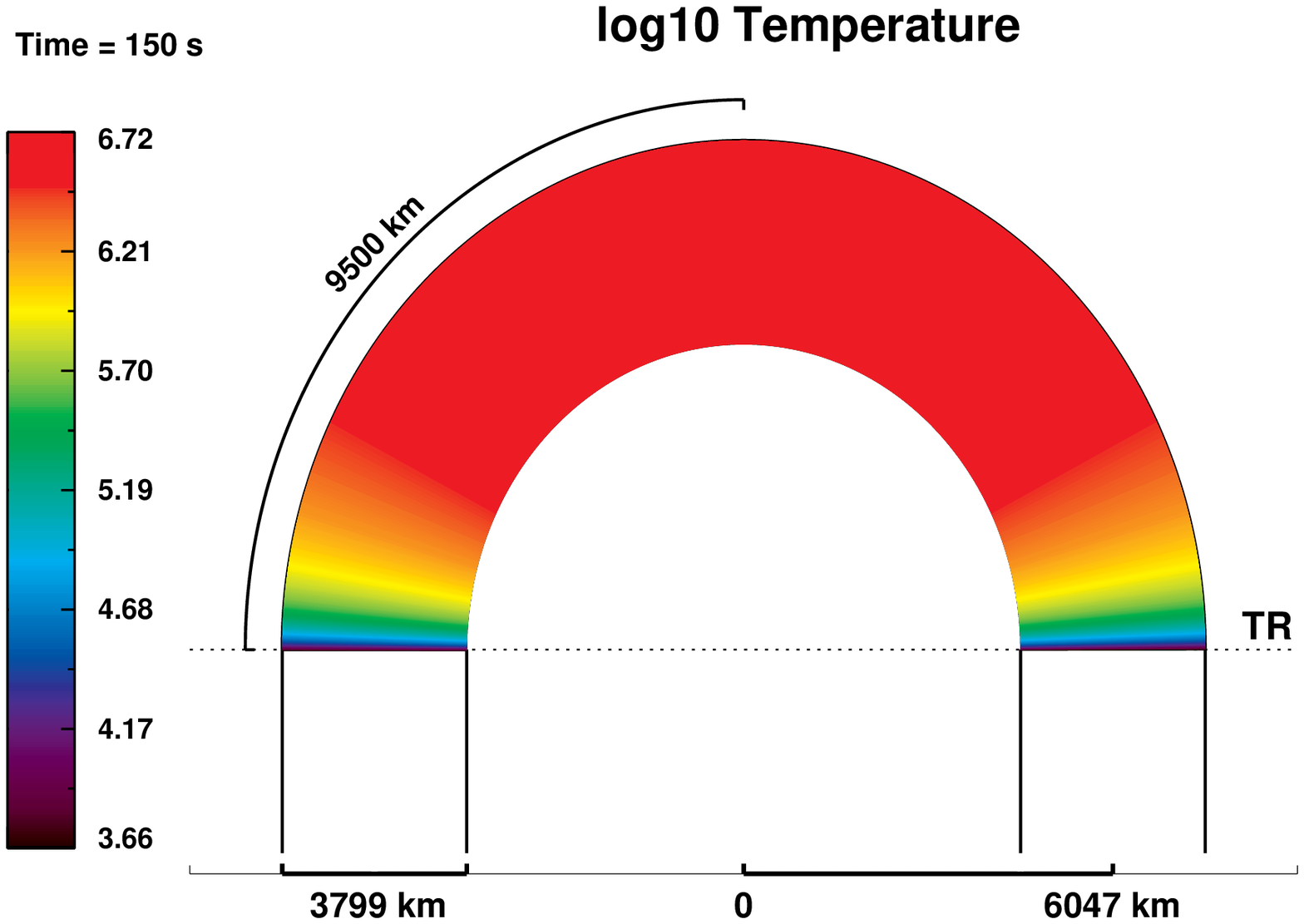}
\includegraphics[angle=0,width=7.6 cm]{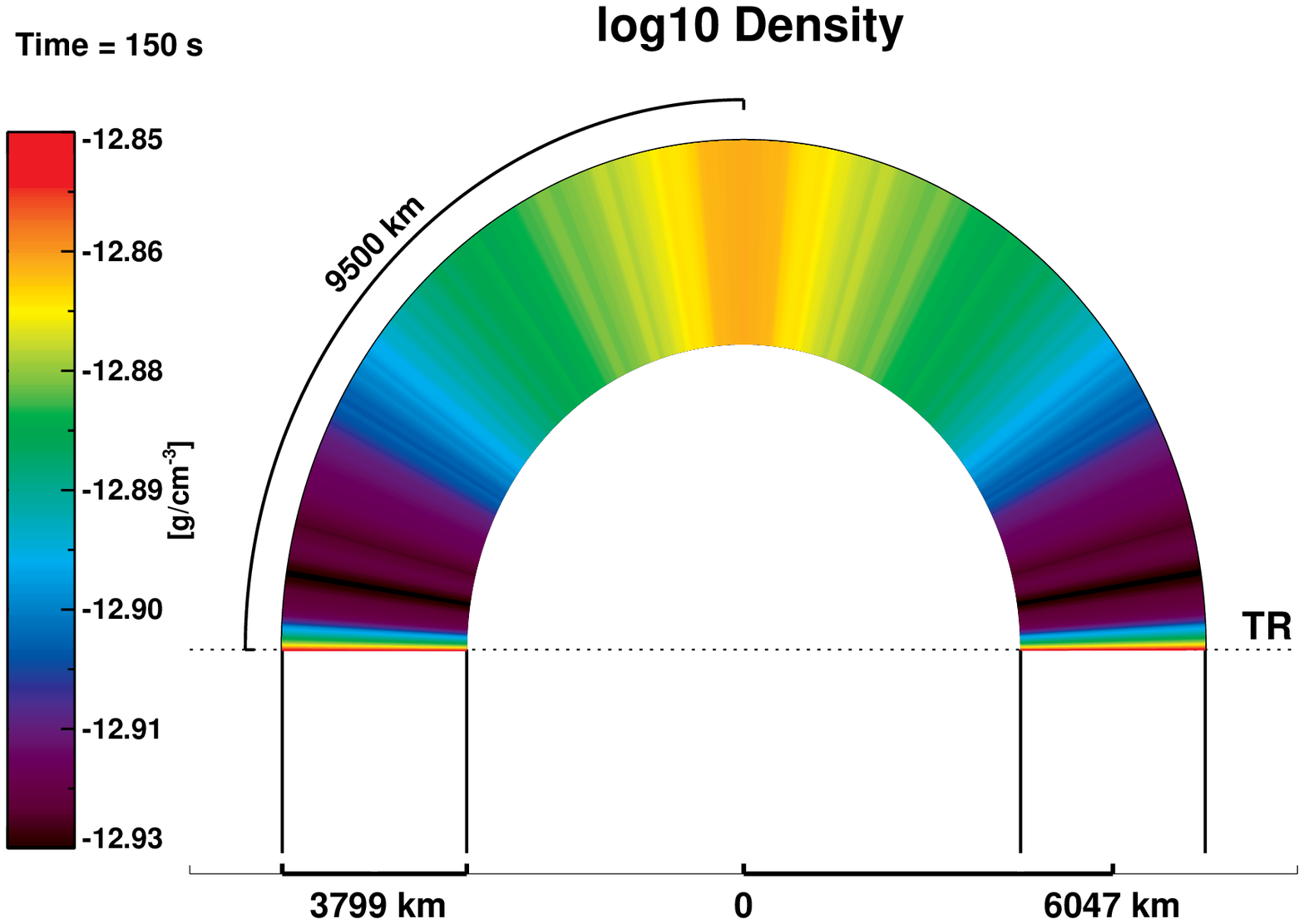}
\includegraphics[angle=0,width=7.6 cm]{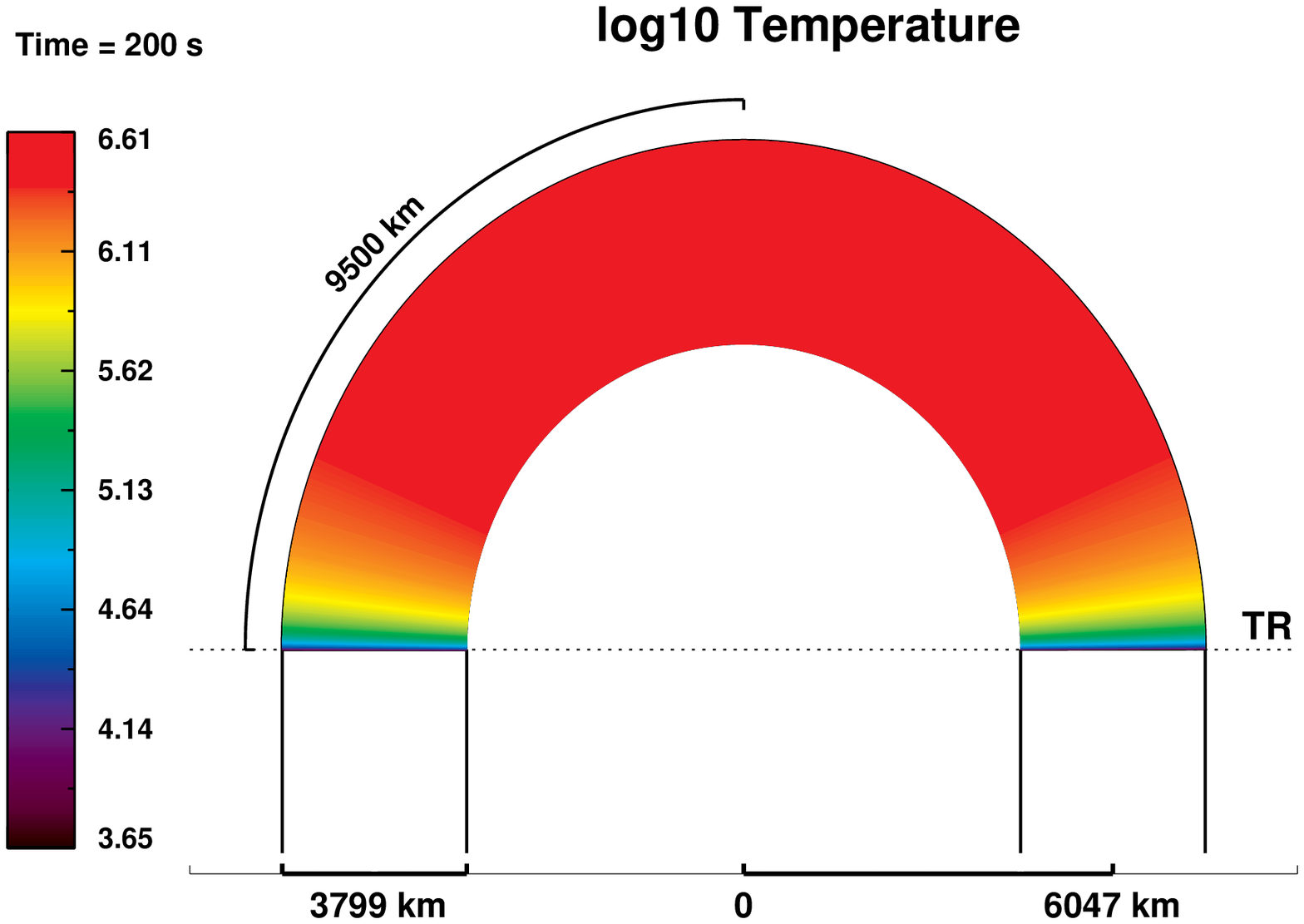}
\hspace{1.00 cm}
\includegraphics[angle=0,width=7.6 cm]{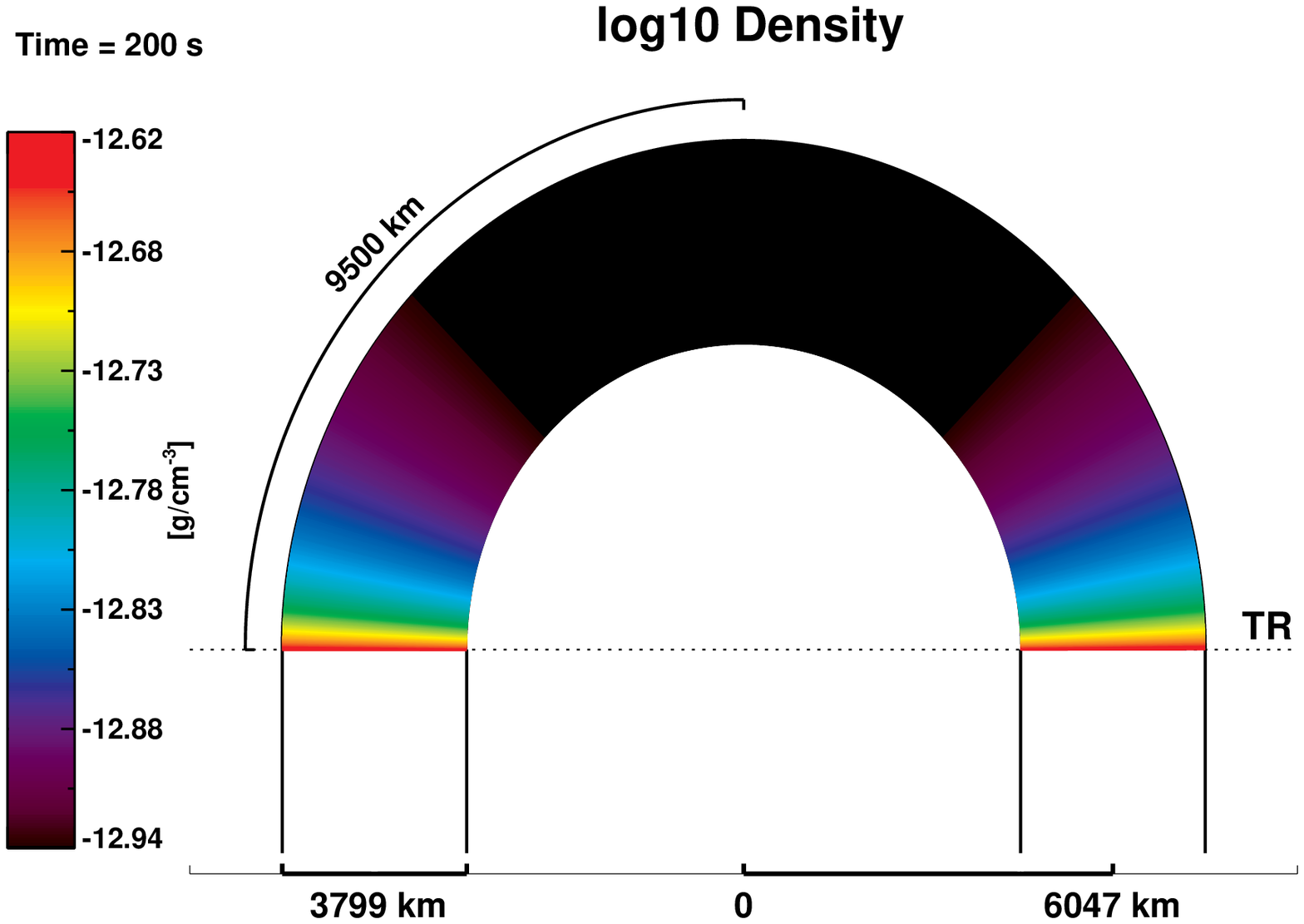}
\vspace{-0.5 cm}
\caption{The spatial distribution of plasma temperature (left-column) and mass density (right-column) in the 1D model in which the energy is deposited at the height of $h=1.24$ Mm at three selected moments: $t=50$ s (top panels), $t=150$ s (middle panels), and $t=200$ s (bottom panels).}
\label{Fig9}
\end{figure*}

\subsection{Flare with energy deposition above the transition region}

In this part of the paper we consider the flare with an energy deposition region located in the lower corona, $h=1.74$ Mm above the temperature minimum (cf. Fig.~\ref{Figatm}). It is heated in the same way as in the model described in Section 4.1. Spatial distributions of the temperature and mass density are presented in Figs.~\ref{Fig10} and~\ref{Fig11} as well as representative values of the main physical parameters of the plasma are presented in Table~\ref{Tab2}. Resulting X-ray fluxes in the 0.5-4 \AA\ and 1-8 \AA\ bands as well as derived \textit{GOES}-like temperatures and densities are presented in Fig.~\ref{Fig12}. The beginning of the noticeable X-ray emissions of the loop in the 1-8 \AA\ band occurs in all models (IMHD, VDC and NRL) on much faster time-scales than in the case of the models with the heating region located at the height of $h=1.24$ Mm above the TR, namely at $\rm \rm t_{IMHD}=9$ s for the IMHD model, $\rm t_{VDC}=8$ s for the VDC model, and at $\rm t_{NRL}=5$ s for the NRL model (cf., Fig.~\ref{Fig12}, right panel).

\begin{figure*}[h!]
\includegraphics[angle=90,width=7.6 cm]{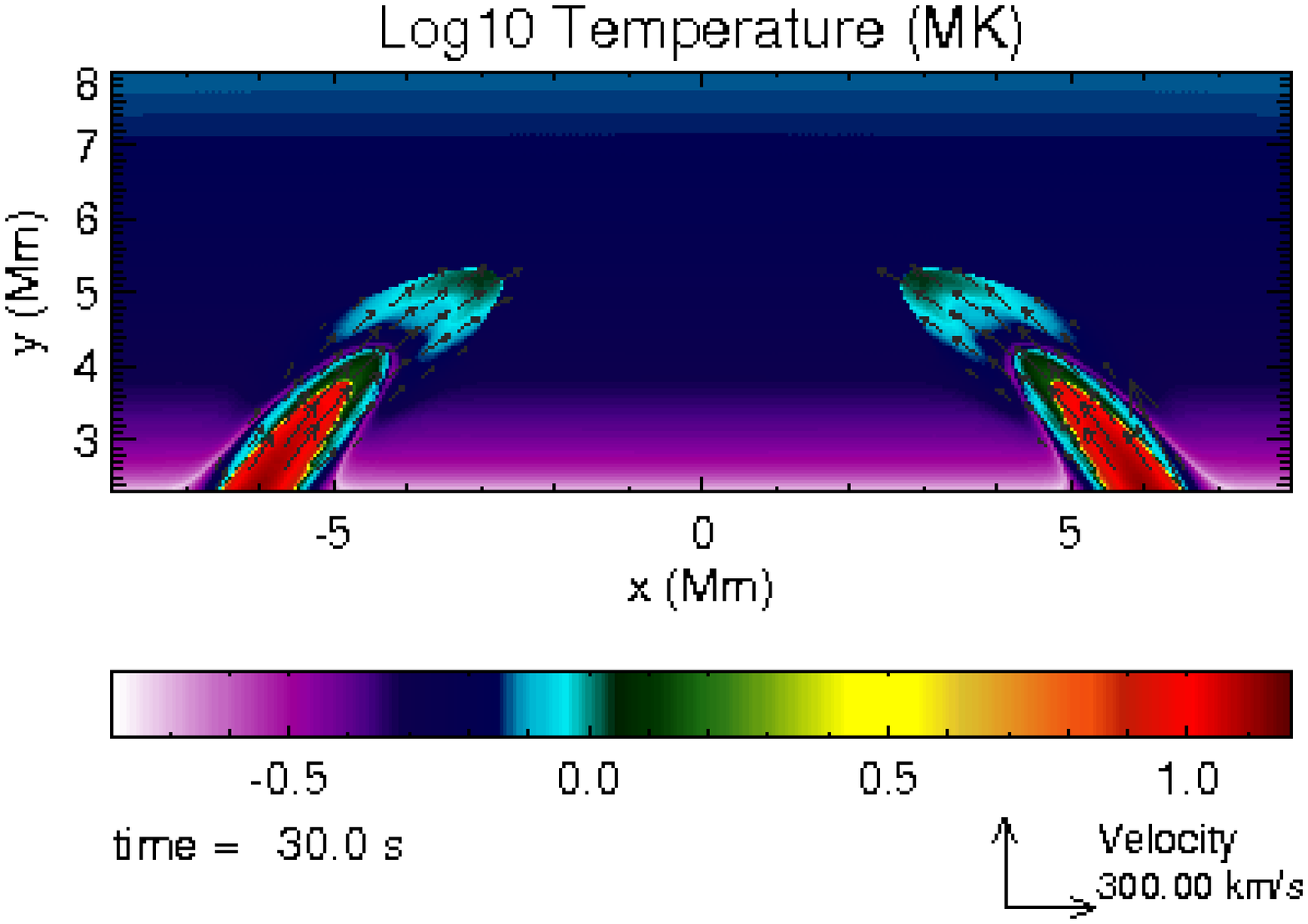}
\includegraphics[angle=90,width=7.6 cm]{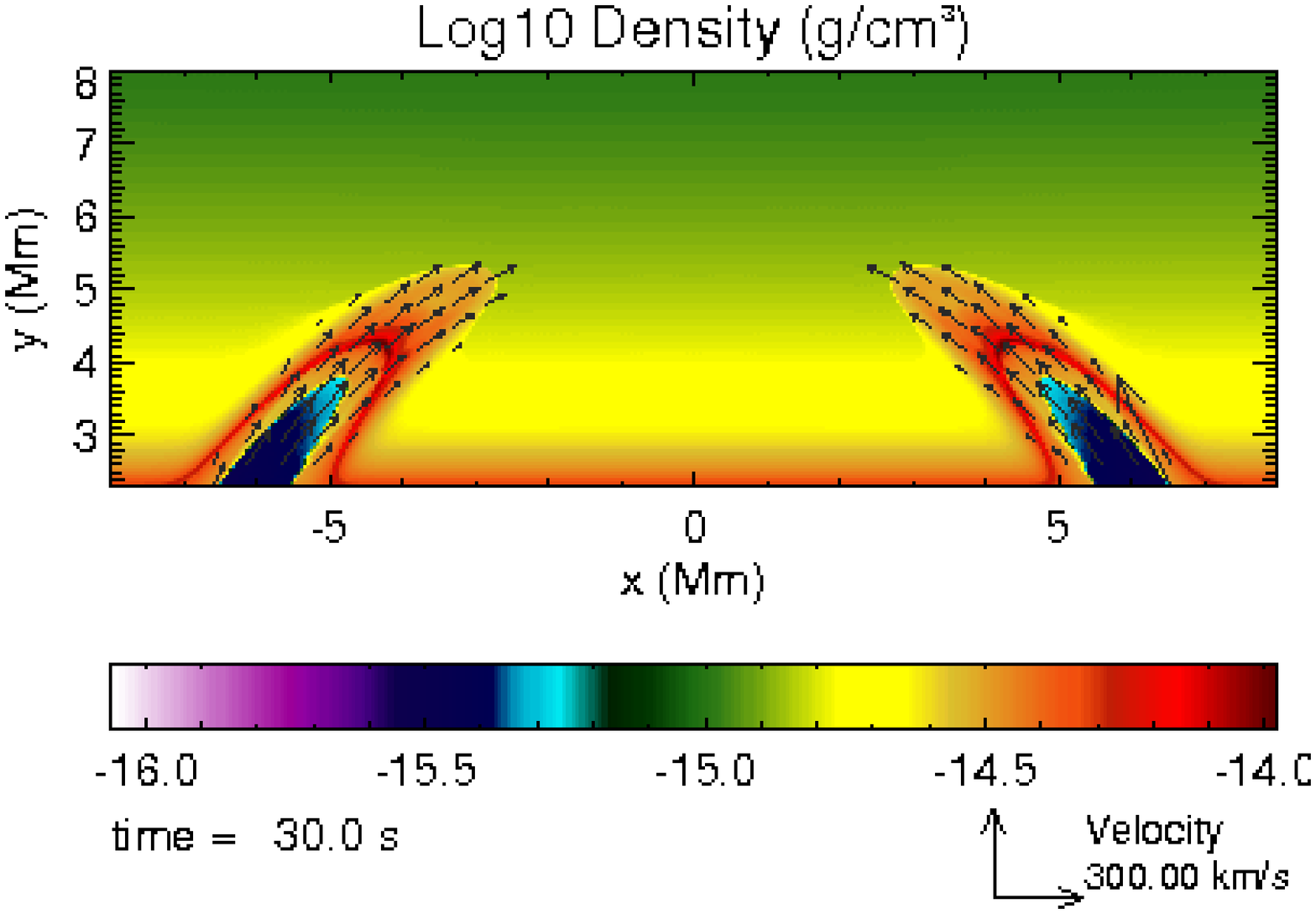}
\includegraphics[angle=90,width=7.6 cm]{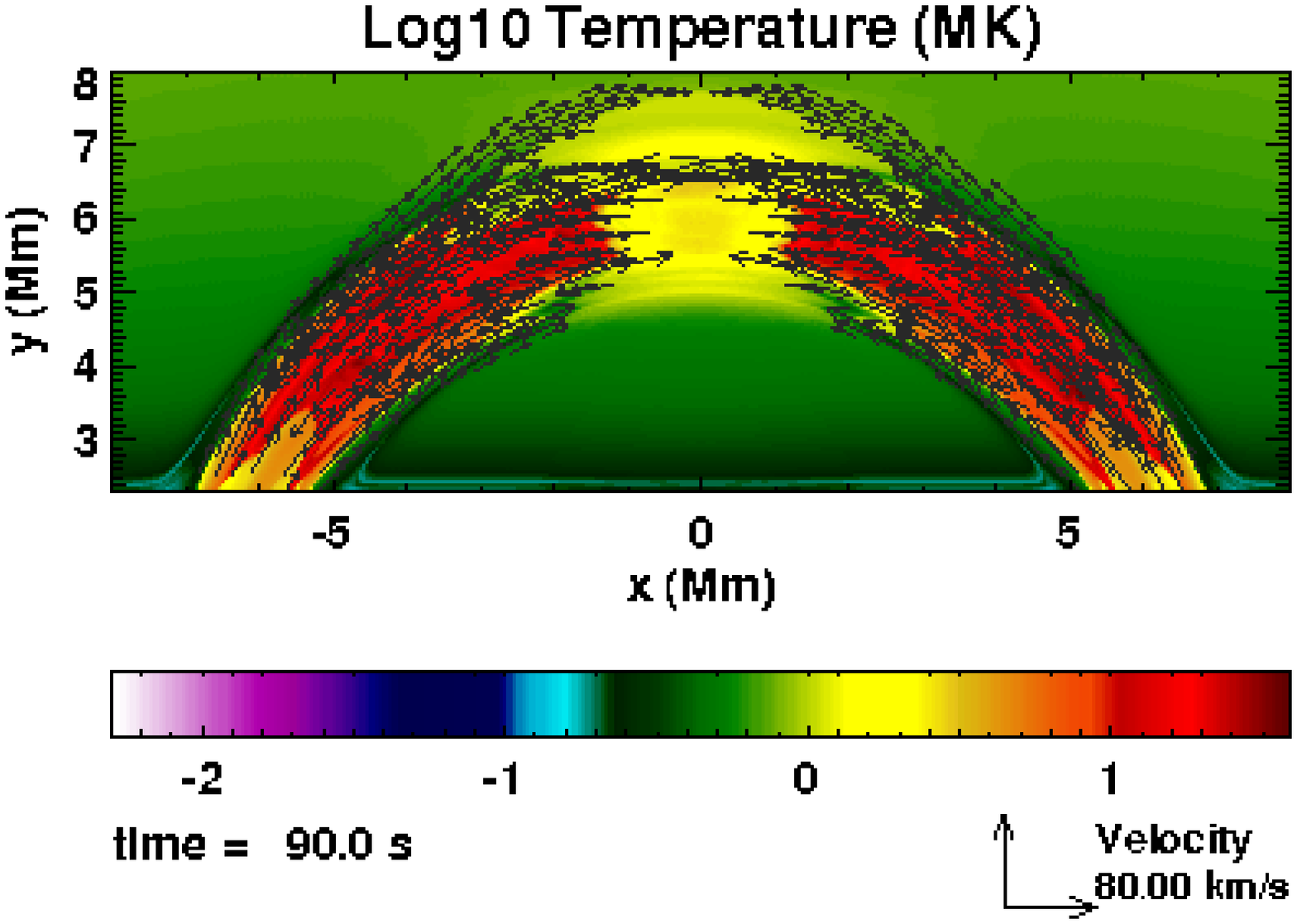}
\includegraphics[angle=90,width=7.6 cm]{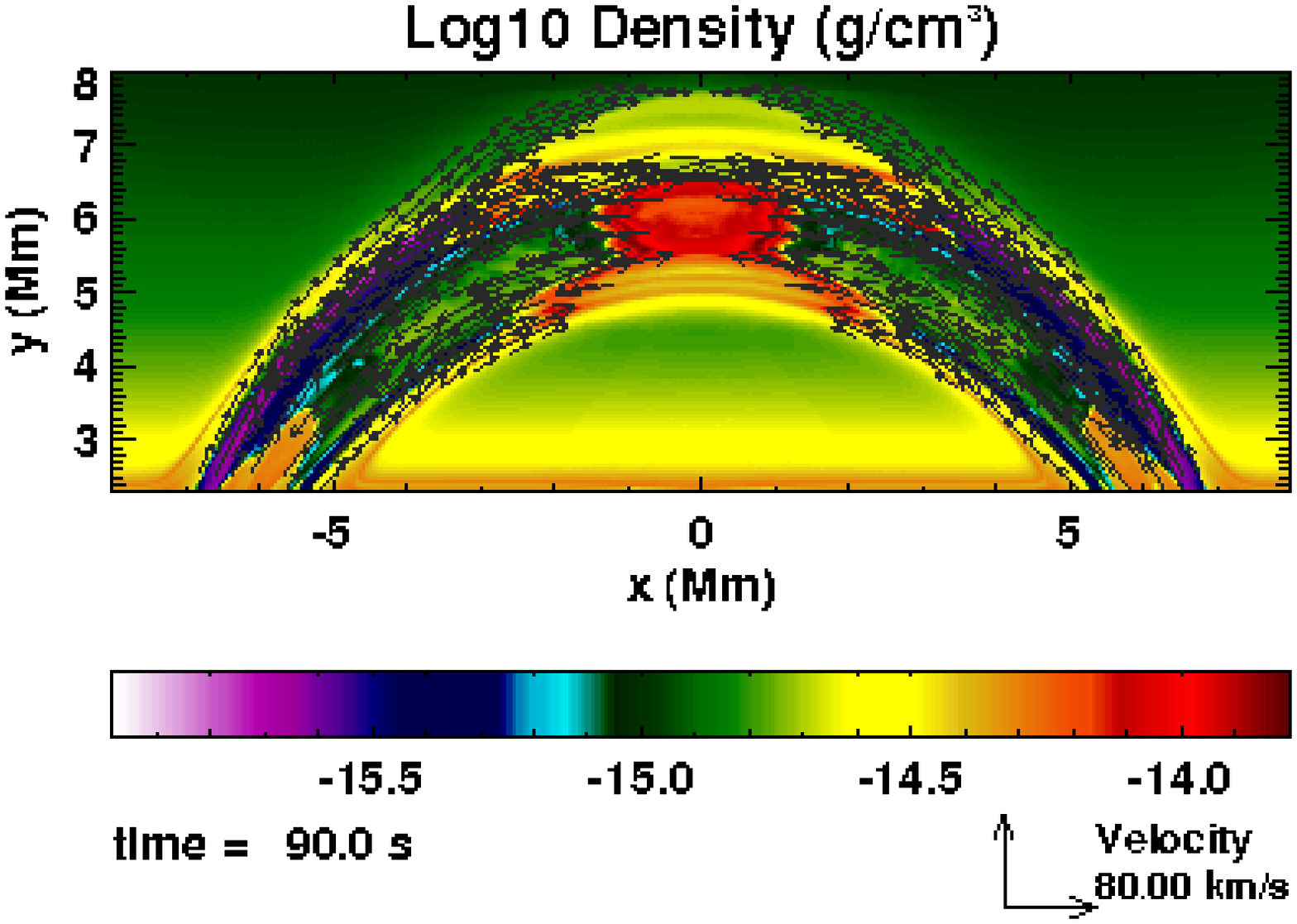}
\includegraphics[angle=90,width=7.6 cm]{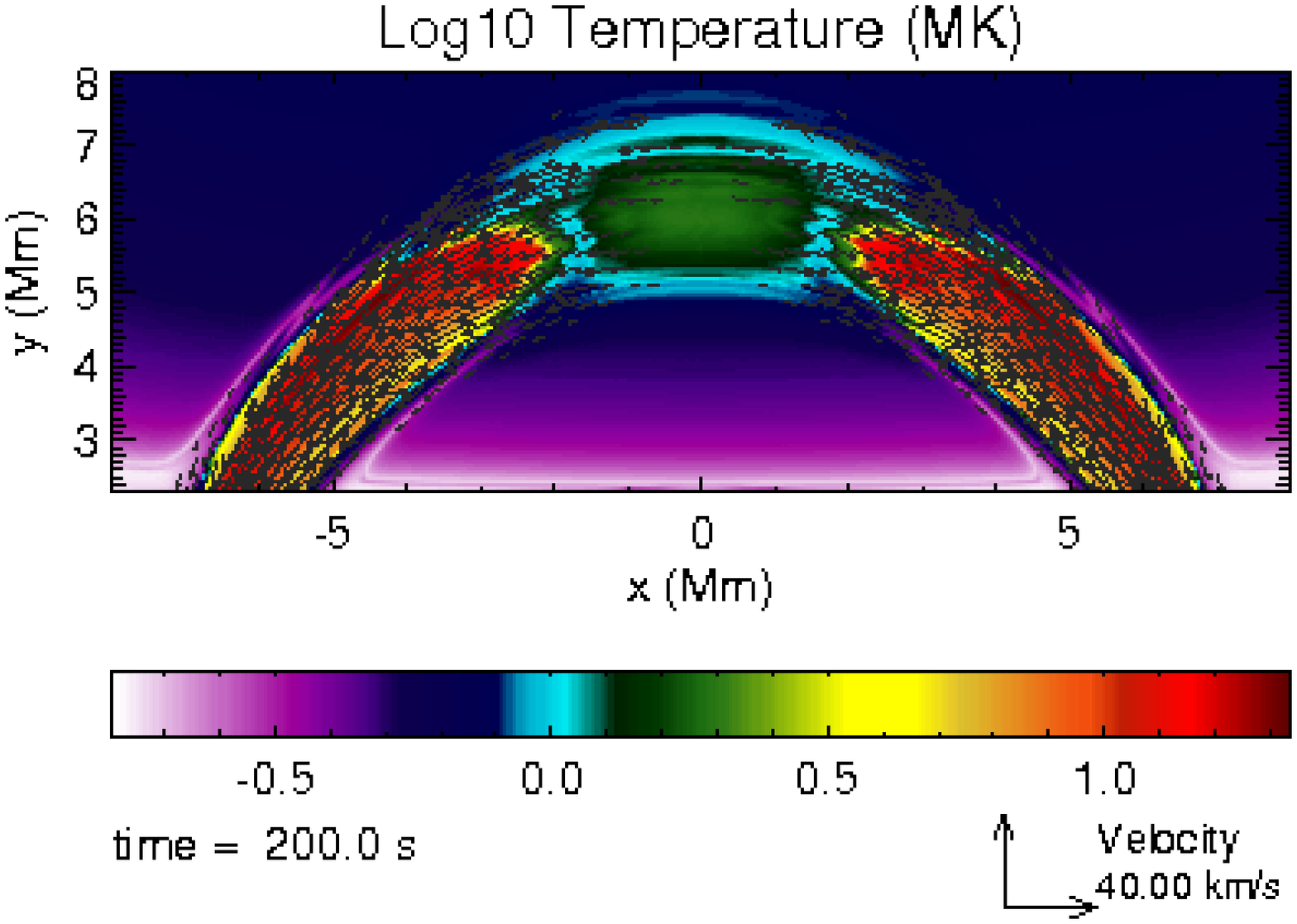}
\hspace{1.00 cm}
\includegraphics[angle=90,width=7.6 cm]{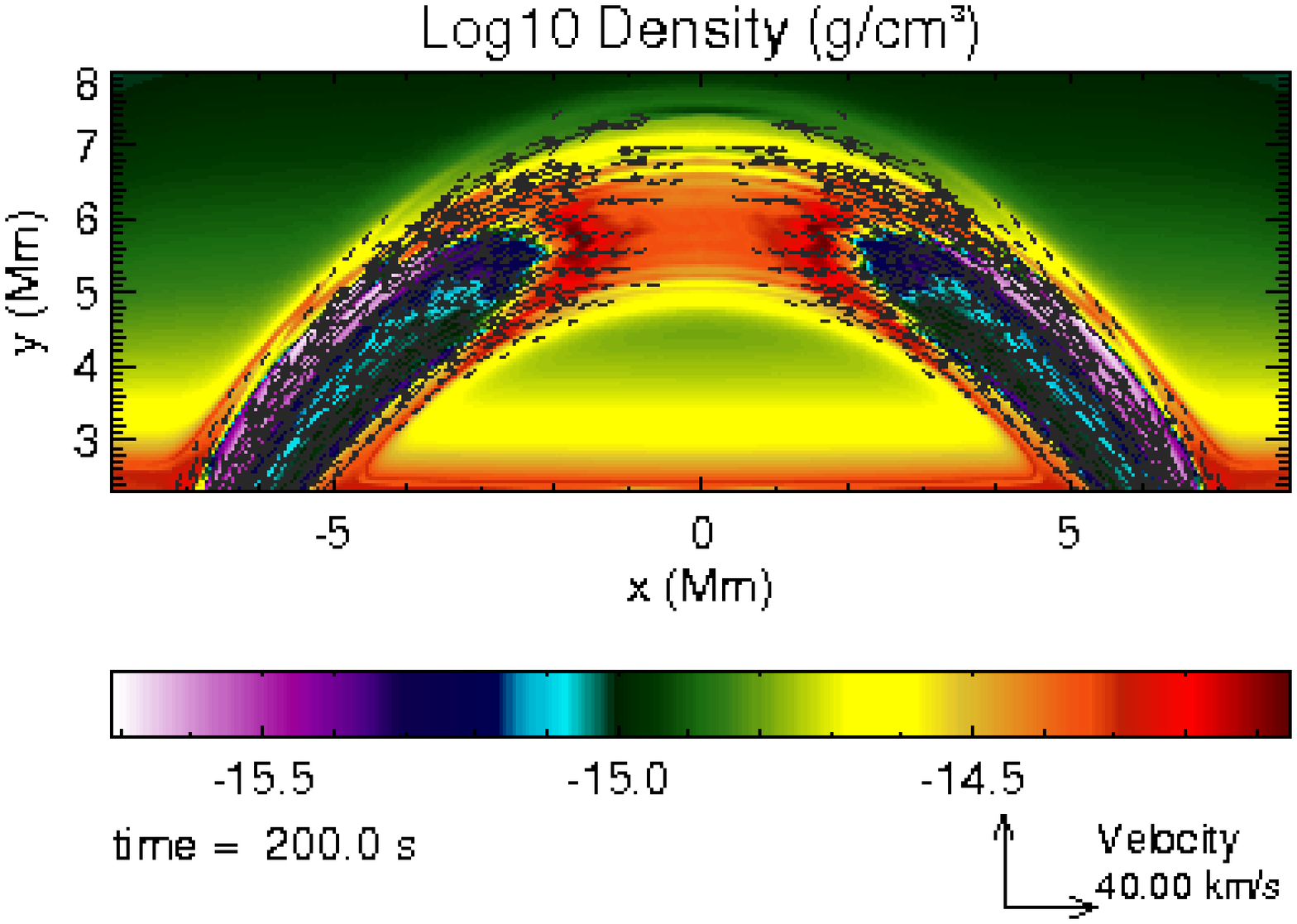}
\vspace{-0.5 cm}
\caption{Temperature (left-column) and mass density (right-column) maps in the frame-work of the 2D IMHD model in which the energy is deposited at the height of $h=1.74$ Mm above the temperature minimum at three selected moments: $t=30$ s (top panels), $t=90$ s (middle panels), and $t=200$ s (bottom panels). Compare with Fig.~\ref{Fig7}.}
\label{Fig10}
\end{figure*}

\begin{figure*}[h!]
\includegraphics[angle=90,width=7.6 cm]{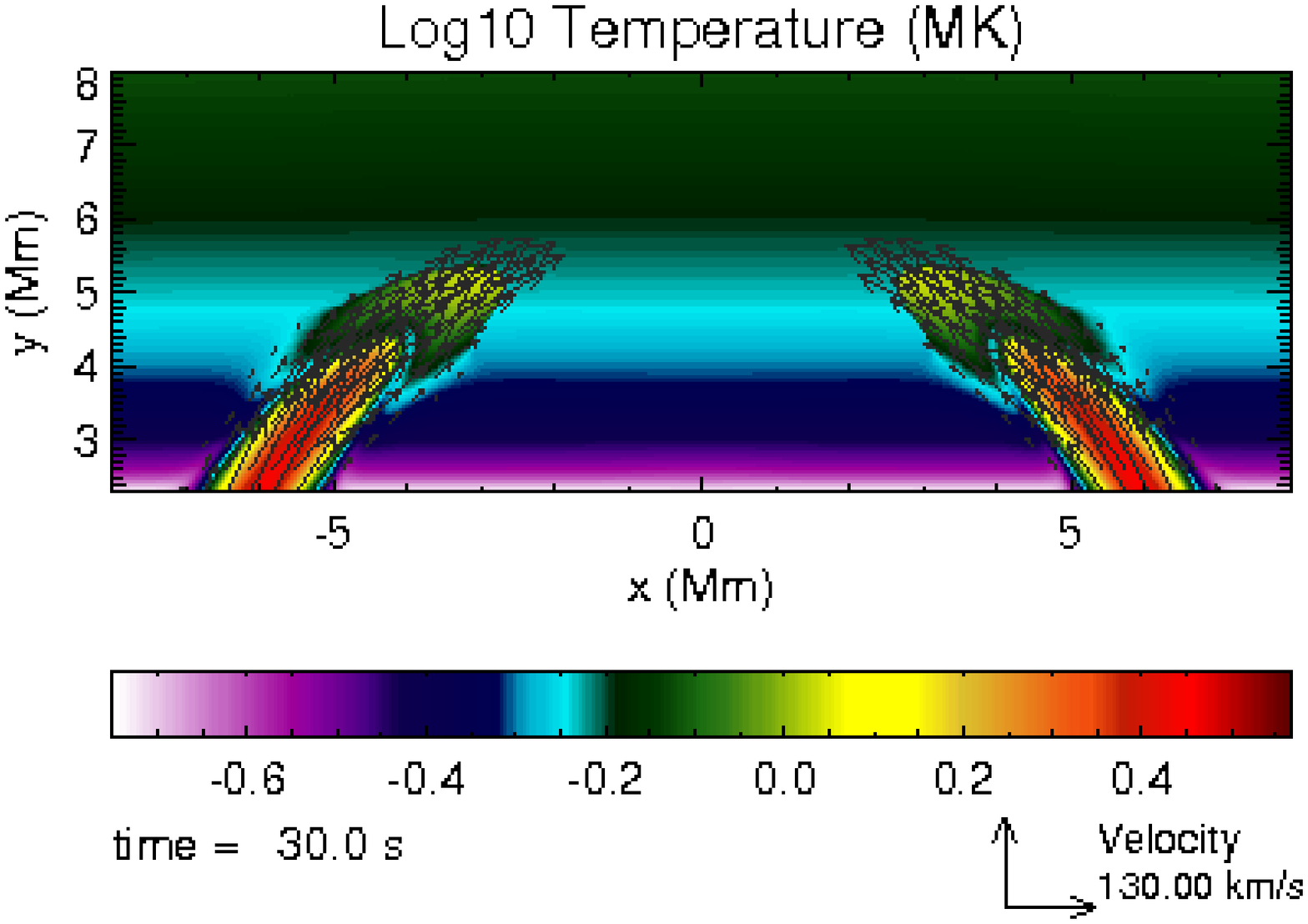}
\includegraphics[angle=90,width=7.6 cm]{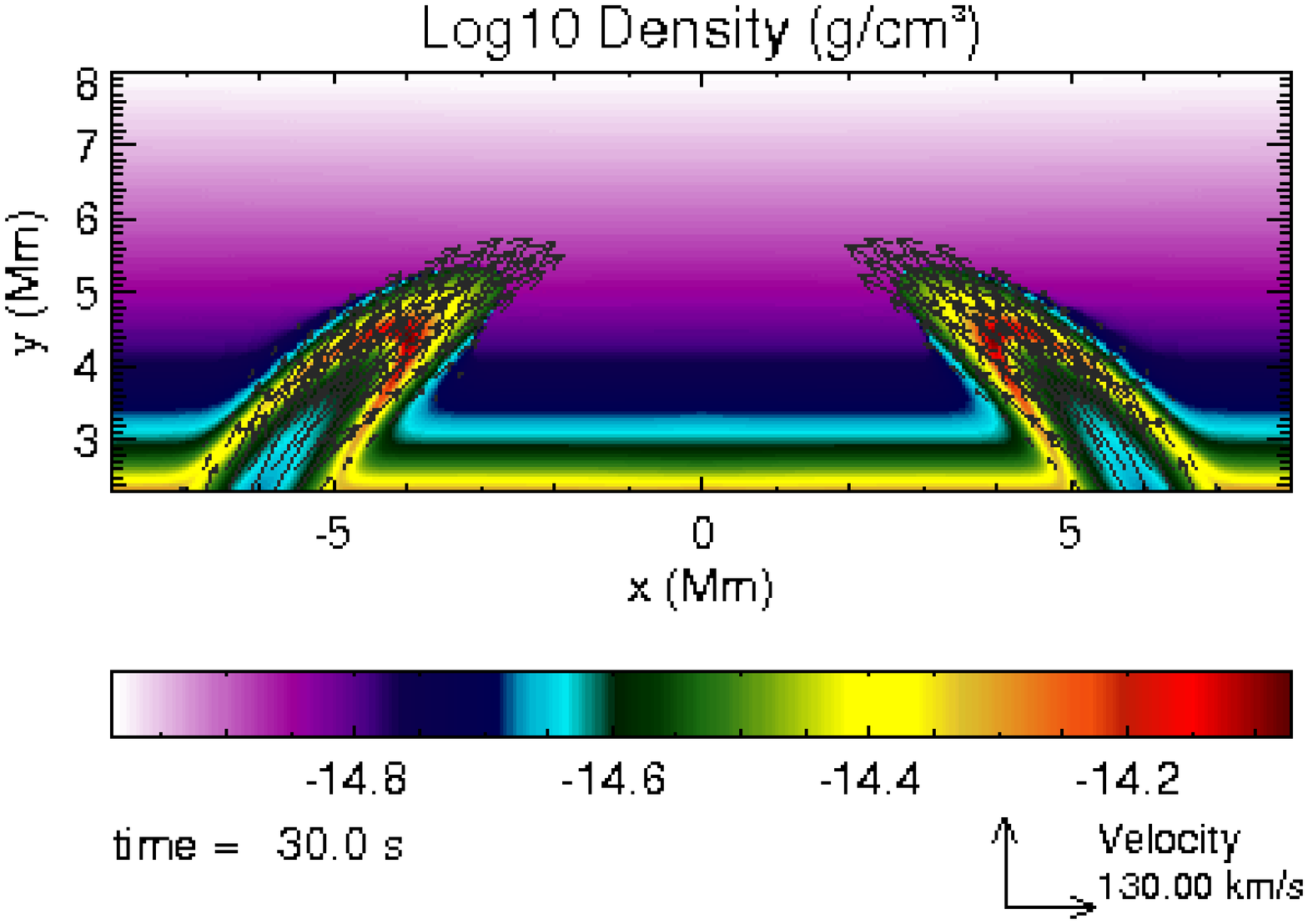}
\includegraphics[angle=90,width=7.6 cm]{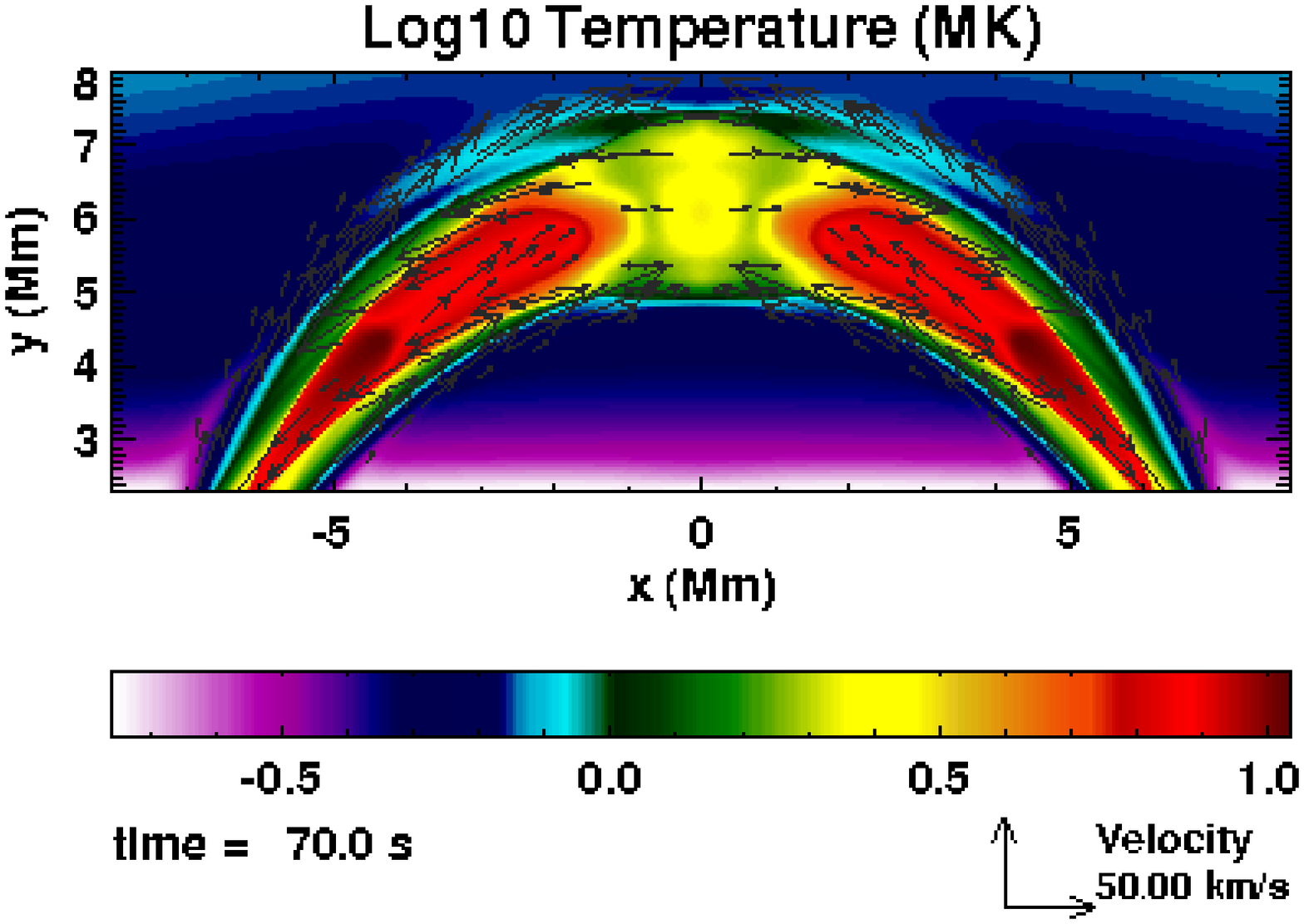}
\includegraphics[angle=90,width=7.6 cm]{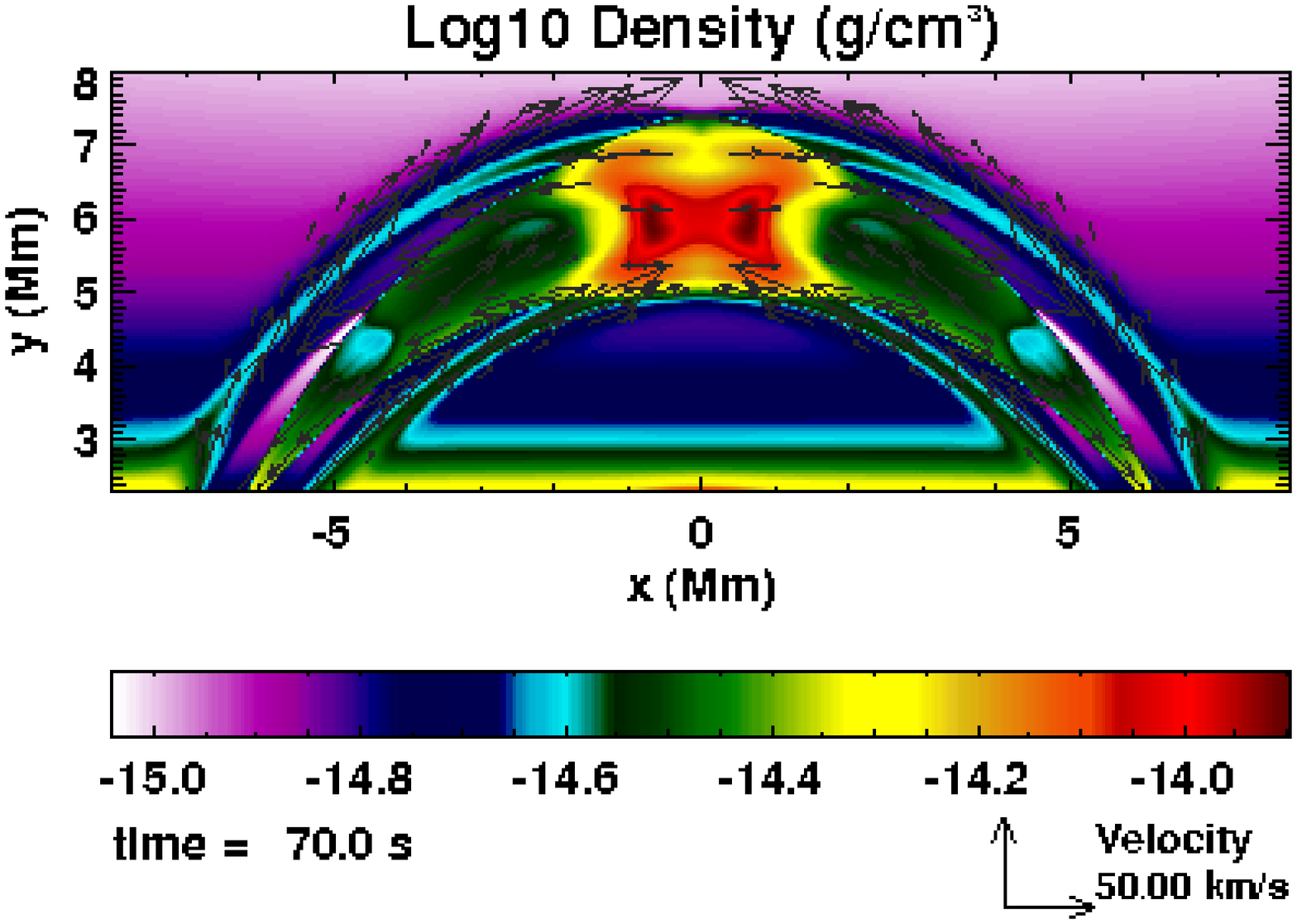}
\includegraphics[angle=90,width=7.6 cm]{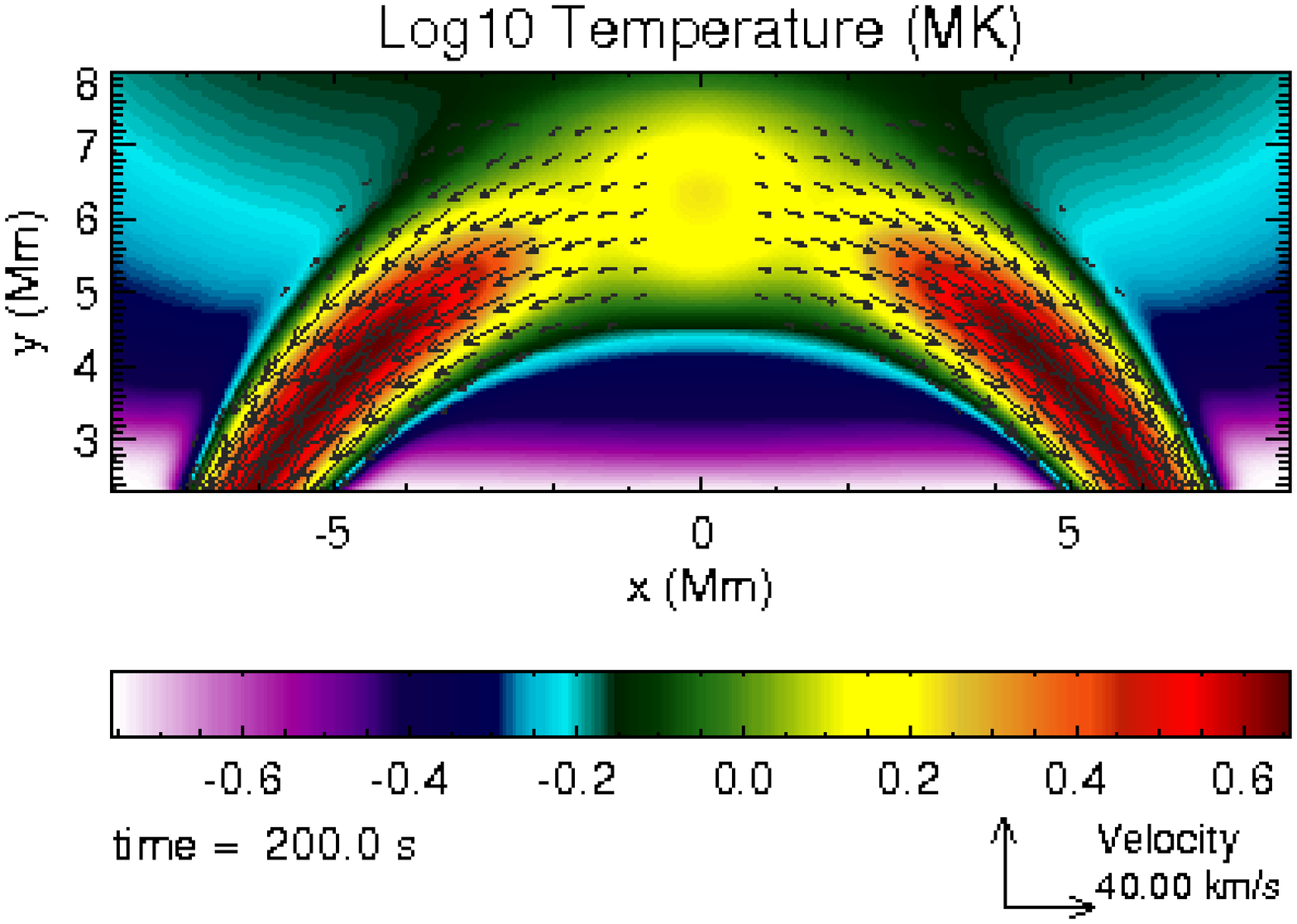}
\hspace{1.00 cm}
\includegraphics[angle=90,width=7.6 cm]{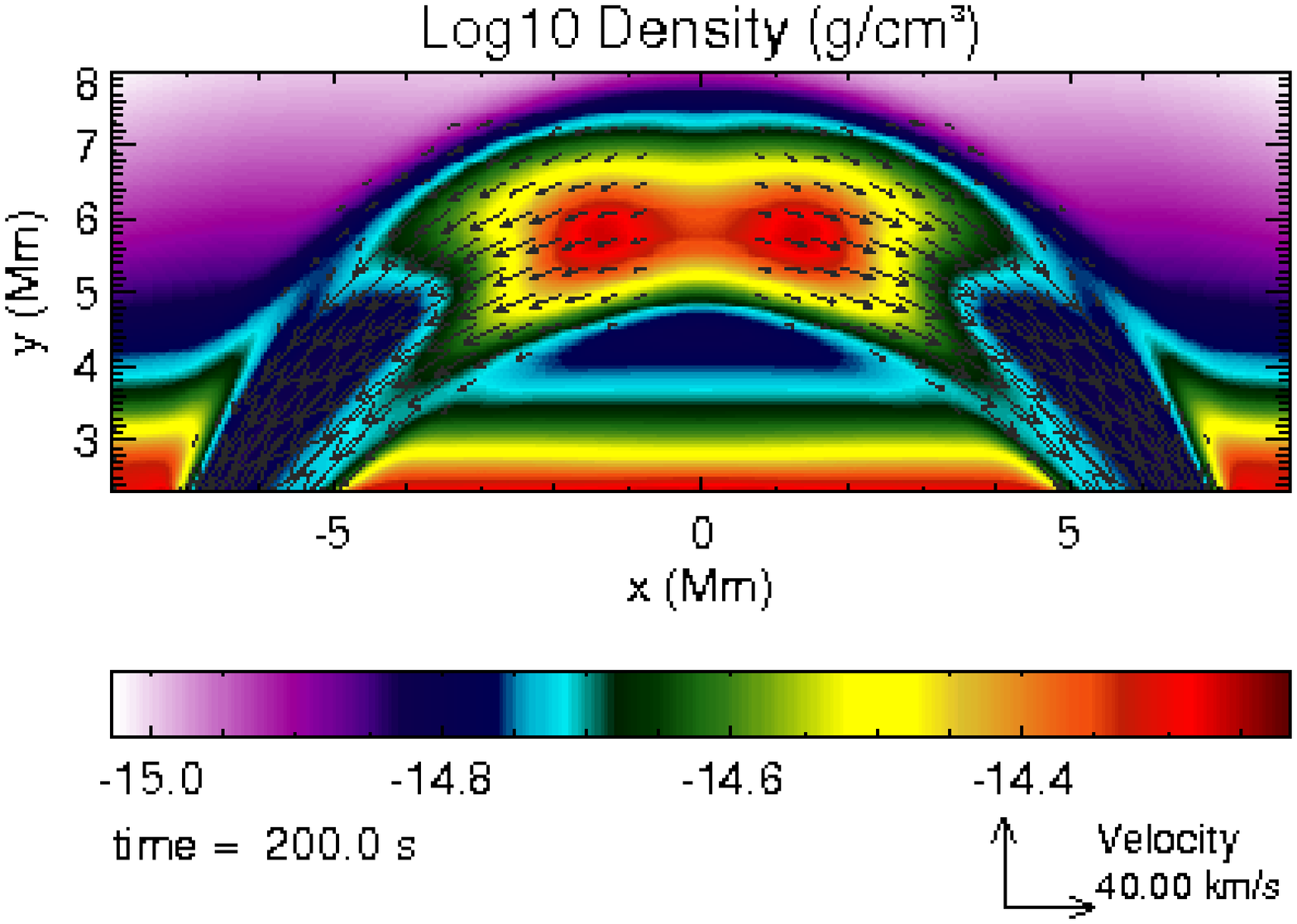}
\vspace{-0.5 cm}
\caption{Temperature (left-column) and mass density (right-column) maps in the frame-work of the 2D VDC model in which the energy is deposited at the height of $h=1.74$ Mm above the temperature minimum at $t=30$ s (top panels), $t=70$ s (middle panels), and $t=200$ s (bottom panels). Compare with Fig.~\ref{Fig7}.}
\label{Fig11}
\end{figure*}

\begin{figure*}[h!]
\includegraphics[angle=0,width=7.6 cm]{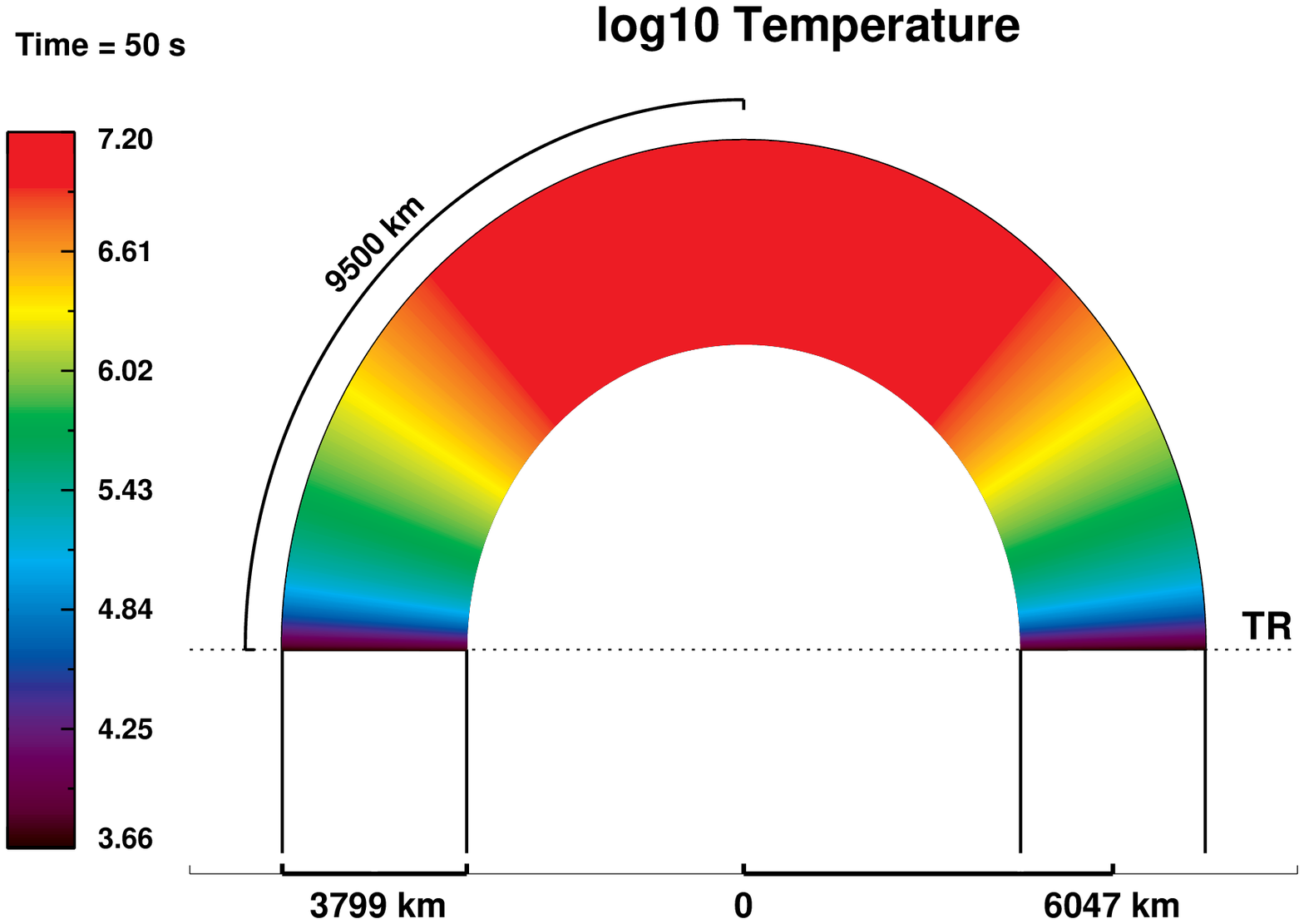}
\includegraphics[angle=0,width=7.6 cm]{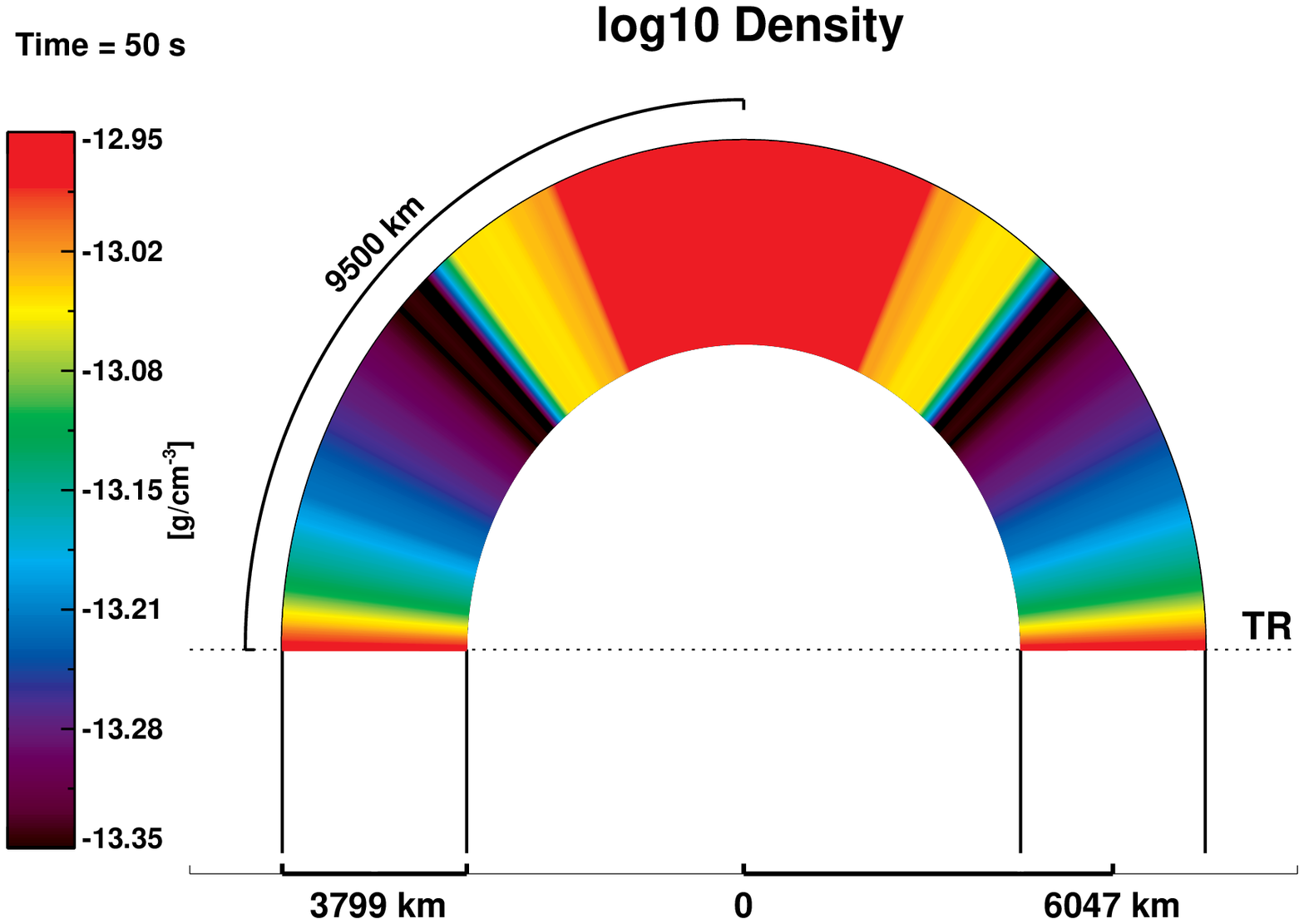}
\includegraphics[angle=0,width=7.6 cm]{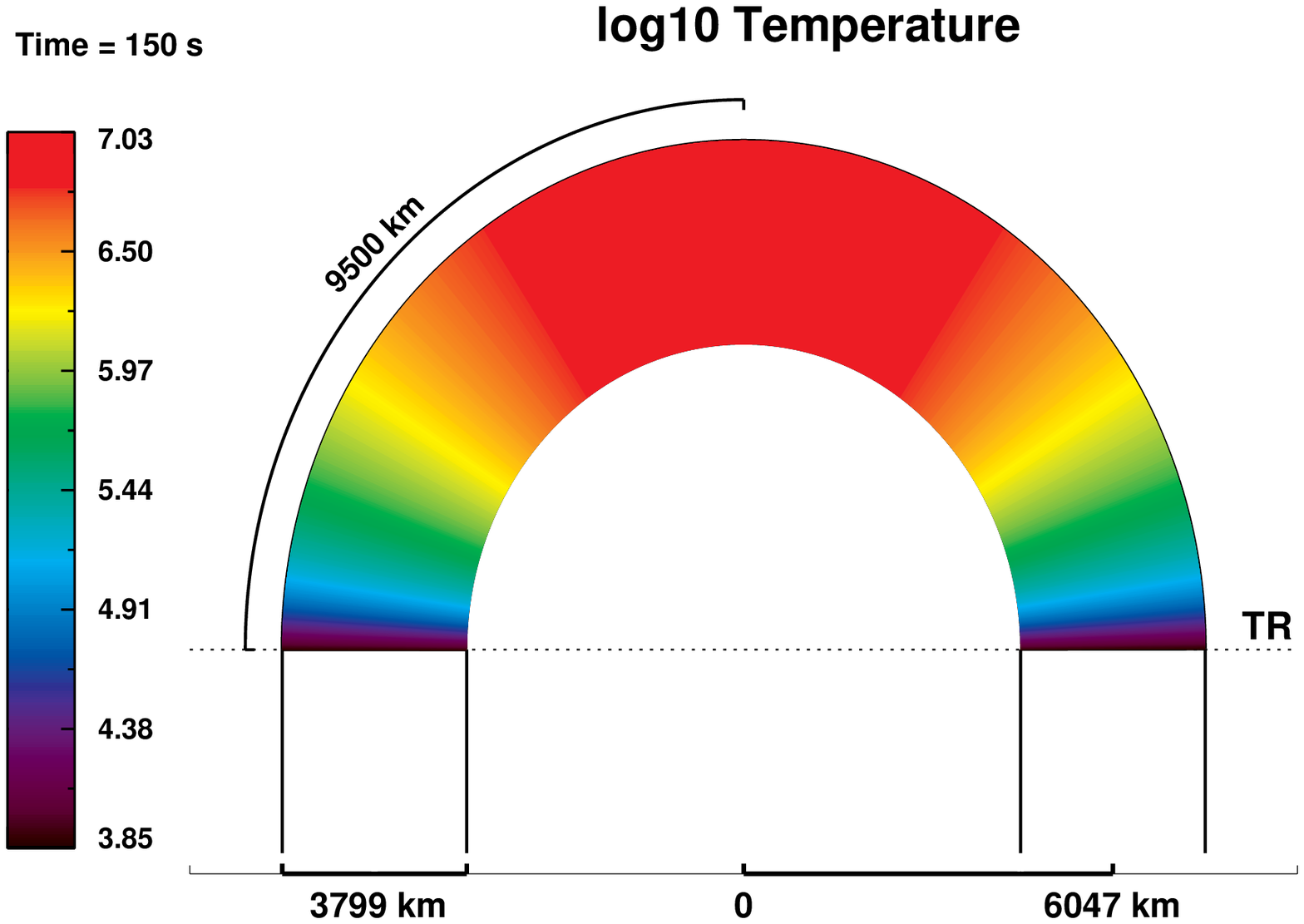}
\includegraphics[angle=0,width=7.6 cm]{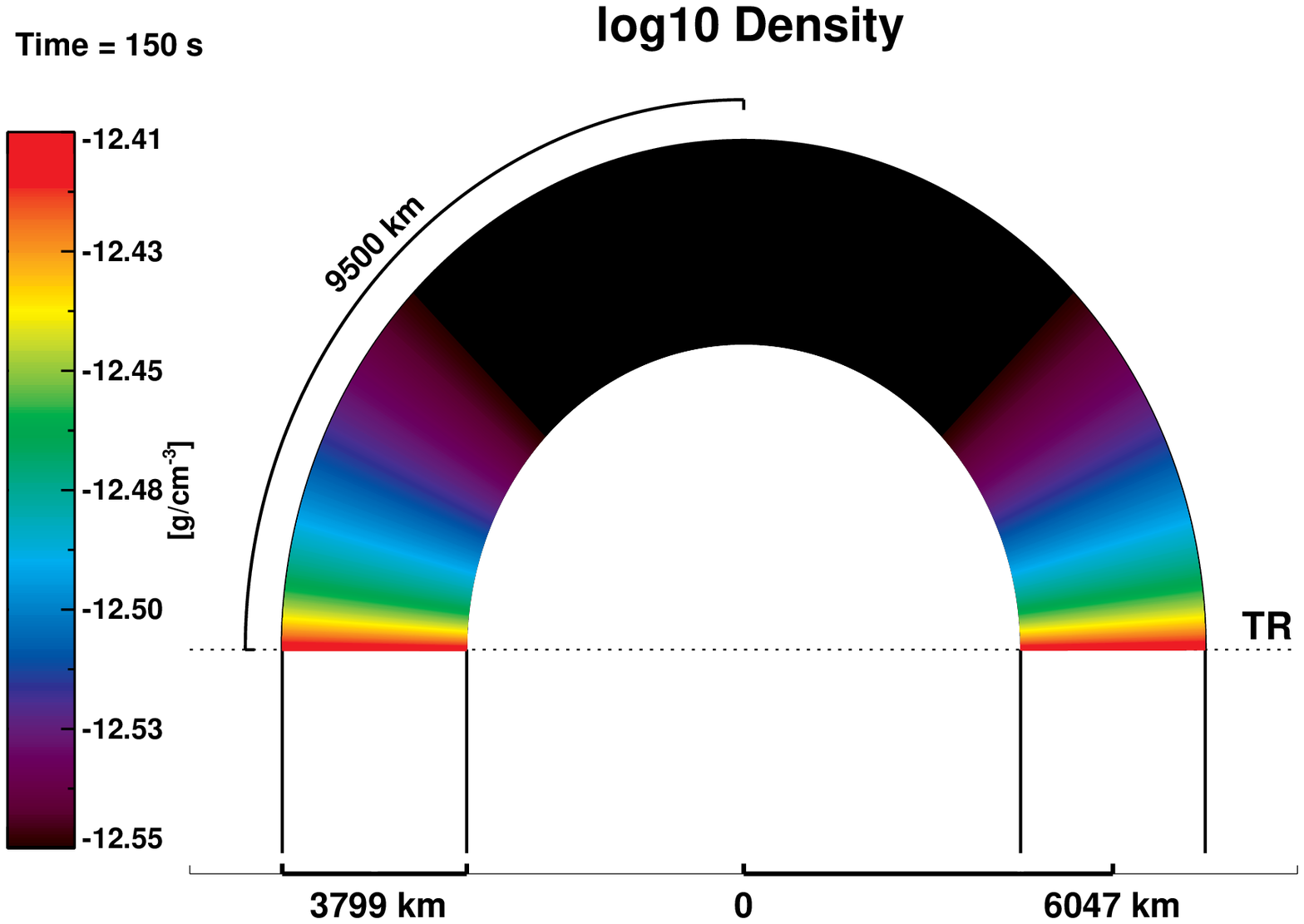}
\includegraphics[angle=0,width=7.6 cm]{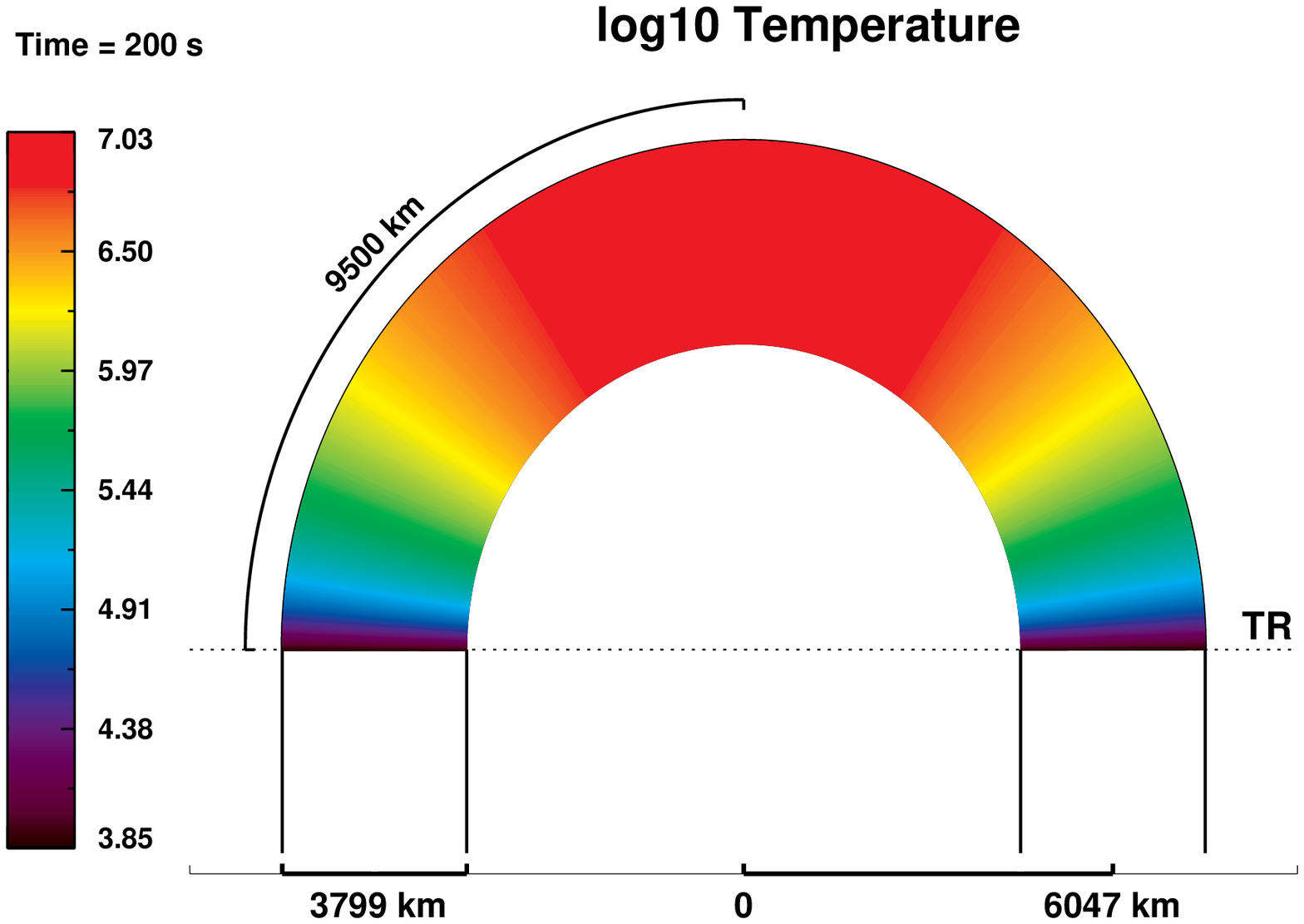}
\hspace{1.00 cm}
\includegraphics[angle=0,width=7.6 cm]{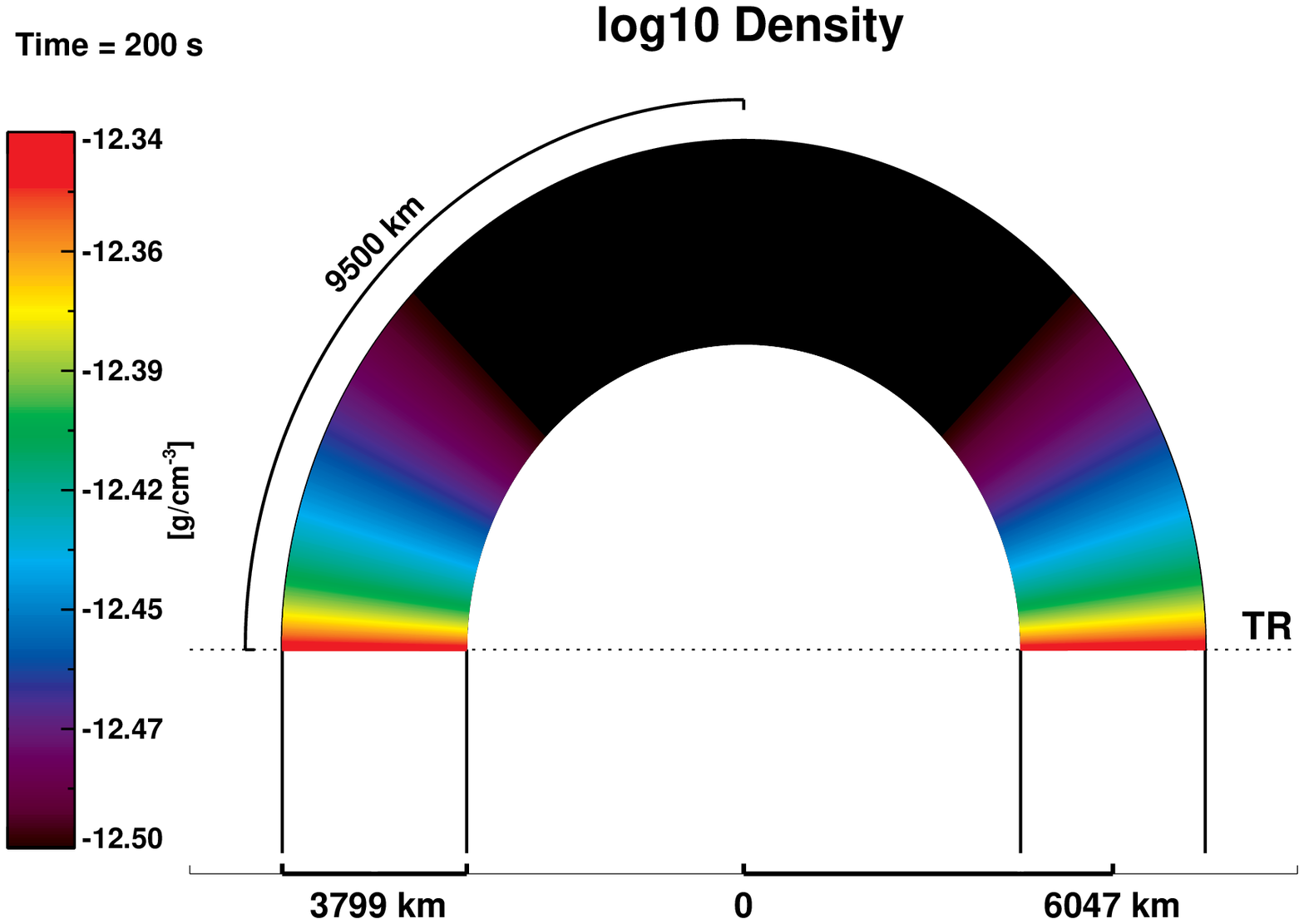}
\vspace{-0.5 cm}
\caption{The spatial distribution of plasma temperature (left-column) and mass density (right-column) in the 1D model in which the energy is deposited at the height of $h=1.74$ Mm at three selected moments: $t=50$ s (top panels), $t=150$ s (middle panels), and $t=200$ s (bottom panels).}
\label{Fig91}
\end{figure*}

The global evolution of the main physical parameters of the plasma confined in the flaring loop occurs in the same way in all models with the heating region at $h=1.74$ Mm above the temperature minimum. During the early phase of the flare, the evaporating plasma rises toward the loop-top, forming in 2D models the columns of very hot plasma in cores and much cooler plasma in the outside layers. When the chromospheric evaporation is fully developed (again, about 20 s after the onset of the noticeable X-ray emissions of the loop in the 1-8 \AA\ band), the highest temperature of the evaporating plasma at the representative height of $h=1.5$ Mm above the TR is noted for the IMHD model, being equal to $\rm T_{IMHD}=10$ MK, while for the VDC model, the temperature is $\rm T_{VDC}=2.1$ MK. In the case of the NRL model, temperature of the plasma at $h=1.5$ Mm is equal to the intermediate value of $\rm T_{NRL}=5.6$ MK.

Significant differences between the temperatures calculated in various models result from application or neglecting various physical mechanisms of energy redistribution such as viscosity and thermal conduction. For instance, exclusion of thermal conduction results in the strong local overheating of the loop core in the IMHD model. The representative temperatures of the outside layer of the loops in the IMHD and VDC models are $\rm T_{IMHD}=0.1$ MK, $\rm T_{VDC}=0.5$ MK, respectively. The very hot plasma in the core of the loop of the IMHD model at $h=1.5$ Mm is at the same time very rarefied with $\rm \varrho_{IMHD}=5.6\times10^{-16}\, g\,cm^{-3}$. The cold and dense envelopes at the edge of the loop resulted from the applied model of plasma heating, with Gaussian-like spatial distribution of the impulse, where plasma on the edges is less heated and expands less than in the central part of the loop. However, an exclusion of thermal conduction in the IMHD model prevents any energy transfers perpendicular to the loops axis and even more exaggerates local differences in temperature. Much colder plasma in the core of the loop of the VDC model has $\rm \varrho_{VDC}=2.5\times10^{-15}\, g\,cm^{-3}$, but the densest plasma occurs in the NRL model at the same stage of the evolution, reaching $\rm \varrho_{NRL}=7.6\times10^{-14}\, g\,cm^{-3}$. Note that for the IMHD and VDC models at the same stage of the evolution massive up-flows of the evaporating plasma take  place, while in the NRL model the plasma is still static at the reference level of $h=1.5$ Mm above the TR, contemporarily moving up in the lower parts of the loop. Conversely to the models described above, for the 2D models the rising pillars of the plasma do not cause any important compression of the plasma present in the upper parts of the loop, where loop-tops remains moderately cold, with  $\rm T_{IMHD}=T_{VDC}=0.5$ MK. As a result of plasma substantial compression, the top of the loop in the NRL model is much hotter, with $\rm T_{NRL}=10.6$ MK. This is consistent with the observational Soft X-ray Telescope (SXT/Yohkoh) data, which show the temperature of the loop-top sources calculated from the ratio of images in the two thickest filters typically $\lesssim\,15$ MK \citep{McTier93, Dos99}.

The maxima of the X-ray 1-8 \AA\ emissions occurs at $\rm t_{maxIMHD}=90$ s in the IMHD model, at $\rm t_{maxVDC}=70$ s in the VDC model and at $\rm t_{maxNRL}=150$ s in the NRL model. The substantial difference between relatively fast occurrence of the maxima in the IMHD and VDC models and delayed maximum in the NRL model results mostly from very extended in time but rather slow increase of the 1-8 \AA\ emissions in the NRL model, which is still lasts between t=90 s and t=150 s. However, during the early phases of the evolution of plasma in the framework of all three models, the calculated 1-8 \AA\ fluxes rise in a very similar way and up to roughly t=70 s. The difference is caused by differences in the dynamics of the plasma motions and densities along the loops in 1D and 2D models with the mass densities of the plasma in the loop-top are equal to $\rm \varrho_{IMHD}=8\times10^{-15}\, g\,cm^{-3}$, $\rm \varrho_{VDC}=1\times10^{-14}\, g\,cm^{-3}$ and $\rm \varrho_{NRL}=2.6\times10^{-13}\, g\,cm^{-3}$. In the case of the IMHD model, the plasma flows downward along the central part of the loop, but it still rises in the ambient layer toward the top of the loop. In the VDC model, the  plasma flows down through the whole legs of the loop, while in the NRL model it moves towards the loop-top.

For both 2D models, the calculated plasma temperatures at the loop-top are equal to $\rm T_{IMHD}=1.8$ MK and $\rm T_{VDC}=2.5$ MK and they are much lower than the temperatures derived from the calculated synthetic X-ray fluxes emitted by the modeled loops. For example, the GOES-like temperatures are equal to $\rm T_{IMHD-GOES}=13$ MK and $\rm T_{VDC-GOES}=7$ MK for the IMHD and VDC models, respectively. For the NRL model, the temperature of the loop-top plasma is equal to $\rm T_{NRL}=11.5$ MK at $t=150$ s, and it almost fits the \textit{GOES}-like temperature of $\rm T_{NRL-GOES}=10.5$ MK. The \textit{GOES}-like temperature peaks in the NRL model about 100 s before the maximum of the emitted 1-8 \AA\ flux, reaching $\rm T_{NRL-GOES}=15$ MK at $\rm t=55$ s.

\begin{figure*}[h!]
\begin{center}
\includegraphics[angle=90,width=6.5 cm]{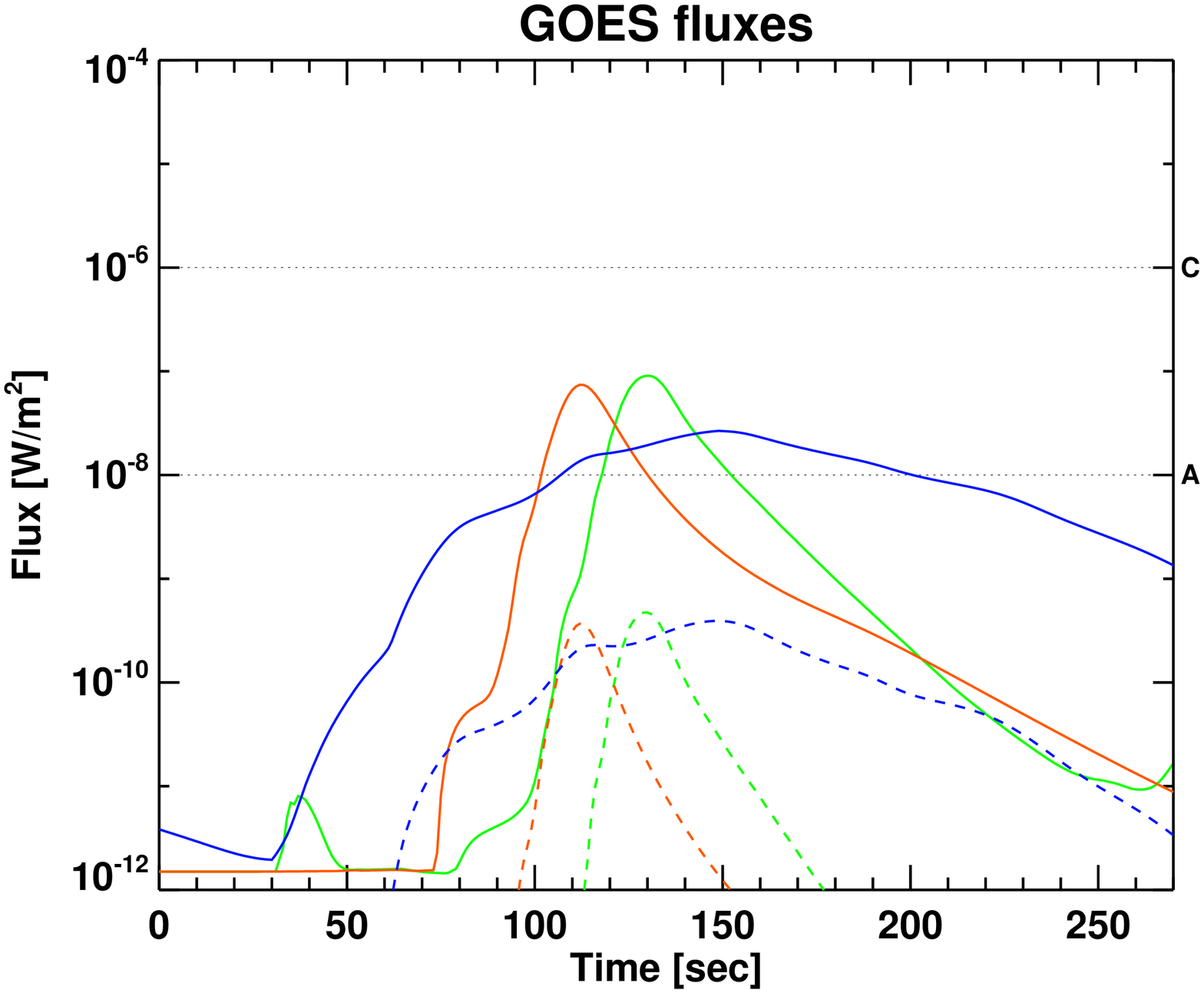} 
\includegraphics[angle=90,width=6.5 cm]{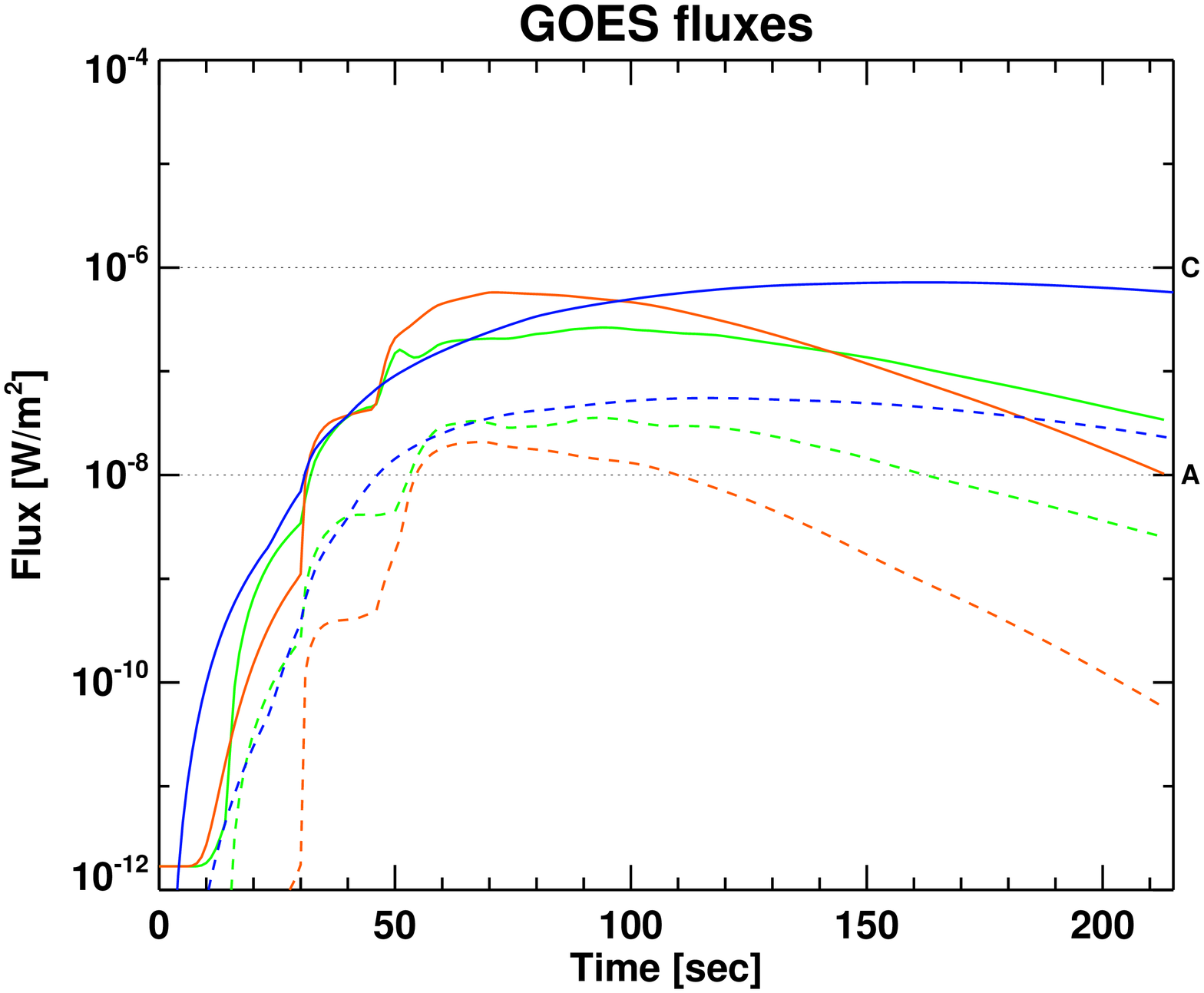} 
\includegraphics[angle=90,width=6.5 cm]{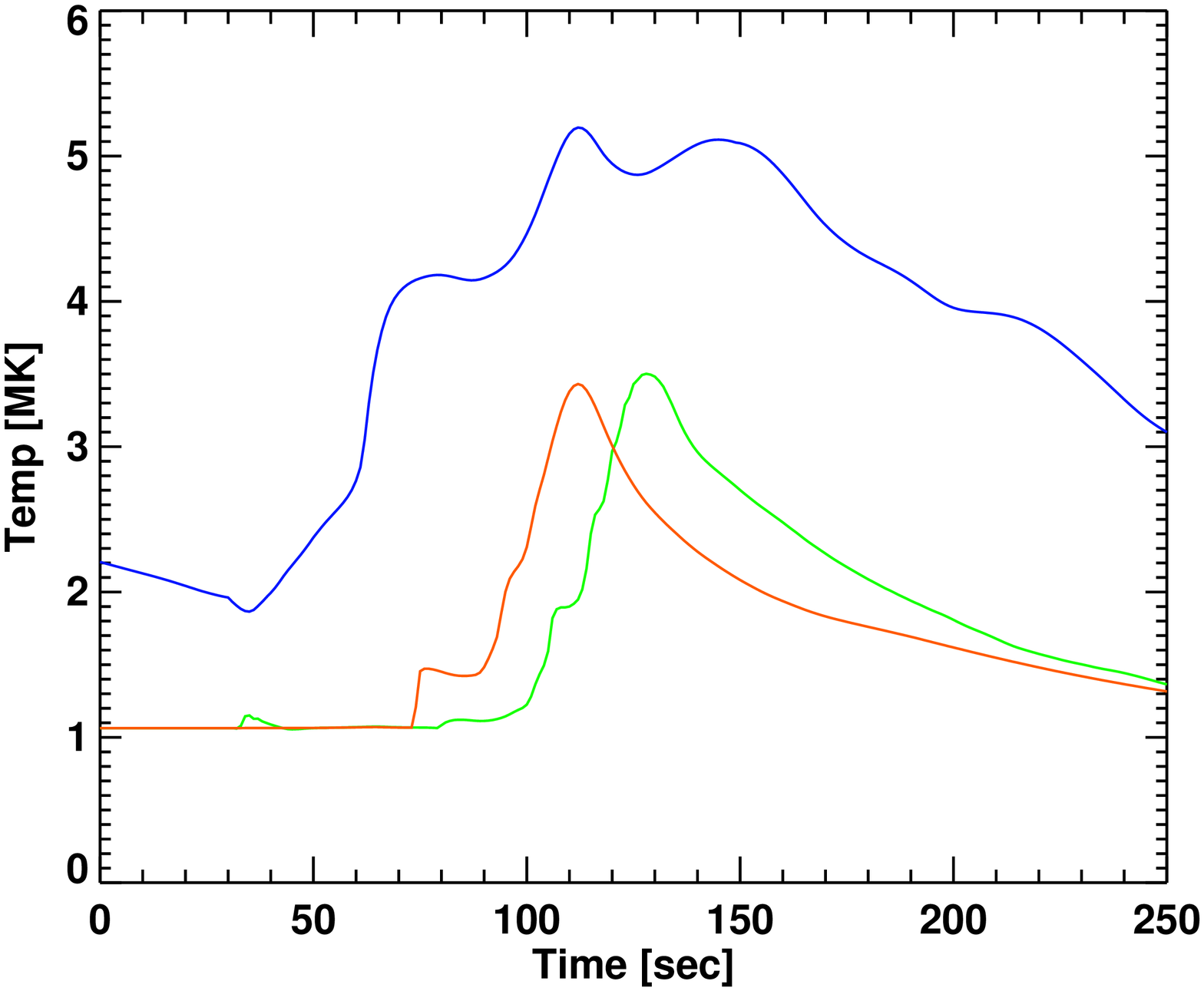} 
\includegraphics[angle=90,width=6.5 cm]{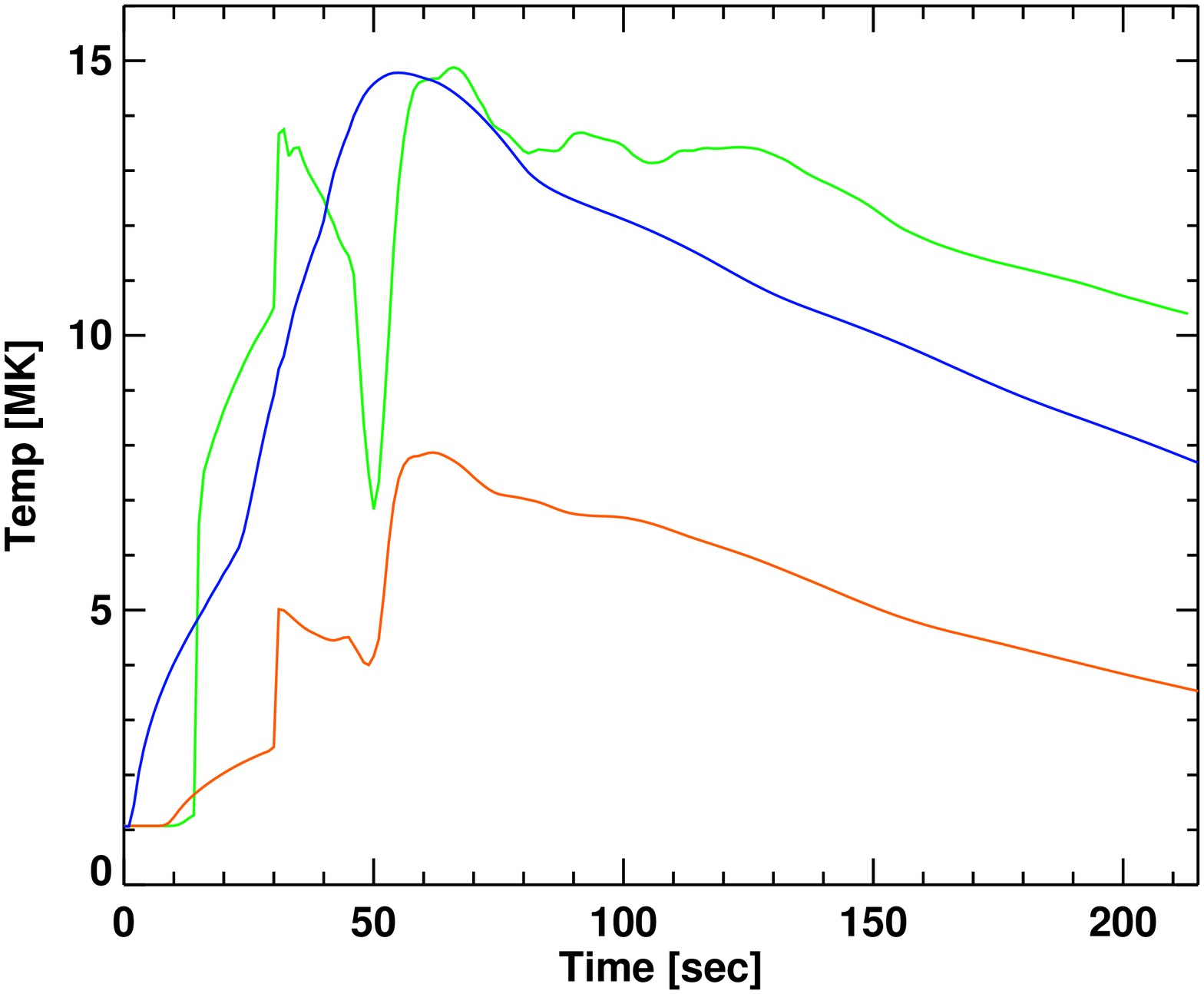} 
\includegraphics[angle=90,width=6.5 cm]{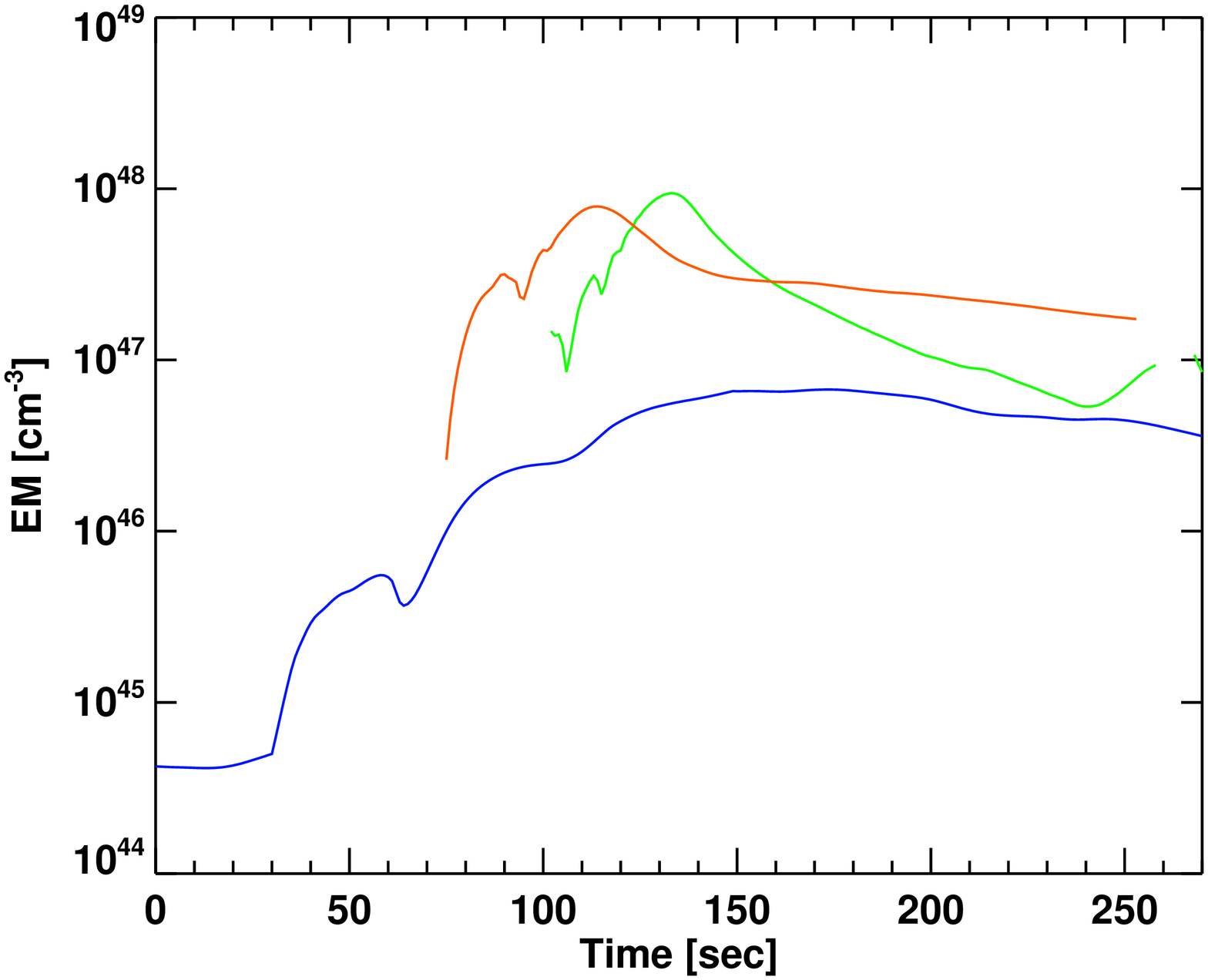} 
\includegraphics[angle=90,width=6.5 cm]{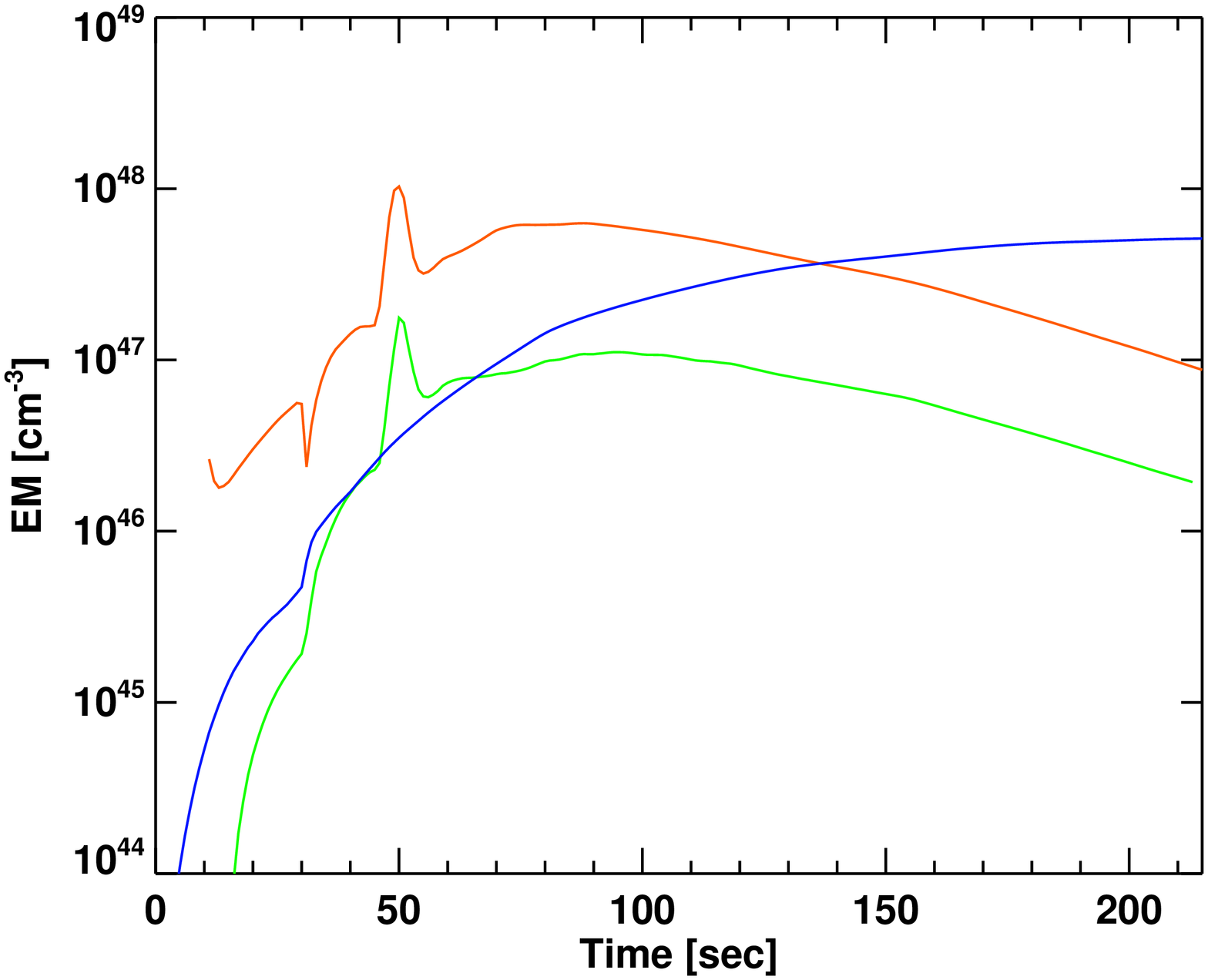} 
\end{center}
\vspace{-0.5 cm}
\caption{Comparison of 1D and 2D models in which the energy is deposited at the height of $h=1.24$ Mm (on the left) and for the height of $h=1.74$ Mm (on the right) above the temperature minimum. The upper panels present the calculated GOES fluxes in the 0.5-4 \AA\ band with dashed lines and 1-8 \AA\ fluxes with solid lines. The middle panels show the temperature and lower panels the emission measure. The color codes are: 1D model--blue, 2D IMHD model--green, 2D VDC--red.}
\label{Fig12}
\end{figure*}

During the late, gradual decay phase of the evolution, at t=200 s in the case of the IMHD and VDC models and t=250 s in the case of the NRL model, the loop-top regions expand, and the plasma flows down along the legs of the loops in all the models. The temperature of the expanding plasma remains on the constant level of $\rm T_{IMHD}=1.8$ MK in the IMHD model, or slowly decreases, reaching at that moment $\rm T_{VDC}=1.4$ MK and $\rm T_{VDC}=7.3$ MK for the VDC and NRL models, respectively. Again, the difference in the temporal variations of the loop-top temperatures results from differences in the dynamics of the plasma motions and mass densities along the loops in the 1D and 2D models. At the representative height of $h=1.5$ Mm above the TR, the plasma temperature in the axial part of the loops is of the order of $\rm T_{IMHD}=10$ MK in the case of the IMHD model and only of $\rm T_{VDC}=4.5$ MK in the case of the VDC model. For the NRL model the temperature at the same height is equal to $\rm T_{NRL}=4.1$ MK. Plasma densities are: $\rm \varrho_{IMHD}=1\times10^{-15}\, g\,cm^{-3}$, $\rm \varrho_{VDC}=1.6\times10^{-15}\, g\,cm^{-3}$, and $\rm \varrho_{NRL}=4.8\times10^{-13}\, g\,cm^{-3}$. Besides, the representative temperatures of the ambient plasma in all the 2D models remain relatively high, and they are equal to $\rm T_{IMHD}=3.1$ MK and $\rm T_{VDC}=1.2$ MK, respectively.

\section{Discussion and conclusions}

 In the case of a flare with the heating region located in the upper chromosphere the main observational characteristics of the flare \emph{e.g.} soft X-ray light curves, derived \emph{GOES}-like temperatures, and emission measures, differ significantly in various models. The noticeable X-ray emissions of the loops in the 1-8 \AA\ band start to be seen in the 1D case just after $t=30$ s, while in the 2D models about $t=70-80$ s after the onset of the heating, if not taking into account a small ,,precursor" as noticed in the 2D VDC model. The precursor present in the VDC model is caused by emission of the uprising relatively dense and cold plasma, which initially is hot and dense enough to be detectable in the GOES range (Fig.~\ref{Fig12}), but stretched and thus decompressed temporarily to disappear. This appears again when the gradual increase of mass density and continuous heating trigger the noticeable emissions again (cf., Fig.~\ref{Fig13}).

\begin{figure}[h]
\includegraphics[angle=90,scale=0.235]{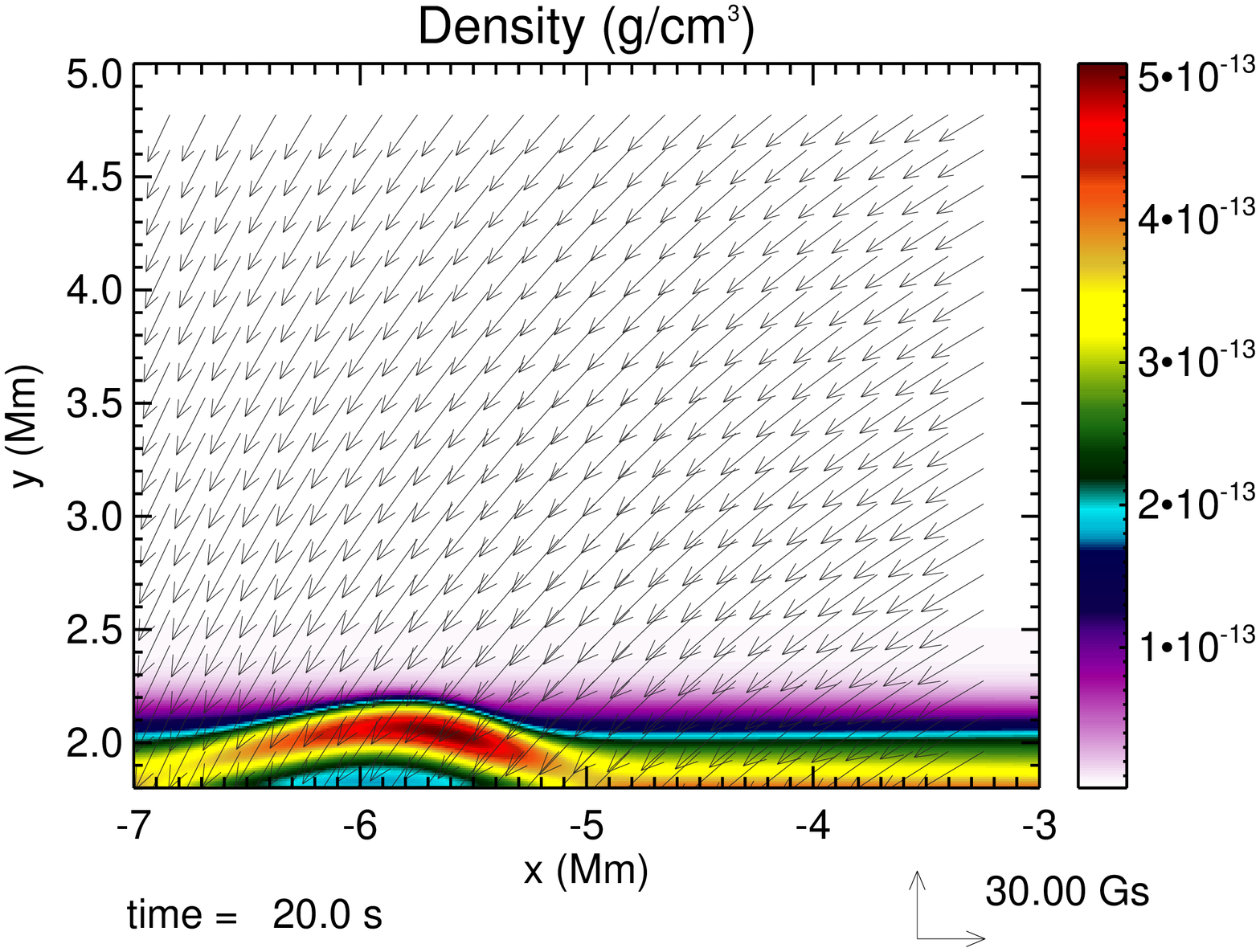} 
\includegraphics[angle=90,scale=0.235]{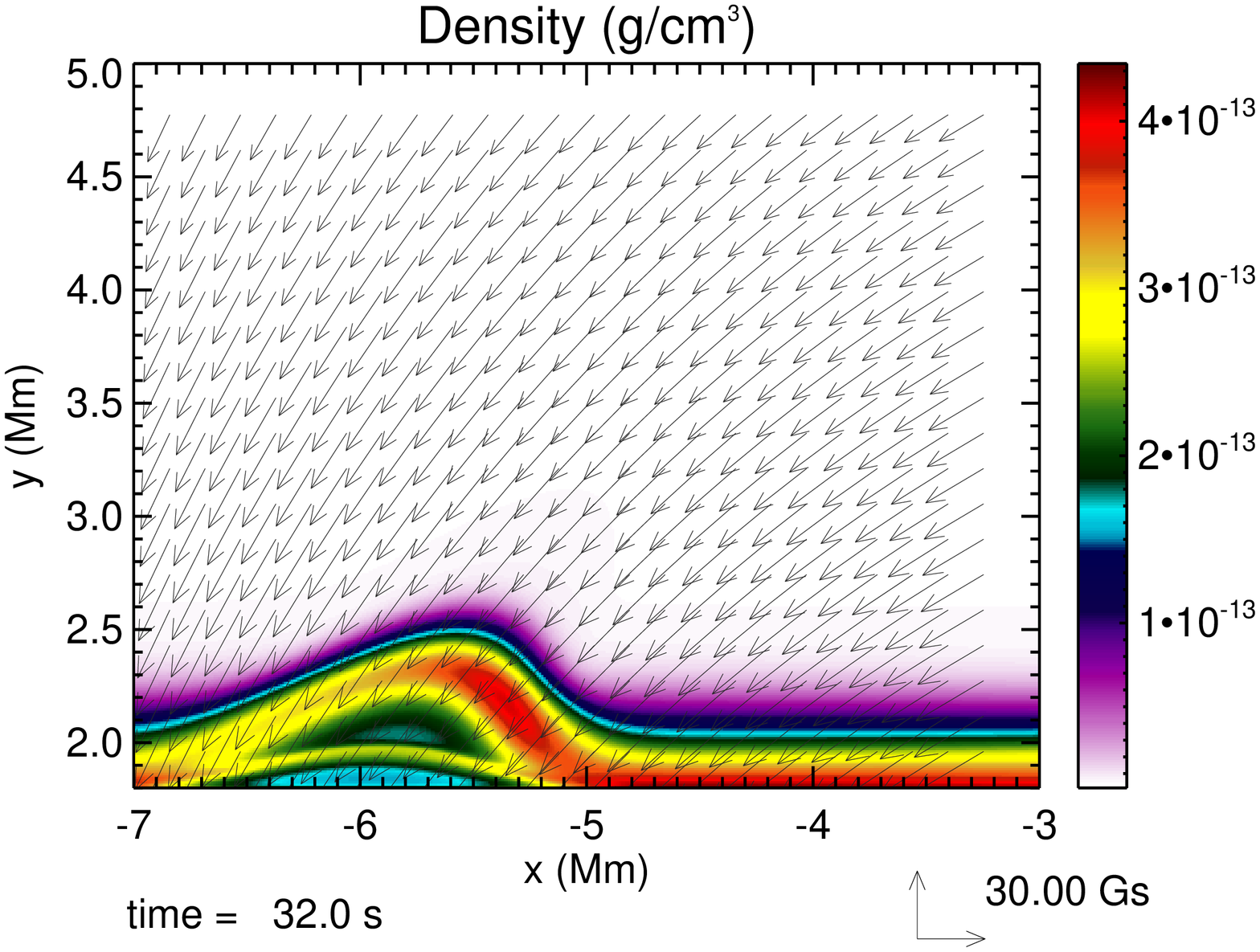} 
\includegraphics[angle=90,scale=0.235]{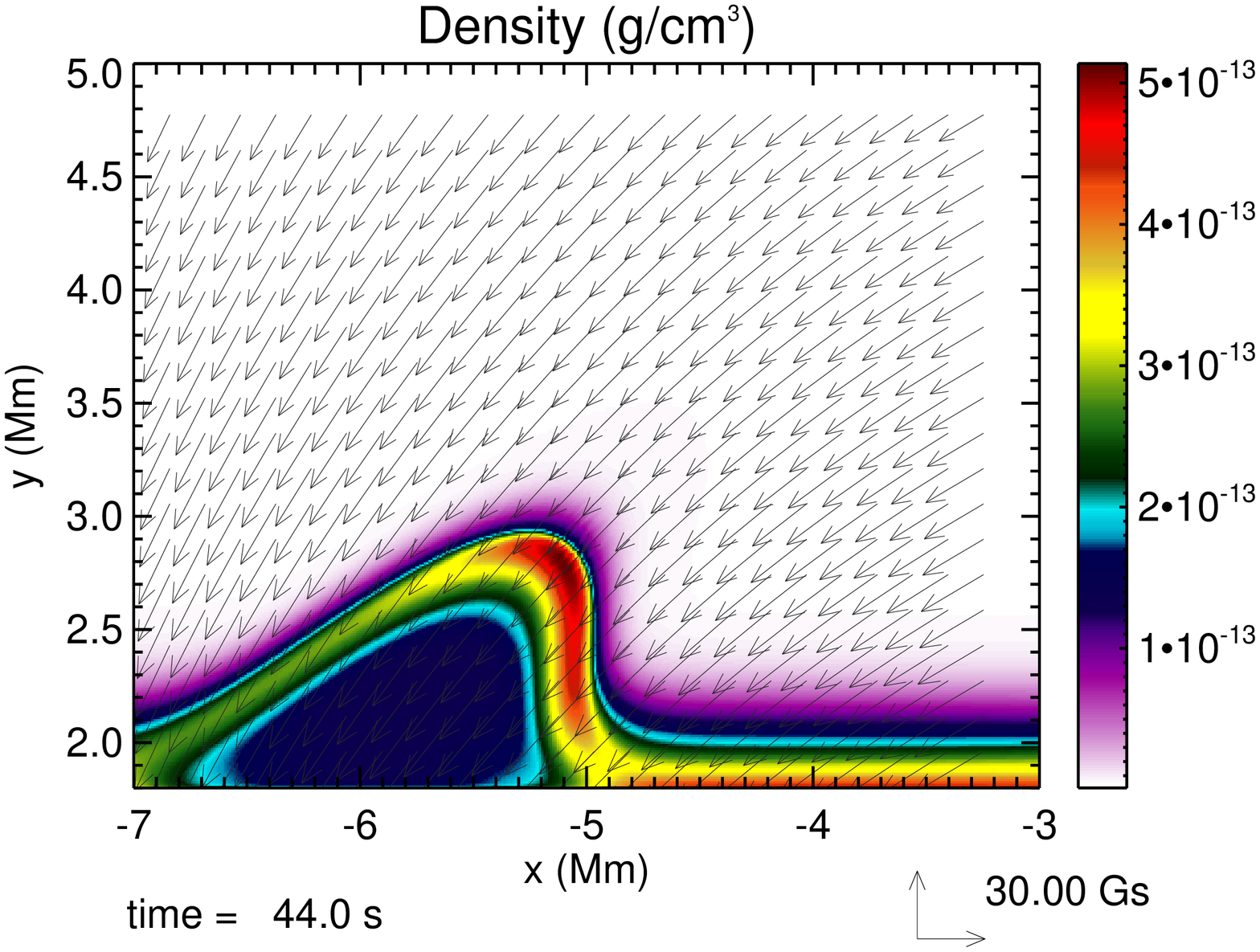} 
\includegraphics[angle=90,scale=0.235]{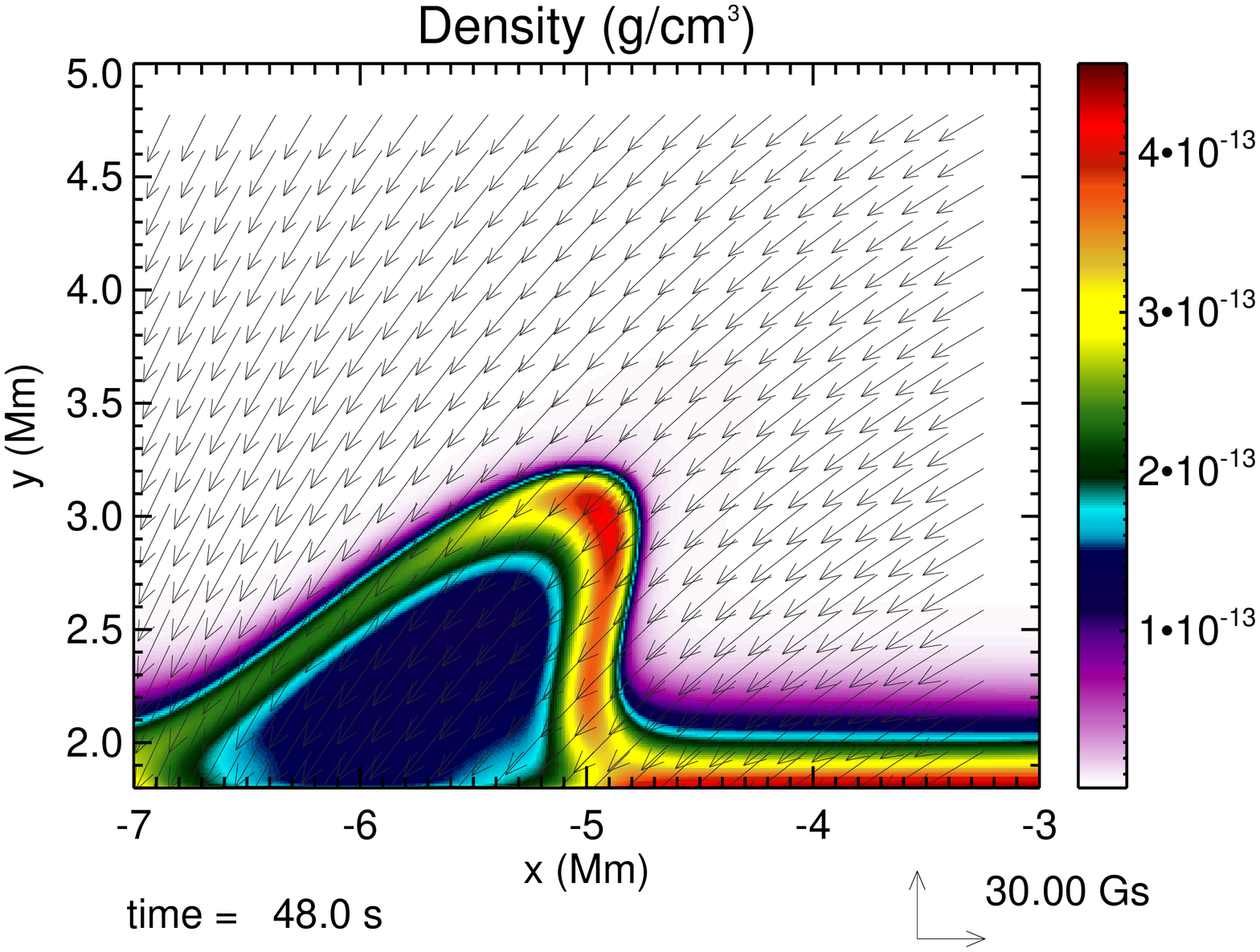} 
\includegraphics[angle=90,scale=0.235]{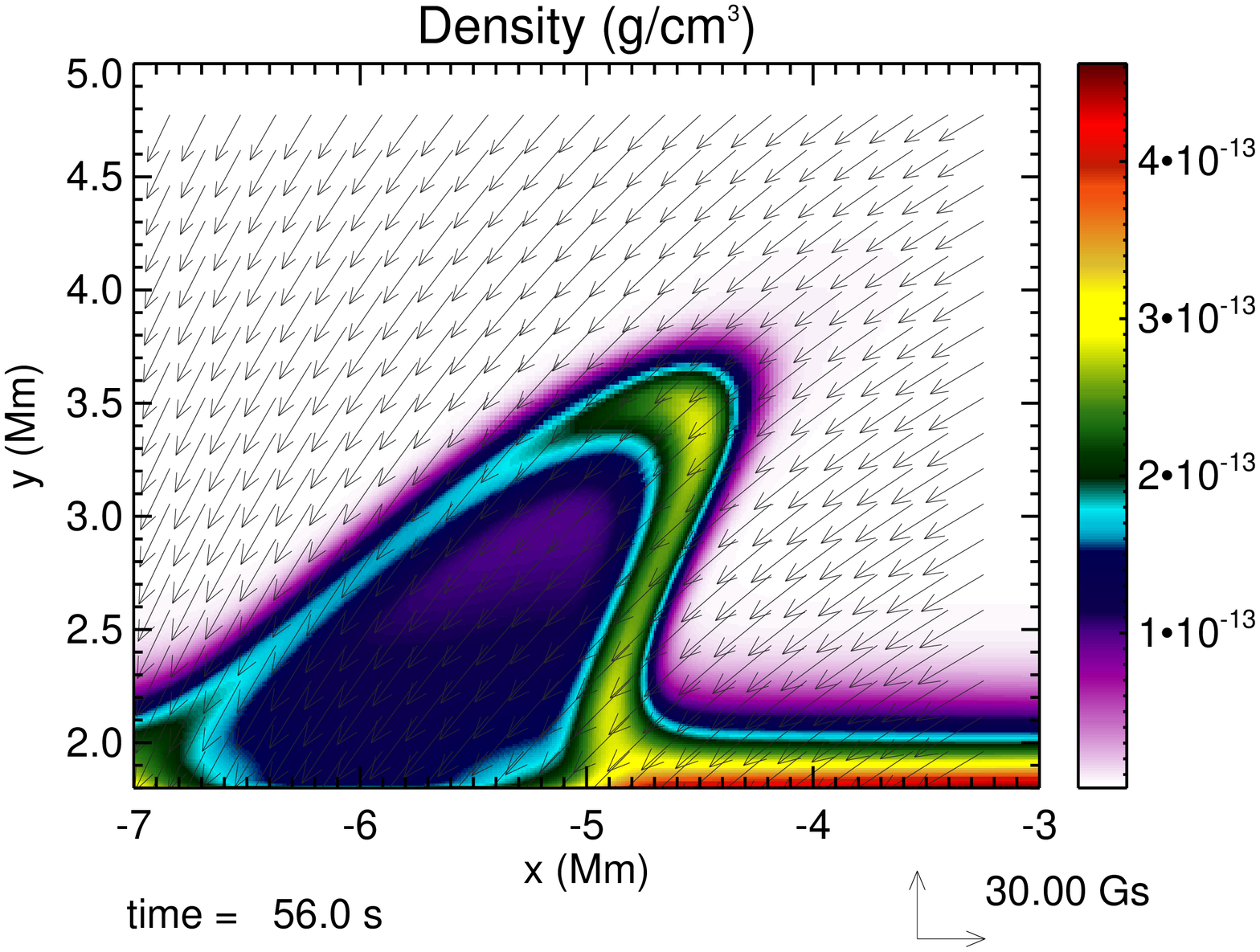} 
\includegraphics[angle=90,scale=0.235]{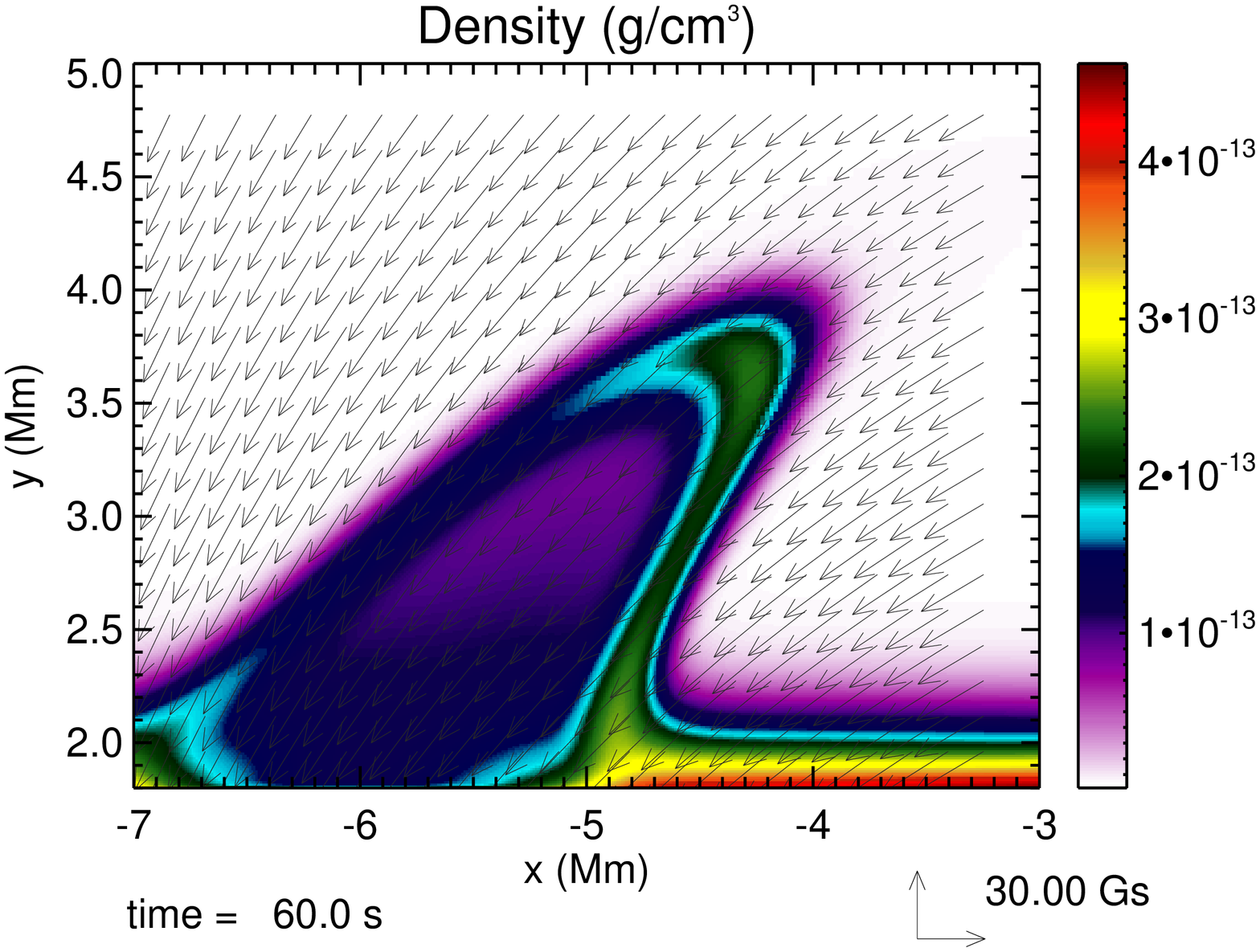} 
\includegraphics[angle=90,scale=0.235]{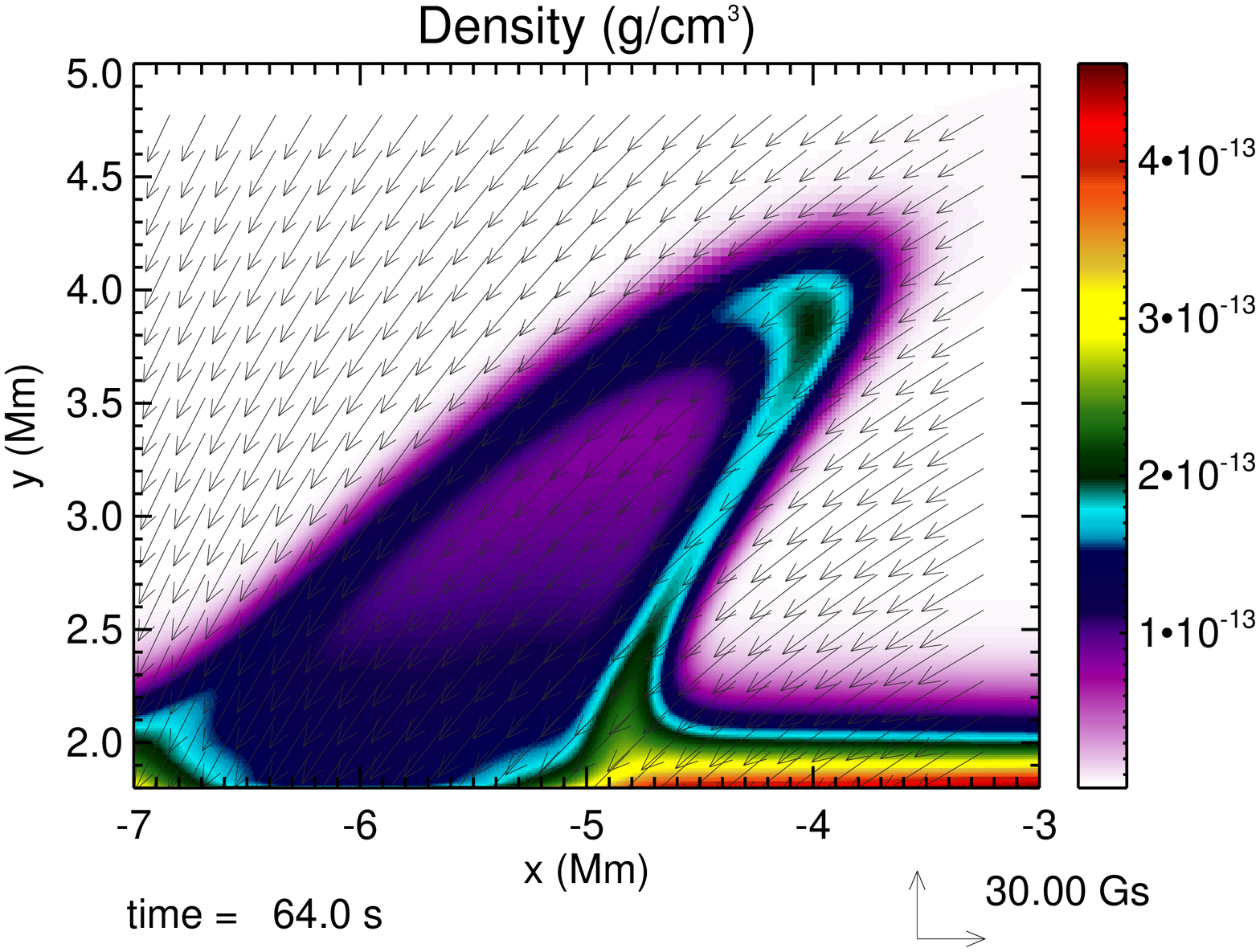} 
\includegraphics[angle=90,scale=0.235]{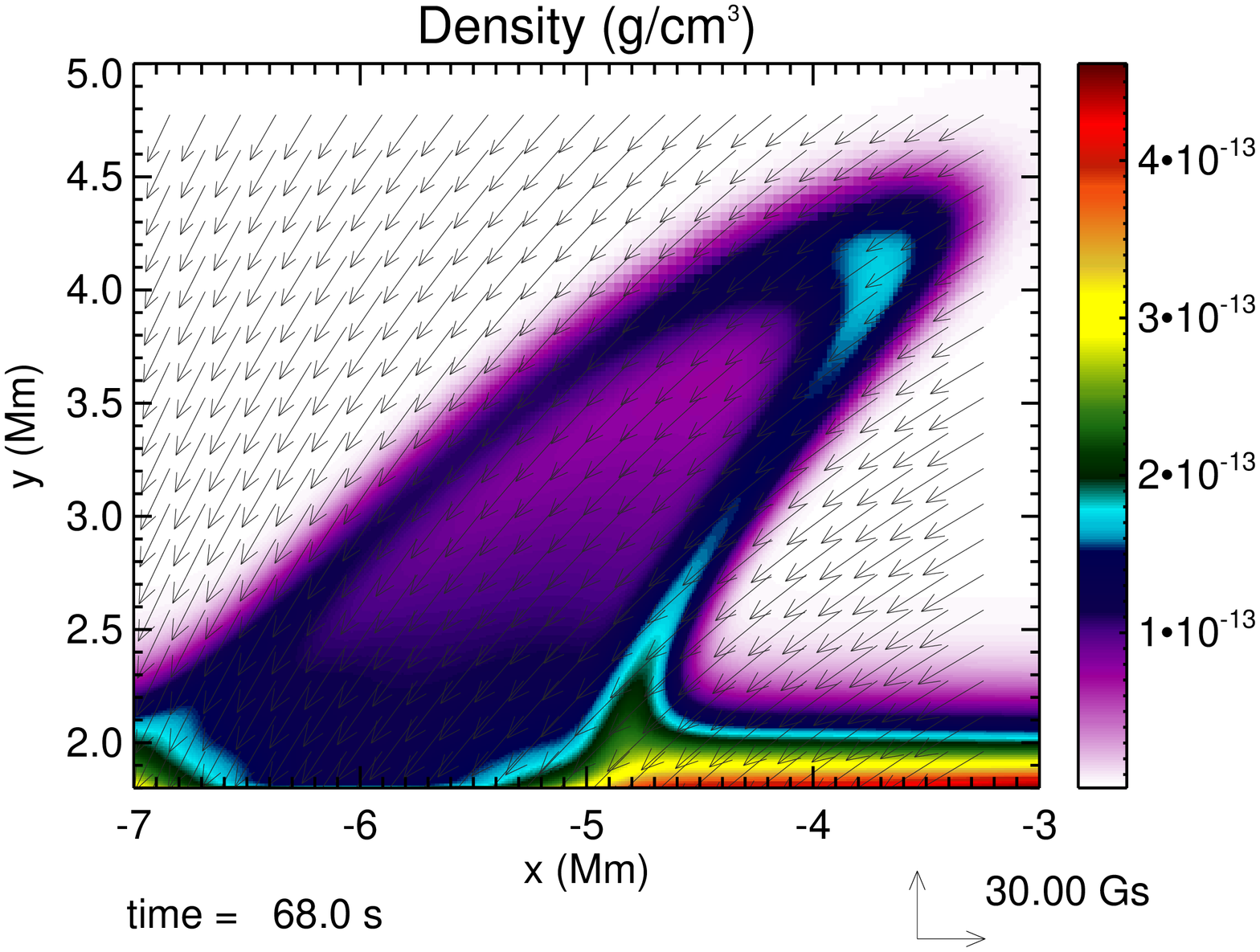} 
\includegraphics[angle=90,scale=0.235]{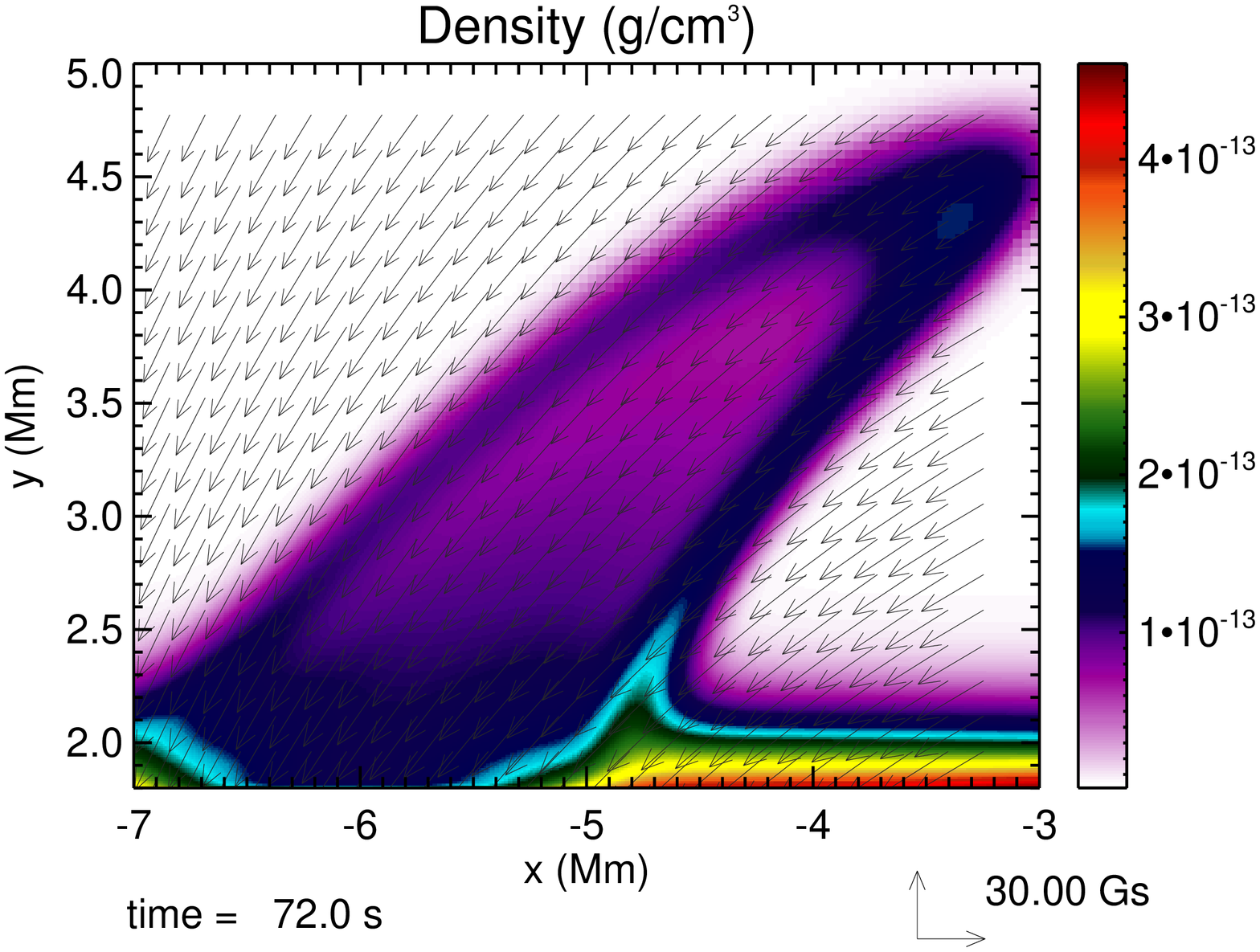} 

\vspace{-0.5 cm}
\caption{The mass density maps of the 2D VDC model at the for beginning phase of the system evolution. The sequence of images shows the situation which explains the temporal disappearance of the precursor observed in SXR. This precursor results from the emission of the uprising relatively dense and cold plasma, which initially is hot and dense enough to emit in the GOES range, but stretched and thus decompressed temporarily to disappear. The energy is deposited at the height of $h=1.24$ Mm above the temperature minimum.}
\label{Fig13}
\end{figure}

Global evolution of the main physical parameters of the plasma confined in the flaring loops course roughly in the same way in the 2D VDC and 1D models. During the early phase of the flare, for all 2D VDC and 1D models, the evaporating plasma rises towards the loop-tops and compresses the plasma contained in the upper parts of the loops, forming hot and dense loop-top X-ray bright sources. For the 2D VDC models the rising pillars of plasma are heterogeneous, with hot and dense cores and relatively cooler and rarefied ambient layers. Obviously, such important stratification of the plasma parameters across the loop cannot be modelled within a 1D framework. For the IMHD models the temperature differences across the loop are exaggerated by exclusion of thermal conduction, which prevents any energy transfers perpendicular to the loops axis. Thus, due to basic assumptions and limitations of the IMHD models, their properties are not consistent with observations and with the other models. The temperature of the evaporating plasma at the representative altitude of $h=1.5$ Mm above the transition region is roughly similar for the NRL model (about $\rm T_{NRL}=2.5$ MK) and in the axial part of the loop for the 2D IMHD model (about $\rm T_{IMHD}=1.3$ MK), but in the 2D VDC model the temperature is lower by one order, reaching merely $\rm T_{VDC}=0.1$ MK. These substantial differences in temperatures exerts a strong influence of a thermal conduction and viscose processes on the momentary and local plasma thermodynamic parameters during the evaporation stage of the flare. The hottest plasma in all models occurs in the compressed loop-top regions of the loops, being roughly similar in all models, i.e., $T_{\rm IMHD}=3.5$ MK in the IMHD model, $\rm T_{VDC}=1.5$ MK in the VDC model and $T_{NRL}=5.2$ MK in the NRL model. For both the 2D models, the calculated temperatures of the loop-top plasma are in good agreement with the temperatures derived from the calculated synthetic X-ray fluxes emitted by the modeled loops, \textit{i.e.} \textit{GOES}-like temperatures. However, the plasma densities in 2D models are one order lower than in the 1D model, causing substantial differences in timings of the maxima of the X-ray 1-8 \AA\ simulated \textit{GOES} soft channel emissions. The X-ray 1-8 \AA\ simulated emission peaks after 70 s in both 2D models and just after 120 s in the NRL model, these timings are given after the onset of the detectable soft X-ray emissions. During the late, gradual decay phase of the evolution, the loop-top sources expanded and the plasma flows down along the legs of the loops in the NRL model as well as along the central parts of the legs in the 2D models with the velocity of range $20-40$ km\,s$^{-1}$. However, in the 2D models the plasma in the outside layer of the loops still rises toward the top of the loops, which can not be modeled within a framework of a 1D model.

Due to over-simplified plasma hydrodynamics in the 1D model in which natural gradients of the physical parameters are absent across the loop, plasma reacts quicker on the abrupt heating, filling the whole loop by dense and hot matter, which results in the fast growth of the soft X-ray emissions. However, in the 2D model the increased local pressure pushes from below the overlaying layers of the relatively dense plasma, but the energy is not enough to heat plasma to high temperatures. The uprising motion of the pushed pillar of the plasma is initially so slow that a substantial volume of the dense plasma flows down along the outskirts of the loop back to the chromosphere.

In the case of a flare with the heating region located above the transition region, the global evolution of the main physical parameters of the plasma confined in the flaring loops also course in the same way in all the models, both 2D and 1D. Taking into account that in the 2D models the evaporating plasma is very hot in cores and much cooler in outside layers while in the 1D models the plasma is isothermal, the hottest evaporating plasma at a representative height of $h=1.5$ Mm above the transition region is noted in the IMHD model ($\rm T_{IMHD}=10$ MK), while for the VDC model the temperature is of the order of $\rm T_{VDC}=2.1$ MK only. In the case of the NRL model, the temperature of the plasma reaches an intermediate level of $\rm T_{NRL}=5.6$ MK. The significant differences of the calculated temperatures in various models are caused directly by differences in application or neglecting various physical mechanisms of the energy redistribution resulting from the presence of viscosity and thermal conduction; namely, exclusion of the thermal conduction results in strong local overheating of the loop core in the 2D IMHD model. For both IMHD and VDC 2D models we see massive up-flows of the evaporating plasma, while in the NRL model the plasma is still static, apparently due to simplified hydrodynamics of the plasma applied in the NRL model.

For all models with the heating region located above the TR the beginning of the noticeable X-ray emissions in the 1-8 \AA\ band occurs a few seconds after the beginning of the heating process only, peaking after 70-90 s in the 2D models and after 150 s in the NRL model. During the early phases of plasma evolution for all models the calculated 1-8 \AA\ fluxes rose in very similar way up to roughly $t=70$ s. Later on, the different scenario results from the differences in dynamics of the plasma motions and mass densities along the loops in the 1D and 2D models. For both 2D models the calculated temperatures of the loop-top plasma, $\rm T_{IMHD}=1.8$ MK and $\rm T_{VDC}=2.5$ MK, are much lower than the temperatures derived from calculated synthetic X-ray fluxes emitted by the modeled loops, \textit{i.e.} \textit{GOES}-like temperatures, $\rm T_{IMHD-GOES}=13$ MK and $\rm T_{VDC-GOES}=7$ MK, respectively. For the NRL model the temperature of the loop-top plasma at $t=150$ s is equal to $\rm T_{NRL}=11.5$ MK, being almost equal to the \textit{GOES}-like temperature of $\rm T_{NRL-GOES}=10.5$ MK.

\subsection{Differences between the IMHD and VCD models}

A comparative analysis of the IMHD and VDC models reveals an overall similarity of temporal variations of the most basic flare parameters, like: \textit{GOES} light curves, derived temperatures and emission measures. In the case of models witch are heated in the upper chromosphere, the IMHD model leads to high temperature apex volume and relatively cold upper parts and hot lower parts of the legs. The temperature of the loop-top for the VDC model is similar to that for the IMHD model, but the temperatures along whole legs are essentially homogeneous, due to effective thermal conduction. Temperature gradients for the IMHD model are inevitably very large, while these gradients for the VDC model are smoothed as a result of the thermal conduction. Numerous small-scale variations of the plasma temperature observed for the IMHD model at the legs of the flaring loop are not seen for the VDC model, being smoothed out efficiently by viscosity and thermal conduction. Comparing both models at the same evolution time, the volume of the apex X-ray source for the VDC model is smaller than for the IMHD model. Local mass density in the loop-top source is relatively low but with abrupt spatial variations for the IMHD model, while for the VDC model the mass density is higher, and all small-scale variations are smoothed by viscosity. These differences are particularly well seen for time  between $\rm t=60$ s and $\rm t=90$ s.

In the case of the models with heating above the TR temperature gradients in the IMHD model are inevitably very strong, while these gradients in the VDC model are smoothed as a result of thermal conduction action. Similar differences could be noticed also for mass densities, which shows abrupt local variations for the IMHD model but it has a relatively smooth distribution for the VDC model. The solely strong mass density variations for the VDC model during the late phase of the evolution occur on the forehead edge of the expanding apex kernel, where the matter is moving down. In both models the loop-top sources exhibit similar volumes.

\subsection{The applied model of plasma heating}

The applied model of local plasma heating mimics positions and limited volumes of plasma in the legs of the flaring loop, where the non-thermal electrons effectively deposit their energy in real flares. However, in our models energy is not deposited in any other part of the flaring loop, including also the loop-top region. Temporal variations of the deposited energy flux are modeled by variations of the local pressure amplitude in rough accordance, and at least qualitatively, with a typical time-profile of the hard X-ray emission observed for a typical solar flare.

Despite local variations of the main physical parameters of the plasma, like a temperature, actual local mass density etc. discussed above, and plasma parameters derived from simulated X-ray emissions of the flare, i.e. \emph{GOES}-like temperatures and emission measures, the most important difference between modeled and real solar flares is the higher brightness of the loop legs in X-ray than the brightness of the loop-top sources - quite the opposite of the real observations (see Fig.~\ref{Fig14}). The reason is the absence of a heating component of the loop-top source, originated from non-thermal electrons, which was not included in the simulation and low plasma mass density despite its relatively high temperature. This striking difference could be attributed to an absence of heating of the plasma deposited along the whole loop, especially in the loop-top region.  The modeled densities of the apex volumes are also relatively low, being, despite relatively high temperatures, not very efficient source of X-rays. However, the model also appropriately fits to the compact loop flares, where the flare is triggered near the loop's feet during low atmospheric reconnection in TR or inner corona, and heats the plasma to excite its evolution and multiwavelength emissions (e.g., EUV, SXR, \textit{etc.}) \citep[e.g.,][]{Kumar11}.

\begin{figure*}[t!]
\begin{center}
\includegraphics[angle=90,width=8.2 cm]{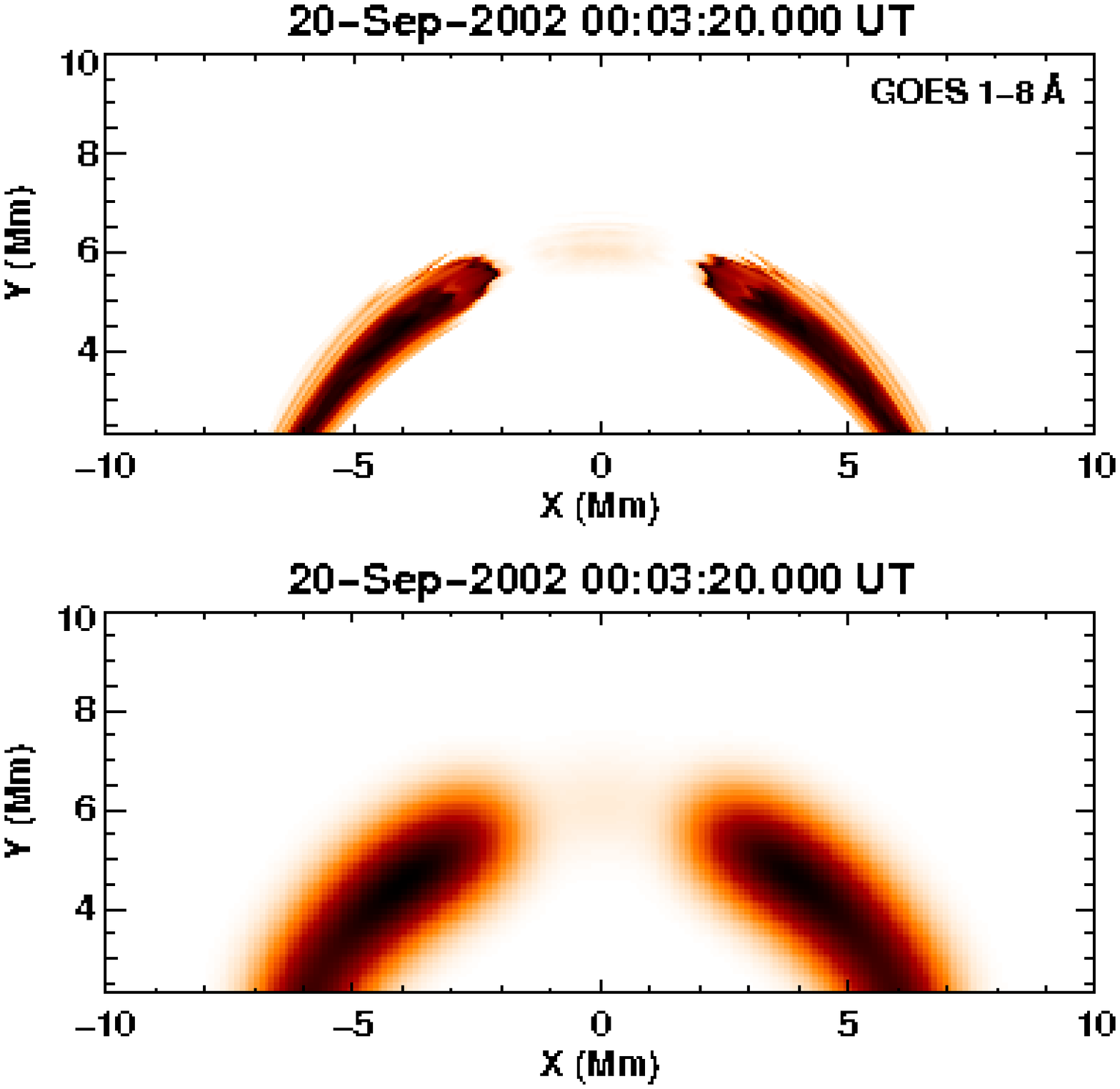} 
\includegraphics[angle=90,width=8.2 cm]{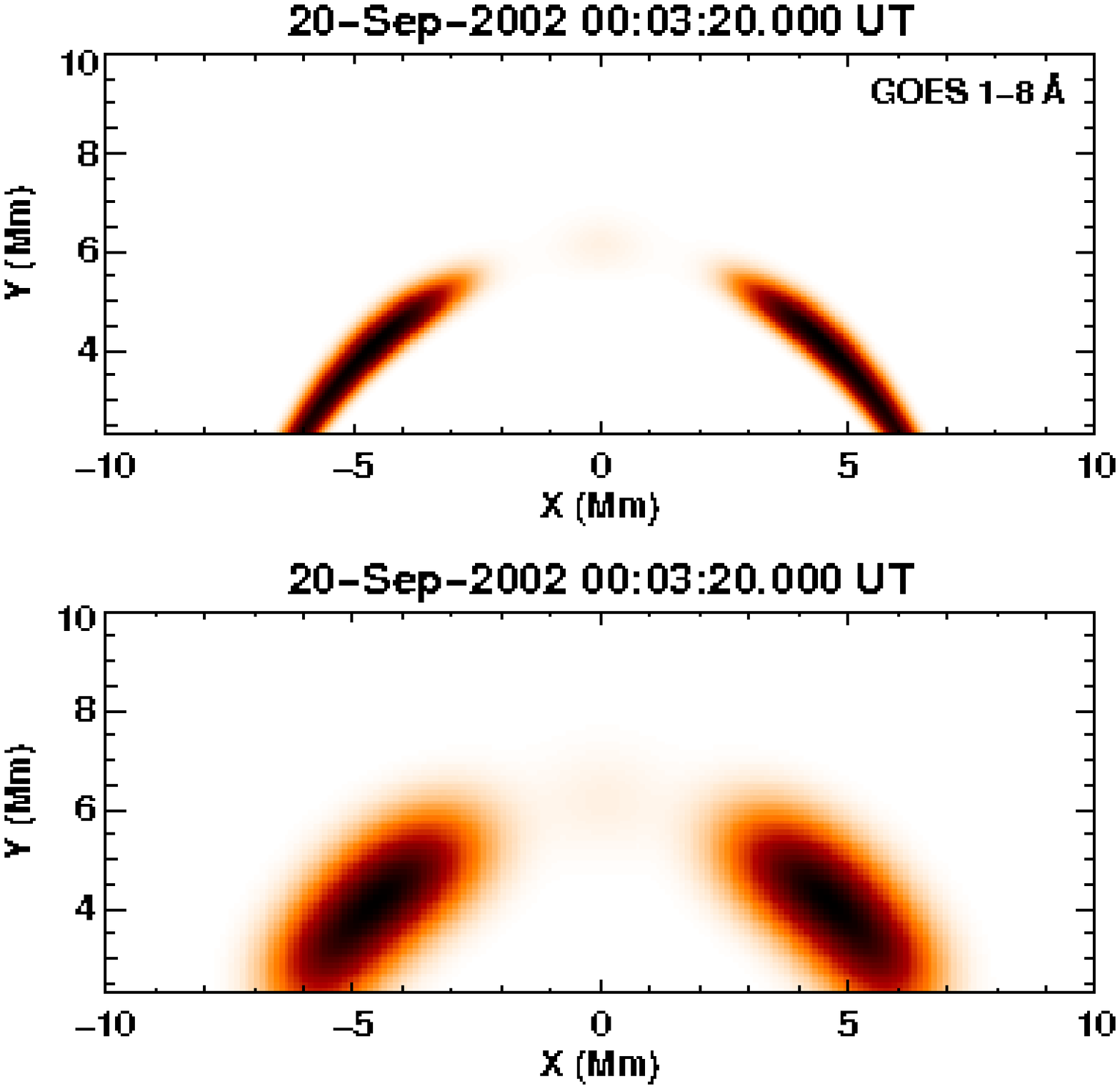} 
\end{center}
\vspace{-1.0 cm}
\caption{SXR emissions for 2D models in which the energy deposit area is at a height of $h=1.74$ Mm: for IMHD (on the left) and for VDC (on the right).The lower panels show the same images, however, blurred using Gauss function having half-width $FWHM=2.5$ arcsec (image comparable with the observations).}
\label{Fig14}
\end{figure*}

Specific spatial distribution of the applied heating model, being a reasonably good model of the spatial distribution of the energy flux inside a beam of the non-thermal electrons and automatically mimicking a multi-thread structure of the real flaring loops, causes that the core parts of the modeled loops to be usually hotter and denser than ambient regions. In the case of other specific spatial distribution of the heating energy, the resulting courses of the modeled flares could be substantially different.

\subsection{The role of waves in plasma heating}

The heating of the spatially limited volume of the plasma and during finite periods leads to the evolution of the heating pulses, we presented here as Gaussian. The implementation of such pulses in the full MHD model results in their propagation in the stratified atmosphere, and under suitable ambient conditions these thermal perturbations generate associated slow waves which can be dissipated by thermal conduction \citep{Ofman02}. It should be noted that the release of plasma heating results in perturbed gas pressure or temperature and mass density. The evolution of Alfv\'en waves are unlikely not allowed in the considered model. The combination of gas pressure and magnetic perturbations during the flare process may trigger fast magnetoacoustic-gravity waves. Within the given time scale of a few tens of seconds these waves leave the simulation domain, thus essentially not contributing to energy transfer on the given time scale of few hundred seconds. In the low $\beta$ plasma slow magnetoacustic waves are guided along magnetic field lines and they may contribute to the partial energy fulfillment of the localized plasma of the medium, where they evolve \citep[cf., theoretical papers by][]{De03,Dwi08}. For instance, in the present flaring loop system,  where the plasma pressure is perturbed by flare energy release, the evolution of slow magnetoacustic waves is likely along the magnetic field lines, and the presence of thermal conduction and viscosity results in their dissipation \citep{Dwi08} leading to the plasma heating. Under suitable conditions, some of these waves may steepen in the form of shocks, which propagate along the magnetic field lines \citep[cf.,][]{Sriv12}. Therefore, in conclusion, slow magnetoacoustic waves may dissipate and heat the plasma.

\subsection{Conclusions}

Numerous previous hydrodynamic simulations of flaring loops failed to reproduce high temperatures and mass densities typical for gradual phase of the solar flares. In particular, during the gradual phase of the flare a much slower decay of the SXR flux is usually observed than could be expected from analytical estimation of energy losses of the flaring plasma by radiation and by conduction \citep{Car95} or from HD numerical modelling of the flares \citep[e.g.,][]{Reale97}, proving that during this phase of the flares significant heating is also present. Similar conclusions were already drawn by \citet{McTier93} and \citet{Jiang06}. In order to solve this problem, \citet{Warr06} applied multi-thread, time-dependent 1D hydrodynamic simulations of solar flares as a multitude of tiny, parallel loops for investigations of the so-called Masuda flare which took place on January 13, 1992. He showed that modelling a flare as a sequence of independently heated threads instead of as a single loop may resolve the discrepancies between the simulations and observations, but by cost of some ad-hoc parameters (number of loops, energy share by particular loops etc.). Additionally, \citet{Fal11} and \citet{Fal14} showed that energy delivered by non-thermal electrons was fully sufficient to fulfill the energy budgets of the plasma during the pre-heating and impulsive phases of analyzed flares as well as during the decay phase. They concluded that in the case of the investigated flares there was no need to use any additional heating mechanisms other than heating by NTEs. These two results allows one to model a flaring loop as a macroscopic loop having a multi-thread internal structure and heated by a variable in time energy flux transported to the feet of the loop by NTEs only. However, the multi-thread internal structure of the flaring loop is poorly resolved by contemporary instruments, and there are no means to investigate individual heating episodes and thermodynamic evolution of the individual threads. These observational limitations cause fundamental problems in numerical modelling of multi-thread flaring loops, including a proper selection of a relevant structure and number of treads, \textit{i.e.} filling-factor of the observed macroscopic loop, as well as a proper model of spatial and temporal energy deposition in the threads.

In this work we analyzed thoroughly a broad set of 1D HD and 2D MHD models of a solar flare, derived  in order to compare and understand the energy transfer, plasma redistribution, and physical conditions of the various models of the flaring loops. Initial physical and geometrical parameters of the flaring loop, applied as a working model of the typical solar flare, are set according to the observations of a M1.8-flare recorded in the AR10126 active region on September 20, 2002 between 09:21 UT and 09:50 UT. The modeled loop was embedded into the realistic temperature model of the solar atmosphere, smoothly extended downward below the photosphere. The non-ideal 1D models include thermal conduction and radiative losses of the optically thin plasma as energy loss mechanisms, while the non-ideal 2D models includes viscosity and thermal conduction as energy loss mechanism only. The global evolution of the main physical properties and associated parameters of the plasma confined in the flaring loops occurred in the same way in all the models. The thorough comparison of the main physical characteristics of the 1D HD and 2D MHD models of the flare shows that the basic properties of the flaring plasma are acceptably well reproduced using both models, ensuring that results obtained already using a simplified 1D approach remain valuable for understanding the chief physical properties of the solar flares.

The IMHD models, often applied in modelling of the solar flares, apply very simplified approximation of the real processes occurring during the flares as well as of local and macroscopic properties of flaring plasma, if compared with much more realistic VDC models. Some non-realistic properties of plasma modelled in the IMHD models are thus the direct effect of basic assumptions and limitation of the IMHD methodology. However, IMHD models can be applied, with great success, for investigations of some problems, such as interactions of the waves in the loops having a non-stiff magnetic field or plasma-waves interactions.

We show that simplified 2D models, having a continuous distribution of the physical parameters of the plasma across the loop and powered by a heating flux variable in time as well as along and across the loop, are an extreme borderline case of a multi-thread internal structure of the flaring loop with a filling factor equal to one. Therefore such models might mimic to some extent the subtle structure and variations of the plasma parameters better than their 1D counterparts, revealing processes which are inherently absent in 1D models. However, the whole complexity in time and space of the processes occurring in the real multi-thread flaring loops is far beyond their scope.

In our approach used in the present work the 2D energy deposition region located inside the limited volume of the magnetic field automatically defines a multi-thread loop (or multi-thread arcade) in a pre-existing magnetic field, filled in various ways by evaporating plasma. In this approach, the only parameters to be selected/determined are the size of heating area (and spatial distribution of the energy flux in a frame of the source, if necessary) as well as the amount of delivered energy.}

\section{Acknowledgments}
The authors acknowledge the \emph{RHESSI} and \emph{SOHO} consortia for providing excellent observational data. The numerical simulations were carried out using resources provided by the Wroclaw Centre for Networking and Supercomputing (http://wcss.pl), grant No. 330. The software used in this work was in part developed by the DOE NNSA-ASC OASCR Flash Center at the University of Chicago. The research leading to these results has received funding from the European Community's Seventh Framework Programme ([FP7/2007-2013]) under grant agreement no. [606862] -- F-CHROMA Project (RF nad PR).


\end{document}